\documentclass[journal]{IEEEtran}
\usepackage[T1]{fontenc}

\usepackage{amsmath,bm}
\allowdisplaybreaks
\usepackage{amsfonts}
\DeclareMathOperator*{\argmin}{arg\,min}

\usepackage{cite}
\usepackage{xcolor}

\usepackage[normalem]{ulem}
\usepackage{graphicx}
\usepackage{booktabs}
\usepackage{subcaption}  
\usepackage{comment}
\usepackage{amsthm}
\newtheorem{assumption}{Assumption}
\newtheorem{lemma}{Lemma}
\newtheorem{corollary}{Corollary}
\newtheorem{theorem}{Theorem}
\newtheorem{definition}{Definition}

\newtheorem{remark}{Remark}

\usepackage{enumitem}
\usepackage{flushend} % balance the two columns on the last page
\allowdisplaybreaks
\newcommand{\jq}[1]{\textcolor{black}{#1}}
\newcommand{\nan}[1]{\textcolor{black}{#1}}

\usepackage[colorlinks = true,
            linkcolor = black,
            urlcolor  = black,
            citecolor = black,
            anchorcolor = black]{hyperref}
\ifCLASSINFOpdf
\else
\fi

\begin{document}
%
% paper title
% Titles are generally capitalized except for words such as a, an, and, as,
% at, but, by, for, in, nor, of, on, or, the, to and up, which are usually
% not capitalized unless they are the first or last word of the title.
% Linebreaks \\ can be used within to get better formatting as desired.
% Do not put math or special symbols in the title.
\title{Competitive Equilibrium for Electricity Markets with Spatially Flexible Loads}
%
%
% author names and IEEE memberships
% note positions of commas and nonbreaking spaces ( ~ ) LaTeX will not break
% a structure at a ~ so this keeps an author's name from being broken across
% two lines.
% use \thanks{} to gain access to the first footnote area
% a separate \thanks must be used for each paragraph as LaTeX2e's \thanks
% was not built to handle multiple paragraphs
%

\author{Nan Gu and
        Junjie Qin% <-this % stops a space
\thanks{N. Gu and J. Qin are with the Elmore Family School of Electrical and Computer Engineering, Purdue University, West Lafayette, IN 47907 USA. E-mail: \texttt{\{gu382, jq\}@purdue.edu}.}% <-this % stops a space
% (e-mail: jq@purdue.edu)
% \thanks{Manuscript received April 19, 2005; revised August 26, 2015.}
}

\maketitle

% As a general rule, do not put math, special symbols or citations
% in the abstract or keywords.
% \begin{abstract}
% Electric-vehicle charging and geo-distributed data centres can relocate demand, tightly coupling power, transportation, and computing networks and creating a bidirectional feedback loop between these agents' decisions and locational marginal prices (LMPs) in the electricity market.
% We model these agents as spatially flexible loads and extend classical competitive equilibrium to a Generalised Competitive Equilibrium (GCE) that that augments classical market clearing with cross-network constraints and an explicit price-to-load mapping for spatially flexible users.
% We derive compact conditions guaranteeing GCE existence, uniqueness, and welfare efficiency, thereby pinpointing when traditional competitive-equilibrium properties carry over to coupled systems. Stylized two-bus examples and large-scale case studies on an IEEE-118-bus grid linked to the Sioux Falls road network and the New York ISO grid with distributed datacenters confirm the theory and quantify welfare impacts. The framework offers a tractable recipe for analyzing and pricing spatial flexibility in emerging, interconnected infrastructures.
% \end{abstract}
\begin{abstract}
Electric vehicle charging and geo-distributed datacenters introduce spatially flexible loads (FLs) that couple power, transportation, and datacenter networks. These couplings create a \jq{closed-loop feedback} between locational marginal prices (LMPs) and  decisions of the FL systems, challenging the foundations of  conventional competitive equilibrium \jq{(CE)} in electricity markets. This paper \jq{studies a notion of} generalized competitive equilibrium (GCE) that \jq{aims to capture} such price-demand interactions across the interconnected infrastructures.  
We establish structural conditions under which the GCE preserves key properties \jq{of the conventional CE} including existence, uniqueness, and  \jq{efficiency}, without requiring \jq{detailed }knowledge of  decision processes within \jq{individual FL systems}. The framework generalizes to settings where the grid is coupled with multiple FL systems. Stylized examples and case studies on the New York ISO grid coupled with the Sioux Falls transportation and distributed datacenter networks 
\jq{demonstrate the use of our theoretical framework and illustrate the mutual influence among the grid and the studied FL systems.}
%validate the theory and demonstrate their mutual influence.
\end{abstract}

% Note that keywords are not normally used for peerreview papers.
\vspace{-0.04cm}

% For peer review papers, you can put extra information on the cover
% page as needed:
% \ifCLASSOPTIONpeerreview
% \begin{center} \bfseries EDICS Category: 3-BBND \end{center}
% \fi
%
% For peerreview papers, this IEEEtran command inserts a page break and
% creates the second title. It will be ignored for other modes.
\IEEEpeerreviewmaketitle

\vspace{-0.30cm}
\section{Introduction}
% The very first letter is a 2 line initial drop letter followed
% by the rest of the first word in caps.
% 
% form to use if the first word consists of a single letter:
% \IEEEPARstart{A}{demo} file is ....
% 
% form to use if you need the single drop letter followed by
% normal text (unknown if ever used by the IEEE):
% \IEEEPARstart{A}{}demo file is ....
% 
% Some journals put the first two words in caps:
% \IEEEPARstart{T}{his demo} file is ....
% 
% Here we have the typical use of a "T" for an initial drop letter
% and "HIS" in caps to complete the first word.
With the electrification of transportation and the rapid expansion of large-scale, energy-intensive computing infrastructures for Artificial Intelligence (AI), power systems are becoming increasingly complex and more tightly integrated with other sectors.  
In 2023, electric vehicle (EV) electricity consumption \nan{in the U.S.} surged to 7,596 GWh, nearly five times the level in 2018 \cite{ev-increase}.  
Simultaneously, datacenters accounted for approximately 4\% of U.S. electricity demand in 2022, a share projected to reach nearly 6\% by 2026 \cite{dc-increase}.
However, the impact of these new loads extends far beyond this sheer increase in demand. Their nature differs fundamentally from conventional demand, which is fixed at a single location. In contrast, both EVs and geo-distributed datacenters can shift their energy consumption across different locations and this distinctive characteristic defines them as \jq{\emph{spatially flexible loads}} (FLs).
% original: \jq{Furthermore, the collective FLs may depend on external infrastructure networks, such as the transportation network for EVs and compute/communication infrastructure for datacenters, leading to the notion of \emph{FL systems} that interconnect with the grid.  }
\jq{Furthermore, the decision makers behind these loads, such as EV drivers or datacenter operators, referred to as \emph{FL agents}, make decisions that may depend on their common external infrastructure networks, such as the transportation network for EVs and compute/communication infrastructure for datacenters, leading to the notion of \emph{FL systems} that interconnect with the grid.  }

\jq{With the growing prevalence of FLs, understanding their system-wide impacts is essential. In the classical setting with exogenous nodal demand, locational marginal prices (LMPs) at competitive equilibrium (CE) are the benchmark signals aligning private supply/demand choices with social welfare.  The existence, uniqueness, and efficiency of the classical CE are well understood. Spatial flexibility complicates this picture: (i) FLs re-site consumption in response to LMPs, and the resulting spatial load profile feeds back into those very LMPs; and (ii) FL systems often comprise many self-interested decision makers coupled through non-electric networks (e.g., transportation, communications). This raises two questions: \emph{Can CE be extended to capture this two-way coupling between the grid and FL systems? Under what conditions do existence, uniqueness, and efficiency persist?} %Addressing these questions calls for a unified framework that couples power system operation with the spatial load-shifting behavior of FLs such as electrified transportation and geo-distributed datacenters.
}

\subsection{Contributions and Paper Organization}

\jq{To address these questions, we develop a unified framework that couples power-system operation with the spatial load-shifting behavior of FLs such as electrified transportation and geo-distributed datacenters. We first introduce an abstract FL model and define the \emph{generalized competitive equilibrium} (GCE) for a wholesale electricity market coupled with an FL system (Section~\ref{sec:unifying_framework}), then analyze its structural properties (Section~\ref{sec:general_results}). We specialize the framework to EV charging (Section~\ref{sec:transportation}) and datacenter loads (Section~\ref{sec:datacenter}), and demonstrate a case where the grid is \emph{simultaneously} coupled to both FL systems (Section~\ref{sec:case_study}). Our contributions are:
\begin{enumerate}[label=\roman*), leftmargin=*]
\item \jq{A modeling framework for GCE that unifies spatial flexibility across heterogeneous FL decision models (non-atomic games, finite-player games, and centralized control) and accommodates coupling through external shared infrastructures (e.g., transportation, communication);}
% \item \jq{General conditions for existence, uniqueness, and efficiency of GCE, which translate into concrete, easily verifiable criteria for FL systems for EVs and datacenters;}
\item An exact characterization of the GCE through low-dimensional aggregate FL-system representations, abstracting away individual FL-agent decisions and yielding general conditions for existence, uniqueness, and efficiency, instantiated as provable efficiency bounds for FL systems for EVs and datacenters;
\item \jq{An extension to multiple, concurrent FL systems, providing (to our knowledge) the first rigorous tools to analyze the couplings among the grid and multiple FL systems in a single equilibrium framework.}
\end{enumerate}}

\vspace{-0.36cm}
\subsection{Related Literature}
% \jq{a) I think ``work'' is uncountable in this context. b) perhaps only cite and review papers that treat the FL \emph{system} + grid. For example, for EVs, focusing on those that model both transportation and grid. }
% The literature that jointly models FL systems and the power grid can be broadly classified into two application domains: EV charging and geographically distributed datacenter operation.
\nan{
%Existing studies typically model the interaction between the power grid and a specific class of FL systems, such as transportation or datacenter networks. 
\jq{Existing studies typically model the interaction between the power grid and a specific class of FLs, without accounting for the mutual impact these FL agents' decisions have on each other due to the external infrastructure network that they share. Very few studies treat the coupling between the grid and an  FL \emph{system} (mostly electrified transportation), offering theoretical insights into efficiency or uniqueness properties that are highly problem-dependent.}
%While a few studies offer theoretical insights into efficiency or uniqueness properties, these analyses remain highly problem-dependent and application-specific, offering limited generality.
Consequently, the literature lacks a unified theoretical framework that explains the structural coupling and equilibrium properties shared across different FL systems.} 
\jq{Market equilibrium with multiple, diverse FL systems connected to the grid simultaneously has not been studied. We  survey application-specific literature next.  }
% \begin{enumerate}[label=\roman*), leftmargin=10pt]
%   \item \textit{Coupled Transportation System and Power System}:

  Studies on power-transportation coupling generally fall into two categories.
  The first assumes a central coordinator, which minimizes combined transportation and power system costs \cite{zhang2020power}, guides EV routing to alleviate grid congestion \cite{guo2014rapid}, or reduces overall operational costs of EVs \cite{8737720}. 
  The second extends classical transportation models by considering a decentralized setting, where numerous EV owners individually select routes and charging locations to minimize travel costs, thereby influencing grid operation. 
  While some studies omit charging costs in route decisions \cite{8445618} or treat charging prices as exogenous retail rates \cite{7572952}, Alizadeh \textit{et al.}~\cite{alizadeh2016optimal} model EV charging within a competitive electricity market and capture the bidirectional coupling between LMPs and EV behavior, a setting conceptually closest to ours.
  This model is also applied in \cite{7967870,  9891825} to the distribution power system studies.
  Building on these models, subsequent research designs charging pricing mechanisms to reduce grid peaks \cite{tan2015real}, maximize charging station profitability \cite{9330803, 8304623}, or align user behavior with social welfare \cite{ 10644866, sheng2021coordinated, 9748107}.

  % \item \textit{Coupled Datacenter and Power Systems:} 
  Research on datacenter-power system coupling has largely been unidirectional.
  The first direction models the grid's impact on datacenters, focusing on how LMPs influence datacenter workload allocation across geographical locations to minimize electricity costs \cite{li2024towards, 6322266, 9963634} or carbon emissions \cite{hall2024carbon, 9770383}.
  The second direction models the datacenter's impact on the grid, particularly how spatial flexibility aids demand response \cite{wierman2014opportunities, fridgen2017shifting, 7517380}, or affects market outcomes \cite{ZHANG2020106723, zhang2022remunerating}.
% original: While a few recent studies address bidirectional interactions, they generally operate outside the framework of competitive wholesale markets.
% original: These studies employ price signals other than LMPs to model the datacenter's response, including Stackelberg game formulations \cite{10636017}, or auction-based market designs \cite{8540414} and flexibility pricing schemes \cite{werner2021pricing}.
  Recent studies address bidirectional interactions outside competitive wholesale markets. They model price-making datacenters \cite{brenner2026strategicdatacenterload}, grid-datacenter coordination \cite{10636017}, or consider alternative market designs, including auction-based mechanisms \cite{8540414} and flexibility-pricing mechanisms \cite{werner2021pricing}.
% \end{enumerate}

% Therefore, prior studies largely examine EVs and datacenters separately \jq{confusion: only study EVs or datacenters, v.s. studying individual EVs and datacenters. Note that the first meaning is somewhat unsuitable as the feature to handling different systems together is a small feature of ours, perhaps not the first thing to highlight.}, without a unified framework for embedding FLs into competitive electricity markets, which is the key focus of this work.
%\jq{this statement is not sharp enough. consider a few options: a) prior work does not treat EVs and datacenters as FL systems, b) it does not study the interactions between the grid and FL systems in a general way, and c) it does not examine the generalized CE concept with a focus on electricity market. 
% Also note that you are putting the literature section before the contribution section, which requires the statement here to be understandable without reading the contribution section. 

\section{\nan{Unified} Framework}
\label{sec:unifying_framework}
For a transmission network with $N$ buses and $M$ lines, we use $i\!\in\! \mathcal{N}\!:=\! \{1,\! \cdots,\! N\}$ to index the buses.
Our framework considers a competitive electricity market with three types of participants. Generators and FL agents are modeled as price takers responding to the LMPs, denoted as $\lambda_i$ at each bus $i$ and in vector form as $\boldsymbol{\lambda}\in \mathbb{R}^{N}$. The FLs collectively form an FL system, in which the decisions of individual FL agents impact each other through shared external infrastructure such as a transportation or datacenter network. Each generator seeks to maximize its own profit, and FL agents make decisions internal to the FL system, which induce the aggregate consumption of the FL system. The system operator (SO) determines the dispatch and the LMPs by solving the economic dispatch problem given the loads. This section also introduces equilibrium notions that formalize when the dispatch and LMPs determined by the SO are consistent with the self-interested behavior of the generators and the FL agents.

% original title (2026-09-07): \subsection{Models of Generators and Stationary Loads}
\subsection{Models of Generators and Loads}
\label{sec:gen-load}
We consider a power system in which each bus is equipped with a generator and a stationary load. 
The generation at bus \( i \) is denoted by \( g_i \), with feasible set \( \mathcal{G}_i \subseteq \mathbb{R} \) being non-empty, compact and convex. 
Let \( \mathbf{g} \in \mathbb{R}^N \) denote the vector of all generator outputs.
Each generator optimizes its output in response to the LMP \( \lambda_i \) at its bus. 
With a strictly convex cost function \( c_i(g_i) \), the cost-minimization problem for the generator at bus \( i \) is given by: % in Eq. (\ref{generator}).
%\jq{note the grammatical error in this sentence.}
\begin{equation}
   \min_{g_i \in \mathcal{G}_i} \   c_i(g_i) - \lambda_i g_i.
\label{uni-generator}%
\end{equation}%
\jq{We denote }the feasible set \jq{for the collection of generators by} \( \mathcal{G} := \{\mathbf{g} \in \mathbb{R}^N : g_i \in \mathcal{G}_i, \ i \in \mathcal{N}\} \), and the total generation cost \jq{by} \( C(\mathbf{g}) := \sum_{i \in \mathcal{N}} c_i(g_i) \).

% In parallel \jq{I would only use this phrase when the two cases connected by it is similar in terms of complexity. how about ``meanwhile''.}, 
Meanwhile, each bus also hosts a stationary load with consumption level \( \ell_i \), which is modeled to be inelastic; let \( \boldsymbol{\ell} \in \mathbb{R}^N \) denote the vector of all such loads. % original (2026-09-08): In addition, FLs draw power from the grid; we denote their aggregate net power consumption at the buses by $\mathbf{s}\in\mathbb{R}^{N}$, whose model is deferred to Section~\ref{mode:sfl}.
In addition, FLs' aggregate impact on the power grid is represented by their power consumption, which is denoted by a vector $\mathbf{s}\in\mathbb{R}^{N}$, whose model is deferred to Section~\ref{mode:sfl}.

% MOVED (2026-09-07): the Economic Dispatch subsection was moved up from after the SFL model, and Definition 1 with its discussion was moved into it from the former Subsection II-D, so that the electricity market given $\mathbf{s}$ is complete before the FL model is introduced.
\vspace{-0.27cm}
% original title (2026-09-07): \subsection{Economic Dispatch}
\subsection{Economic Dispatch and Classical Competitive Equilibrium}
% After generators submit their cost bids\jq{I'd avoid this word, as this is a level of detail that we didn't cover.}, 
The SO solves the following economic dispatch problem to determine both the generation dispatch and the resulting prices, minimizing generator costs:%
% \begin{subequations}
% \begin{align}
% \min_{\mathbf{g}\in \mathcal{G}, \mathbf{p}\in \mathcal{P} } \quad\  & C(\mathbf{g})\\
%  \mathrm{s.t.} \quad\ & \boldsymbol{\lambda}: \mathbf{p} = \mathbf{g} - \boldsymbol{\ell
% } -\mathbf{s},
% \end{align}
% \label{unify:economic dispatch}%
% \end{subequations}%
\begin{subequations}
    \label{unify:economic dispatch}
    \begin{alignat}{2}
    \min_{\mathbf{g}\in \mathcal{G}, \mathbf{p}\in \mathcal{P}}& \quad  && C(\mathbf{g}) \\
     \mathrm{s.t.} \quad &\quad  && \boldsymbol{\lambda}: \mathbf{p} = \mathbf{g} - \boldsymbol{\ell} - \mathbf{s}, 
     \label{ed:balance}
    \end{alignat}
\end{subequations}
% where \(\mathbf{p} \in \mathbb{R}^{N}\) represents the vector form of power injections at each bus, and \(\mathcal{P} \subset \mathbb{R}^{N}\) denotes the set of feasible power injections defined by network constraints\jq{do you have to introduce assumptions on this set?}.
{where \(\mathbf{p} \in \mathbb{R}^{N}\) represents the vector  of power injections at each bus. The feasible set \(\mathcal{P} \subset \mathbb{R}^{N}\) is defined under the standard DC power flow model \cite{4956966} as a convex and compact polyhedral set:
\(
    \mathcal{P} := \left\{ \mathbf{p} \in \mathbb{R}^N \; \middle| \; \mathbf{1}^\top \mathbf{p} = 0,\;-{\mathbf{f}}^{\max} \le \mathbf{H} \mathbf{p} \le {\mathbf{f}^{\max}} \right\},
\)
where the first condition ensures system-wide power balance, and the second enforces line flow limits using the shift-factor matrix \(\mathbf{H} \in \mathbb{R}^{M \times N}\), with \({\mathbf{f}^{\max}} \in \mathbb{R}^M\) denoting the vector of line capacity bounds.} 
% The notation \(:\) \jq{perhaps put this in quotation marks, or directly say $\bm \lambda$ is the dual variable for the constraint to avoid potential interpretation of ``The notation:'' } indicates that the vector of dual variables \(\boldsymbol{\lambda}\) is associated with the power injection constraint, and these dual variables equal the locational marginal prices. 
% The symbol ``\(:\)" indicates that \(\boldsymbol{\lambda}\) is the dual variable associated with the power injection constraint, and corresponds to the LMPs.
The dual variable \(\boldsymbol{\lambda}\) on the nodal power balance constraint corresponds to the LMPs.
% From the SO's perspective, both \(\boldsymbol{\ell}\) and \(\mathbf{s}\) are treated as \nannew{exogenous} demand. 
%, as they do not directly participate in the bidding process.
\begin{remark}[Demand Modeling in Economic Dispatch]
    From the SO's perspective, both $\mathbf{s}$ and $\boldsymbol{\ell}$ enter~\eqref{unify:economic dispatch} as 
{exogenous} demand.
For $\boldsymbol{\ell}$, this is a modeling choice rather than a restriction. Specifically, the load at each bus $i$ could alternatively be modeled as a market participant that maximizes utility net of electricity cost, i.e., $\max_{\ell_i \in \mathcal{L}_i} \; u_i(\ell_i) - \lambda_i \ell_i$,  where $\mathcal{L}_i$ denotes the feasible consumption set.
Correspondingly, the system-level economic dispatch objective can be written as $\min_{\mathbf{g}\in \mathcal{G},\, \boldsymbol{\ell}\in \mathcal{L},\,\mathbf p \in \mathcal P } \; C(\mathbf{g}) - \sum_{i\in\mathcal{N}} u_i(\ell_i).$
With $u_i(\cdot)$ concave under standard conditions, this modification alters the optimality conditions through the derivative of the utility terms but does not fundamentally change the structure of the general results developed below.
For $\mathbf{s}$, however, there is no widely adopted or standardized mechanism that maps FL behavior, potentially involving many heterogeneous FL agents, directly into bidding decisions observable by the SO, without introducing additional layers of market design, aggregation, or coordination.
To focus on the fundamental interaction between the FL systems and the electricity market (rather than the design of their interfaces),
we therefore treat $\mathbf s$ as  inelastic from the SO's perspective and taken as given in the economic dispatch and LMP calculation.
\end{remark}

To formally characterize the market outcome when the FL load is treated as exogenous, we introduce the following definition of \jq{the} \emph{classical CE}.

\begin{definition}[Classical CE given FL Load]\label{def:CE_fixed_s}
Given  aggregate FL power consumption
%an FL vector 
$\mathbf{s}\in\mathbb{R}^N$, a pair $(\mathbf{g}^\star,\boldsymbol{\lambda}^\star)$ constitutes a {classical CE} if the following conditions are satisfied:
\begin{enumerate}[label=\roman*), leftmargin=*]
    \item {Individual rationality for generators:} 
    % \jq{I haven't heard this term before. I would say ``Individual rationality for generators''} 
    For every bus $i\in\mathcal{N}$, the generation decision 
    $g_i^\star$ solves
    \[
        \min_{g_i\in\mathcal{G}_i}\; c_i(g_i)-\lambda_i^\star g_i .
    \]
    \item {Feasibility:} The resulting net injection vector
    $\mathbf{p}:=\mathbf{g}^\star-\boldsymbol{\ell}-\mathbf{s}$ is feasible, satisfying network constraints $\mathbf{p} \in \mathcal{P}$.
    \item {Price optimality:} The price vector $\boldsymbol{\lambda}^\star$ is a dual-optimal solution to the economic dispatch problem~\eqref{unify:economic dispatch} given  $\mathbf{s}$.
%     , i.e., it reflects the marginal cost of supplying power at each bus. %\jq{I can't reason this through being somewhat dizzy on my delayed flight, but I don't think I have seen this condition previously. Is this redundant? Could you please check my paper on coordinated trading with Pravin?}
\end{enumerate}
%\nanre{Conditions (i) and (ii) alone are insufficient to characterize a competitive equilibrium, as any feasible point can satisfy them for some corresponding \(\boldsymbol{\lambda}\), potentially leading to incorrect identification of equilibrium. The version in \cite{qin2017flexible} is as follows:}
%%-----------------------------------------------------------
%\nan{\begin{enumerate}[label=\roman*), leftmargin=*]
%    \item {Individual rationality for generators:} 
%    % \jq{I haven't heard this term before. I would say ``Individual rationality for generators''} 
%    For every bus $i\in\mathcal{N}$, the generation decision 
%    $g_i^\star$ solves
%    \[
%        \min_{g_i\in\mathcal{G}_i}\; c_i(g_i)-\lambda_i^\star g_i .
%    \]
%    \item {ISO optimality:}  
%    The ISO is modeled as a firm that converts injections at one bus into withdrawals at others using the power network. Given prices $\boldsymbol{\lambda}^\star$, the ISO chooses a power injection vector $\mathbf{p}^\star$ by solving:
%    \[
%        \max_{\mathbf{p} \in \mathcal{P}} \; -{\boldsymbol{\lambda}^{\star\top}} \mathbf{p}.
%    \]
%    \item {Market clearing:}  
%    The resulting net injection vector 
%    \[
%        \mathbf{p}^\star := \mathbf{g}^\star - \boldsymbol{\ell} - \mathbf{s}
%    \]
%    coincides with the ISO's optimal injection $\mathbf{p}^\star$ and lies in the feasible set $\mathcal{P}$.
%\end{enumerate}}
\end{definition}
Definition~\ref{def:CE_fixed_s} shows that the LMPs $\boldsymbol{\lambda}^\star$ can be seen in two ways. They are determined by the SO as the dual optimal solution of \eqref{unify:economic dispatch}, which by strong duality also guarantees that no price-taking generator has an incentive to deviate from the SO's dispatch $\mathbf{g}^\star$. They also emerge as equilibrium prices under which supply balances demand at every bus.
%\nanre{I believe the two definitions are equivalent: KKT conditions for the ISO's optimization also yield the correct LMP components, rather than just any dual associated with a feasible point. The first definition may be easier to interpret as it directly links the economic dispatch formulation and how LMPs are computed?}

\subsection{Model of Spatially Flexible Loads}
\label{mode:sfl}
Unlike stationary loads, FLs can shift their electricity consumption across different geographical locations in response to LMPs.
Examples include EVs, which choose routes and charging sites, and datacenters, \jq{whose users may} migrate computational tasks \jq{among} facilities.
% This inherent mobility couples the power grid with external infrastructures, such as transportation or datacenter networks, referred to as the \emph{FL system}.
% MOVED to the Section II opening paragraph (2026-09-07): \nan{These FLs collectively form an \emph{FL system}, \jq{in which decisions of individual FL \blue{agents} impact each other through their impact on shared external infrastructure such as a transportation or datacenter network.}
\nan{
%which is an external infrastructure such as a transportation or datacenter network, that connects to the power grid and thereby couples the two systems.
We focus on a single FL system coupled with the grid in this section, with the extension to multiple FL systems discussed in Section~\ref{sec:general:nfl}.}
%\jq{need to make it clear that FL system is the collection of FL loads that are coupled via an external network system and then connected to the grid; need to make it clear that for now we focus on the case of a single FL system with the case of multiple FL system deferred to section XXX. }
Let $\mathbf{x}\! \in\! \mathcal{X}\! \subset\! \mathbb{R}^{N_\mathrm{L}}$ denote the vector of decisions internal to the FL system, where
the feasible set $\mathcal{X}$ is a convex and compact set that encodes the physical operational limits of the FL system and ${N_\mathrm{L}}$ specifies the decision space dimensionality. For simplicity, we refer to $\mathbf x$ as FL decisions subsequently. 

% original (2026-09-07): The FLs' aggregate impact on the power grid is represented by the vector of net power consumption $\mathbf{s} \in \mathbb{R}^{N}$. 
% MOVED to Section II-A (2026-09-08): The FLs' aggregate impact on the power grid is represented by the vector of net power consumption $\mathbf{s} \in \mathbb{R}^{N}$ introduced in Section~\ref{sec:gen-load}.
To have a general yet tractable model for the FL power consumption $\mathbf{s}$, we make several simplifying assumptions next. Despite  appearing to be restrictive, we will demonstrate that the resulting model can encapsulate important FL systems, including EVs and datacenters.
\begin{assumption}[Linear Decision-Load Mapping]
    \label{ass:linear-map}
The mapping from FL decisions $\mathbf{x}$ to their aggregate grid-level power consumption vector $\mathbf{s}$ is {linear}, modeled as:
\begin{equation}
    \mathbf{s} = \mathbf{A}^\mathrm{FL} \mathbf{x},
    \label{eq:linear-map}
\end{equation} 
for some fixed  $\mathbf{A}^\mathrm{FL} \in \mathbb{R}^{N \times N_\mathrm{L}}$.
Denote the set of all feasible FL load vectors by
\(
    \mathcal{S} := \{ \mathbf{s} \in \mathbb{R}^N \mid \mathbf{s} = \mathbf{A}^\mathrm{FL}\mathbf{x},\ \mathbf{x} \in \mathcal{X} \}.
\)
\end{assumption} 
\begin{remark}[Interpreting $\mathbf{x}$ and Justifying Assumption~\ref{ass:linear-map}]
The vector $\mathbf{x}\in\mathcal{X}$ collects the \emph{internal} decision variables of the FL system, i.e., the representation of how flexible demand is allocated across its own network (e.g., route/charging choices in a transportation network, or workload allocation across datacenter sites).
Each component $x_k$ is a \emph{continuous} variable that measures the amount or intensity of decision option $k$; it may be inherently continuous or arise as a continuous abstraction or relaxation of an underlying physical quantity such as counts or flows.
The matrix $\mathbf{A}^{\mathrm{FL}}$ encodes how these internal decisions map into \emph{nodal} net consumption as observed by the power system.
The linear form~\eqref{eq:linear-map} reflects a general aggregation principle: each decision option $x_k$ is geographically associated with a specific bus at which it draws power, while multiple options may aggregate at the same bus. 
The coefficient $[\mathbf{A}^{\mathrm{FL}}]_{ik}$ captures/approximates the fixed or average power requirement~\cite{Desislavov_2023} per unit of $x_k$ at bus $i$.
Heterogeneous power requirements can be accommodated either by allowing these coefficients to vary across options or by refining the decision set $\mathbf{x}$ into more granular categories.
In Sections~\ref{sec:transportation} and~\ref{sec:datacenter}, we specify $\mathbf{x}$ and $\mathbf{A}^{\mathrm{FL}}$ explicitly for EVs and datacenters.
\end{remark}
% The form of \(\mathbf{A}^\mathrm{sfl}\) is specified in Sections \ref{sec:transportation} and \ref{sec:datacenter}. 

Furthermore, we stipulate that there exists a function $\Phi(\mathbf x)$, referred to as the \emph{collective preference/disutility function for the FL system}, that ranks different FL decisions $\mathbf x$ without accounting for the cost of electricity, with a lower $\Phi$ value  being better. 
% original: Notably, the function $\Phi(\mathbf x)$ may not simply be the sum of the cost/disutility of individual agents in the FL system when multi-agent settings are considered due to their strategic interactions; see Sections~\ref{sec:transportation} and \ref{sec:datacenter} for details.
% original pointer (2026-09-07): ; see Sections~\ref{sec:transportation} and \ref{sec:datacenter} for details.
Notably, the function $\Phi(\mathbf x)$ may not simply be the sum of the cost/disutility of individual FL agents in the FL system due to the externalities among self-interested agents; see Remark~\ref{rem:phi}.
Then, the following model characterizes the aggregate FL behavior given the LMPs. 
\begin{assumption}[Convex Best Response Model]\label{ass:SFL-specific cost}
%    Define an {SFL-specific cost function} that characterizes SFL system preferences, $\Phi(\mathbf{x})$.
%Given a vector of LMPs $\boldsymbol{\lambda}$, the {external cost of electricity} is $\boldsymbol{\lambda}^{\top}\mathbf{A}^\mathrm{sfl}\mathbf{x}$.
The FL system's best response   is to trade off between its internal preference/disutility and the resulting electricity cost:
\begin{equation}
    \mathbf{x}^{\star}(\boldsymbol{\lambda}) = \argmin_{\mathbf{x} \in \mathcal{X}} \left\{ \Phi(\mathbf{x}) + \boldsymbol{\lambda}^\top \mathbf{A}^\mathrm{FL}\mathbf{x} \right\}.
    \label{eq:sfl-response}
\end{equation}
Further, we assume that $\Phi$ is a strictly convex function over the non-empty, compact, and convex set $\mathcal X$. 
\end{assumption}
% \vspace{-0.48cm}
\begin{remark}[Interpreting $\Phi(\mathbf{x})$ and Justifying Assumption~\ref{ass:SFL-specific cost}]
\label{rem:phi}
The optimization form in~\eqref{eq:sfl-response} is a behavioral representation and does {not} necessarily require the existence of a central controller.
Its general form can encapsulate two distinct classes of settings:
(i) {Centralized Coordination}, where a coordinator directly minimizes a system-level FL objective, 
and (ii) {Decentralized Equilibrium}, where many self-interested FL agents interact (through mechanisms like competition for shared resources) and the induced Nash equilibrium coincides with the minimizer of a potential function $\Phi(\mathbf{x})+\boldsymbol{\lambda}^\top\mathbf{A}^{\mathrm{FL}}\mathbf{x}$ (non-potential cases are addressed in Appendix~\ref{app:non-potential}).
In both settings, $\Phi(\mathbf{x})$ encodes preferences that are internal to the FL system and independent of electricity prices.
The term $\boldsymbol{\lambda}^\top\mathbf{A}^{\mathrm{FL}}\mathbf{x}$  represents the electricity payment by the FL system to the grid.
Sections~\ref{sec:transportation} and~\ref{sec:datacenter} detail $\Phi(\mathbf{x})$ for specific sectors.
\end{remark}

While the above model defines the FL system's internal choices, the SO only observes the aggregate power consumption, $\mathbf{s}$. 
Given our grid focus, it is ideal to abstract away internal details of the FL system and create a more succinct model of the FL system  that does not explicitly depend on $\mathbf x$. 
%To create a useful model for the grid, if possible, we must abstract away the internal details of $\mathbf{x}$.
To this end, we define $J(\mathbf s)$ as a value function that maps any given aggregate FL power consumption $\mathbf s$ to the minimum FL preference value $\Phi$ given $\mathbf s$. Mathematically, 
\begin{equation}
J(\mathbf{s}) := \min_{\mathbf{x} \in \mathcal{X}} \left\{ \Phi(\mathbf{x}) \quad \mathrm{s.t.} \quad \mathbf{s} = \mathbf{A}^\mathrm{FL} \mathbf{x} \right\}.
\label{j-phi-re}
\end{equation}

Under Assumption~\ref{ass:SFL-specific cost}, the value function $J(\mathbf{s})$ is also strictly convex over its domain. 
As a result, we can formulate the aggregate FL power consumption profile $\mathbf s$ as a function of the LMPs $\bm \lambda$, denoted by $\bm \sigma: \mathbb R^N \mapsto \mathbb R^N$, taking the form of
\begin{equation}
\boldsymbol{\sigma}(\boldsymbol{\lambda}) := \argmin_{\mathbf{s} \in \mathcal{S}} \left\{ J(\mathbf{s}) + \boldsymbol{\lambda}^{\top}\mathbf{s} \right\}. 
\label{value:sfl}
\end{equation}
This function $\boldsymbol{\sigma}(\boldsymbol{\lambda})$ is the key interface between the FL system and the power grid.

\vspace{-0.24cm}
% original title (2026-09-07): \subsection{{Classical and Generalized Competitive Equilibrium}}
\subsection{Generalized Competitive Equilibrium}
\begin{figure}
    \centering
    \includegraphics[width=1\linewidth]{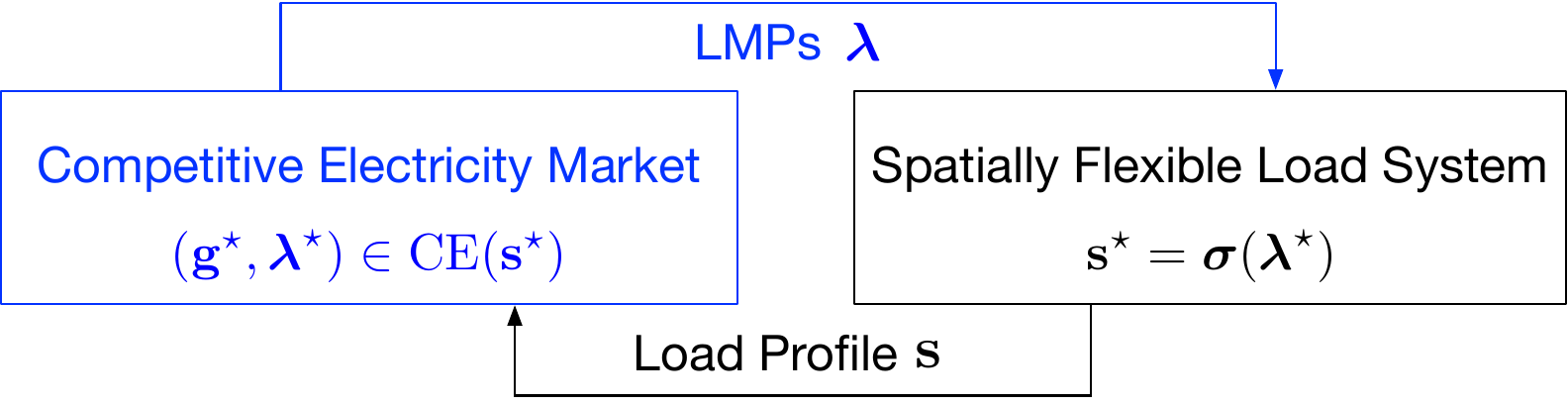}
    \caption{Closed-loop feedback between the grid and FL system. \jq{See Section~\ref{sec:general:nfl} for the extension to multiple FL systems.}}
    \label{fig:GCE}
     \vspace{-1em}
\end{figure}
% \subsection{Generalized Competitive Equilibrium for the Coupled System}
% \label{sec:gce_definition}
% In contrast to CE, where the SFL profile is treated as exogenous, we now endogenize this component by incorporating its dependence on prices. A tuple \((\mathbf{g}^*, \mathbf{s}^*, \boldsymbol{\lambda}^*)\) constitutes a generalized competitive equilibrium (GCE) if the following conditions are satisfied:
% \begin{enumerate}[leftmargin = *, label=\alph*)]
%     \item \textit{CE Conditions:} The pair \((\mathbf{g}^*, \boldsymbol{\lambda}^*)\) satisfies the conditions of a CE under the given spatial load profile \(\mathbf{s}^*\);
%     \item \textit{Spatial Load Equilibrium:} The SFL profile \(\mathbf{s}^*\) represents the equilibrium aggregate response to the price vector \(\boldsymbol{\lambda}^*\), that is, \(\mathbf{s}^* = \boldsymbol{\sigma}(\boldsymbol{\lambda}^*)\).
% \end{enumerate}
% This fixed-point condition couples the power system optimization problem with the collective behavior of spatially-flexible consumers, closing the loop between price formation and responsive demand.
%-----------------------------------------------------------
% Motivation sentence (placed right before the definition):
To capture the closed-loop feedback between the grid and FL system \nan{in Fig. \ref{fig:GCE}}, we extend the classical CE as follows. %\jq{loads}

\begin{definition}[Generalized CE]\label{def:GCE}
A triple $(\mathbf{g}^\star,\mathbf{s}^\star,\boldsymbol{\lambda}^\star)$ is said to constitute a {GCE} if
\begin{enumerate}[label=\roman*), leftmargin=*]
    \item %{Classical CE condition:}  
          The pair $(\mathbf{g}^\star,\boldsymbol{\lambda}^\star)$ constitutes a classical CE given $\mathbf{s}^\star$. 
    \item %{Best Response of FL:}  
    % \jq{trying to see where does this adjective comes from and how to interpret it}
          The FL power consumption \(\mathbf{s}^
          {\star}\) represents the aggregate best response to the LMPs, i.e.,
          \(
              \mathbf{s}^{\star} \;=\; \boldsymbol{\sigma}\bigl(\boldsymbol{\lambda}^\star\bigr).
          \) %\jq{star v.s. asterisk, try to be consistent (note this issues show up later too).}
\end{enumerate}
\end{definition}
%-----------------------------------------------------------

{CE in electricity markets is well-studied and is known to exist under standard convexity assumptions. Moreover, it yields a socially optimal allocation and enjoys uniqueness under appropriate conditions \cite{hsu1997introduction, 4111733}.} 
Building on these classical insights, \jq{we aim} to investigate whether\jq{/when} these desirable properties extend to the GCE framework, which is formalized and analyzed in the following sections.\footnote{The GCE rests on the price-taking premise of the competitive equilibrium setting, in which each FL agent's load is assumed to be small relative to the total system demand so that it lacks market power. If an FL agent instead anticipated how its own consumption affects the LMPs, it would withhold or relocate load strategically to lower the prices it pays. The interaction then becomes a Stackelberg game with the FL agents as leaders and the SO as follower, of Cournot type when several FL agents anticipate prices, e.g., \cite{brenner2026strategicdatacenterload}.}
% \jq{note that this is the only discussion for subsection D and E. shall we merge these two subsections?}

% \input{1-general-model}
% \input{1-individual-model}
\section{General Results}
\label{sec:general_results}
\jq{In this section, we work with general FL systems to establish properties of the GCE. We start by examining the setting with a single FL system in Section~\ref{sec:general:1fl} and then address the case where multiple FL systems are simultaneously connected to the grid in Section~\ref{sec:general:nfl}. }
% \blue{Understanding the structure of the GCE matters for the planning and operation of the coupled infrastructure systems. Existence certifies that the closed loop in Fig.~\ref{fig:GCE} can settle at a consistent outcome rather than oscillating. Uniqueness makes the equilibrium predictable, allowing system operators and urban planners to anticipate the coupling between the power grid and other infrastructures, and renders the subsequent comparison with the social welfare benchmark well posed.}
Understanding the structure of the GCE matters for the planning and operation of the coupled infrastructure systems. Existence certifies that the closed loop in Fig.~\ref{fig:GCE} admits a load profile and LMPs that are mutually compatible. Uniqueness makes the equilibrium predictable, allowing system operators and urban planners to anticipate the coupling between the power grid and other infrastructures, and renders the subsequent comparison with the social welfare benchmark well posed.
Building on the model in Section~\ref{sec:unifying_framework}, 
\emph{Assumptions \ref{ass:linear-map} and \ref{ass:SFL-specific cost} are in force for all results in this section.}

\vspace{-0.24cm}
\subsection{Grid Coupled with Single FL System}\label{sec:general:1fl}
We begin by establishing \jq{a structural characterization of} the GCE, which allows analysis of its existence and uniqueness. 
\begin{theorem}[{Structural Characterization for GCE}]\label{thm:GCE-Structural}
  A triple $(\mathbf{g},\mathbf{s},\boldsymbol{\lambda})$ is a
GCE if and only if it is a primal-dual optimal solution to the following optimization:
\begin{equation}\label{prob:P_GCE}
  \tag{$\mathrm{P}_{\mathrm{GCE}}$}\! 
  \begin{aligned} 
  \min_{\mathbf{s}\in\mathcal{S},\,\mathbf{g}\in\mathcal{G},\,\mathbf{p}\in\mathcal{P}}\quad 
  & J(\mathbf{s}) + C(\mathbf{g}) \\  
  \mathrm{s.t.}\quad \quad\ \ & \boldsymbol{\lambda}: \mathbf{p} = \mathbf{g} - \boldsymbol{\ell} - \mathbf{s}. 
  \end{aligned} 
  \end{equation}
\end{theorem}
Theorem~\ref{thm:GCE-Structural} indicates that the closed-loop feedback between the grid and the FL system can be characterized by a convex optimization problem, and
the equilibrium price \(\boldsymbol{\lambda}^\star\) coincides with the Lagrange multiplier of the nodal power balance constraint in that optimization. 
Under the scenario when \(J(\mathbf{s})\) is available or can be learned from data, \eqref{prob:P_GCE} allows us to abstract away individual decisions from FL agents while preserving their aggregate response to \nan{LMPs}. 
% \jq{``states'' is unclear; change to LMPs?}
% This formulation captures the collective behavior of a large population of FL through a low-dimensional, system-level representation.
This formulation captures the collective behavior of many FL agents of a single type within one FL system, through a low-dimensional, system-level representation.
If $J(\mathbf{s})$ carries an approximation error, its impact on the GCE outcome is quantified in Corollary~\ref{cor:approx-gce} in Appendix~\ref{app:sec3-proofs}.

Leveraging Theorem \ref{thm:GCE-Structural}, we establish the existence and uniqueness of the GCE under the stated assumptions.
\begin{corollary}[Existence and Uniqueness of GCE]\label{thm:gce_existence}
Define the feasible set for independent decision variables
  \(
  \mathcal{F}\!:=\!\left\{(\mathbf{s},\mathbf{g})\!\in\!\mathcal{S}\!\times\!\mathcal{G}:\, \exists\,\mathbf{p}\!\in\!\mathcal{P}\, \mathrm{s.t.}\, \mathbf{p}=\mathbf{g}-\boldsymbol{\ell}-\mathbf{s}\right\}.
  \)
  If $\mathcal{F}\!\neq\!\emptyset$, a GCE exists.
  % and the primal solution $(\mathbf{g}^\star,\mathbf{s}^\star)$ is unique; 
  % previous (2026-09-08): \nan{Furthermore, if linear independence constraint qualification \blue{\cite{nocedal2006numerical}} holds for \eqref{prob:P_GCE}, then the GCE is unique.}
  \nan{Furthermore, if linear independence constraint qualification (LICQ)\footnote{LICQ holds if the set of gradients of the constraint functions active at the optimal solution is linearly independent \cite[Ch.~12, Def.~12.4]{nocedal2006numerical}.} holds for \eqref{prob:P_GCE}, then the GCE is unique.}
  % \jq{shall we just say ``Furthermore, if LICQ holds for \eqref{prob:P_GCE} then the GCE is unique''? the goal here is not talking about the property of the optimization, which should be in the proof, rather is about the GCE (and its components) exists and is unique. }
% Moreover, the price vector $\boldsymbol{\lambda}^\star$
% is unique whenever the linear independence constraint qualification holds for the active constraints defining $\mathbf{p}^\star = \mathbf{g}^\star - \boldsymbol{\ell} - \mathbf{s}^\star$.
\end{corollary}

Given the existence and the uniqueness, we examine the efficiency of the GCE, i.e., whether it supports the social welfare. 
To that end, we define the following \emph{social welfare maximization (SWM) benchmark}.
% Beyond this abstract FL system model, we define the following \emph{social welfare optimization benchmark}. 
In particular, let $\Phi^\mathrm{SW}(\mathbf x)$ denote the \emph{sum of individual disutility functions} of all FL agents in the FL system, which may or may not be the same as $\Phi(\mathbf x)$, especially when FL agents act selfishly.
Then, the SWM can be written as the following optimization problem:
\begin{equation}
  \label{eq:swm_ori}
    \begin{aligned}
    \min_{\mathbf{x}\in \mathcal{X},\,\mathbf{g}\in\mathcal{G},\,\mathbf{p}\in\mathcal{P}}\quad
          & \Phi^\mathrm{SW}(\mathbf{x})+ C(\mathbf{g}) \\
    \mathrm{s.t.}\quad \quad \ & \mathbf{p} = \mathbf{g} - \boldsymbol{\ell} - \mathbf{A}^\mathrm{FL} \mathbf{x},
    \end{aligned}
  \end{equation}
where we optimize the disutility/cost of both the FL system and the power system in a centralized manner.
% To formulate SWM from an ISO's perspective, we can create a more succinct model that replace FL's internal choices by the aggregate FL power consumption $s$:
To facilitate comparison between GCE and SWM, we obtain a characterization of SWM analogous to that of GCE in Theorem~\ref{thm:GCE-Structural}:
%to \eqref{prob:P_GCE}:
\begin{lemma}[{Structural Characterization for SWM}]\label{thm:GCE-SWM}
Let $(\mathbf{s}^\star, \mathbf{g}^\star,\mathbf{p}^\star)$ be an optimal solution to:
\begin{equation}\label{prob:P_SW}
\tag{$\mathrm{P}_{\mathrm{SWM}}$}\!
  \begin{aligned}
  \min_{\mathbf{s}\in \mathcal{S},\,\mathbf{g}\in\mathcal{G},\,\mathbf{p}\in\mathcal{P}}\quad
        & J^{\mathrm{SW}}(\mathbf{s}) + C(\mathbf{g}) \\
  \mathrm{s.t.}\quad \quad\ \ \ & \mathbf{p} = \mathbf{g} - \boldsymbol{\ell} - \mathbf{s},
  \end{aligned}
\end{equation}
where
%defining $J^\mathrm{SW}(\mathbf s)$. 
\(
J^\mathrm{SW}(\mathbf{s}) := \min_{\mathbf{x} \in \mathcal{X}} \left\{ \Phi^\mathrm{SW}(\mathbf{x}): \mathbf{s} = \mathbf{A}^\mathrm{FL} \mathbf{x} \right\}.
%\label{j-phi-re}
\)
Then there exists an $\mathbf{x}^\star$ such that ${A}^\mathrm{FL} \mathbf{x}^\star = \mathbf{s}^\star$, where  $(\mathbf{x}^\star, \mathbf{g}^\star,\mathbf{p}^\star)$ is an optimal solution to \eqref{eq:swm_ori}.
\end{lemma}
%$J^\mathrm{SW}(\mathbf{s})$ is defined analogous to \eqref{j-phi-re}, which is the minimum $\Phi^\mathrm{SW}(\mathbf x)$ value that can be achieved given aggregate power consumption $\mathbf s$.
% Since the outcomes of the GCE and the SWM can be characterized as the solutions to two related, yet distinct, central optimization problems.
%The following corollary provides the necessary and sufficient conditions under which a GCE aligns with the welfare-optimal allocation.
Comparing \eqref{prob:P_GCE} and \eqref{prob:P_SW} in Theorem~\ref{thm:GCE-Structural} and Lemma~\ref{thm:GCE-SWM}, we obtain the following {necessary and sufficient} condition for the GCE outcome to support social welfare. 
\begin{corollary}[Efficiency of the GCE]\label{thm:gce_efficiency}
  Let $(\mathbf{s}^{\mathrm{GCE}},\!\mathbf{g}^{\mathrm{GCE}},\!\mathbf{p}^{\mathrm{GCE}})$ be an optimal solution to \eqref{prob:P_GCE}. 
  Then the GCE is \emph{efficient}, i.e., $(\mathbf{s}^{\mathrm{GCE}},\!\mathbf{g}^{\mathrm{GCE}})$ solves \eqref{prob:P_SW}, if and only if for every \((\mathbf{s},\mathbf{g})\in\mathcal{F}\),  {there exist} subgradients
  $\bm{\mu}\in\partial J^{\mathrm{SW}}(\mathbf{s}^{\mathrm{GCE}})$ and
  $\bm{\nu}\in\partial C(\mathbf{g}^{\mathrm{GCE}})$ such that:
  \begin{equation}\label{eq:vi-eff}
  \langle \bm{\mu},\,\mathbf{s}-\mathbf{s}^{\mathrm{GCE}}\rangle+\langle \bm{\nu},\,\mathbf{g}-\mathbf{g}^{\mathrm{GCE}}\rangle\ \ge 0.
  \end{equation}
% Let $(\mathbf{s}^\mathrm{GCE}, \mathbf{g}^\mathrm{GCE},\mathbf{p}^\mathrm{GCE})$ be an optimal solution to \eqref{prob:P_GCE}, iff for any $(\mathbf{s} \in \mathcal{S}, \mathbf{g} \in \mathcal{G})$ that satisfies $\mathbf{g} - \boldsymbol{\ell} - \mathbf{s} \in \mathcal{P}$,
% \begin{equation}
%   \left(\partial J^{\mathrm{SW}}(\mathbf{s}^\mathrm{GCE}) , \partial C(\mathbf{g}^\mathrm{GCE}) \right) \left(\mathbf{s}- \mathbf{s}^\mathrm{GCE}, \mathbf{g}- \mathbf{g}^\mathrm{GCE}\right)^\top \ge 0,
% \end{equation}
% then \eqref{prob:P_GCE} supports social welfare.
\end{corollary}
Corollary~\ref{thm:gce_efficiency} establishes the necessary and sufficient condition for \eqref{prob:P_GCE} to coincide with the social welfare optimum.\footnote{The low-dimensional aggregation through $J$ and $J^{\mathrm{SW}}$ keeps the system-level comparison concise, but loses the heterogeneity of individual FL agents' decisions; recovering it requires the internal model of the FL system.}
A simple sufficient condition is when \(J=J^{\mathrm{SW}}\), in which the two optimization problems share the same form.
More generally,  condition \eqref{eq:vi-eff} tests, \jq{starting at $(\mathbf{s}^{\mathrm{GCE}},\mathbf{g}^{\mathrm{GCE}})$,} whether any feasible direction can further improve social welfare.
This characterization applies regardless of whether the solution lies in the interior or on the boundary of the feasible set. 
In the latter case, it is possible that the GCE supports the social welfare even if $J \neq J^\mathrm{SW}$; see an example  in Section~\ref{sec:ev:eg}. 
% Even when \(J\neq J^{\mathrm{SW}}\), the two optimizations \eqref{prob:P_GCE} and \eqref{prob:P_SW} can still share the {same} optimal solution if a hard system constraint binds both problems into the same feasible corner, such as in Lemma \ref{lem:num-trans}-ii).  
% \begin{theorem}[Efficiency of Competitive Equilibrium]
%     \label{thm:efficiency_equilibrium}
% If \(J^{\mathrm{UE}}(\mathbf{s})\) equals  \(J^{\mathrm{SW}}(\mathbf{s})\), then the GCE is efficient. This means that the allocation of resources at equilibrium maximizes the total welfare of all participants in the market, including producers and consumers.
% \end{theorem}
\vspace{-0.24cm}
\subsection{Generalization to Multiple Independent FL Systems}\label{sec:general:nfl}
% \jq{
% \begin{remark}
% 	Generalization to multiple independent FL systems.
% \end{remark}
% }
\nan{
  Consider multiple FL systems connected to the same power system, each responding to common LMPs but without direct interaction beyond their shared grid connection, and each satisfying the framework in Section~\ref{sec:unifying_framework}.} 
% \jq{explain ``independent''} 
 \nan{Consider a setting with $W$ FL systems. We begin by redefining the variables:}
Let $\mathbf{S} \in \mathbb{R}^{N\times W}$ denote the matrix of electricity consumption, and its $w$-th column $\mathbf{S}_w$ represents the load vector of the $w$-th FL system, associated with its own feasible set $\mathcal{S}_w$ and value functions $J_w(\cdot)$ and $J_w^{\mathrm{SW}}(\cdot)$. 
% \jq{avoid $T$ and $t$ for things that are not time.}
%  Each FL system is viewed as an independent aggregate FL block that interacts with the grid solely through the shared LMP vector $\boldsymbol{\lambda}$.
\begin{definition}[Stacked FL Representation for Multiple Systems]\label{def:stacked-FL} 
  % \jq{caps}
Consider $W$ FL systems. Define
\begin{subequations}
  \begin{align}
  &\mathbf{s} := (\mathbf{S}_1^\top,\ldots,\mathbf{S}_W^\top)^\top,\quad
  \mathcal{S} := \mathcal{S}_1 \times \cdots \times \mathcal{S}_W,
  \label{eq:stacked-s}\\
  &J(\mathbf{s}) := \sum_{w=1}^{W} J_w(\mathbf{E}_w \mathbf{s}),\quad J^{\mathrm{SW}}(\mathbf{s}) := \sum_{w=1}^{W} J_w^{\mathrm{SW}}(\mathbf{E}_w \mathbf{s}),
  \label{eq:new-def}
\end{align}%
\end{subequations}%
\nan{where $\mathbf{E}_w \in \mathbb{R}^{N\times NW}$ is the block-selection matrix that satisfies $\mathbf{E}_w \mathbf{s} = \mathbf{S}_w$, i.e., $(\mathbf{E}_w)_{i,j}=1$ if and only if $j=(w-1)N+i$, and $0$ otherwise.}
\end{definition}
We have the following property of GCE for this system:
  \begin{theorem}[GCE with Multiple Independent FL Systems]
    \label{thm:GCE-multi}  
    Let $\mathbf{s}$, $\mathcal{S}$, $J(\mathbf{s})$, and $J^{\mathrm{SW}}(\mathbf{s})$ follow~\eqref{eq:stacked-s} and~\eqref{eq:new-def} in Definition~\ref{def:stacked-FL}, and let~\eqref{prob:P_GCE} be posed with these primitives and with the nodal balance constraint \(\mathbf{p} = \mathbf{g} - \boldsymbol{\ell} -  \sum_w\mathbf{E}_w\mathbf{s}\). The following statements hold.
  % \begin{equation}
  %   \mathbf{p} = \mathbf{g} - \boldsymbol{\ell} -  \mathbf{S}\mathbf{1},
  % \end{equation}
  % where $\mathbf{1} \in \mathbb{R}^W$ is the all-ones vector.
  \begin{enumerate}[label=\roman*), leftmargin = 0.5cm]
    \item A triple $(\mathbf{g},\mathbf{s},\boldsymbol{\lambda})$ is a GCE of the coupled system if and only if it is primal-dual optimal to~\eqref{prob:P_GCE}, and the equilibrium price $\boldsymbol{\lambda}^\star$ coincides with the Lagrange multiplier of the nodal balance constraint.
    \item Existence and uniqueness of the GCE follow under the same conditions as in Corollary~\ref{thm:gce_existence}.
    \item The efficiency characterization in Corollary~\ref{thm:gce_efficiency} holds with $J(\mathbf{s})$ and $J^{\mathrm{SW}}(\mathbf{s})$.
  \end{enumerate}
\end{theorem}
\vspace{-0.36cm}
\section{Coupled Power and Transportation Networks}
\label{sec:transportation}
\def\flr{x_r}
\def\fer{y_e(\mathbf{x})}
\def\Fr{\mathbf{x}}
\def\Frs{\mathbf{x}^\mathrm{UE}}
\def\Fone{x_1}
\def\Ftwo{x_2}
\def\pitravel{\pi^\mathrm{travel}}
\def\picharge{\pi^\mathrm{charge}}
\def\piaug{\pi}
\def\nev{N_\mathrm{EV}}
In this section, we specialize our general model presented in Section~\ref{sec:unifying_framework} to an important class of FL systems: electrified transportation systems. The spatial flexibility arises due to the charging location choices of the population of EVs in such systems, which in turn depend on EVs' route choices given traffic conditions. Section~\ref{sec:FLelement:EV} outlines essential elements to model the electrified transportation system as an FL system, which then enables us to apply our general results to examine the existence, uniqueness, and efficiency of the GCE in the coupled power-transportation systems in Section~\ref{sec:resultsEV}. We close the section with an illustrative example of such a system, highlighting a nuanced comparison between GCE and SWM outcomes under various system parameter regions.  

\vspace{-0.36cm}
\subsection{Elements of the FL System for Electrified Transportation}\label{sec:FLelement:EV}
We model the transportation network as a directed graph \(\mathcal{G}^\mathrm{tr}= (\mathcal{V}^\mathrm{tr}, \mathcal{E}^\mathrm{tr})\), where \(\mathcal{V}^\mathrm{tr}\) denotes the set of locations (a subset of which host EV charging stations) and \(\mathcal{E}^\mathrm{tr}\) denotes the set of directed roads connecting these locations.
Let \(N_\mathrm{C}\) denote the number of charging stations. 
Let there be \(K\) origin-destination (OD) pairs, i.e., \(\{(o_1, d_1), \ldots, (o_K, d_K)\}\) with $(o_k, d_k) \in \mathcal V^\mathrm{tr} \times \mathcal V^\mathrm{tr}$.
For each OD pair \(k\), let \(\mathcal{R}_k^\mathrm{tr}\) denote the finite set of feasible routes, each route being an ordered list of edges. Define the set of all routes as \(\mathcal{R}^\mathrm{tr} = \bigcup_{k=1}^K \mathcal{R}_k^\mathrm{tr}\) and the number of routes as \(N_\mathrm{R} = |\mathcal{R}^\mathrm{tr}|\). 
Each route is assumed to include exactly one charging station.\footnote{If a route contains multiple charging stations, we conceptually split it into distinct routes, each associated with a single charging station along the path.} Define the charger-route incidence matrix \(\mathbf{A}^\mathrm{CR} \in \mathbb{R}^{N_\mathrm{C} \times N_\mathrm{R}}\), whose entry \((j, r)\) is 1 if the \(r\)-th route  includes the \(j\)-th charging station, and 0 otherwise. Each column of \(\mathbf{A}^\mathrm{CR}\) contains exactly one nonzero entry.
Similarly, define the charger-bus incidence matrix \(\mathbf{A}^\mathrm{CB} \in \mathbb{R}^{N_\mathrm{C} \times N}\), where entry \((j, i)\) is 1 if the \(j\)-th charging station is connected to bus \(i\). 
% Denote the total number of EVs in the transportation system by $\nev$.
%%% Ori loc:
% For simplicity, each EV is assumed to charge a fixed amount of energy $q > 0$.  

\subsubsection{Collective decision model}
The collective behavior of the electrified transportation system can be modeled via a \emph{non-atomic game}, where the EV drivers act as the agents. % The continuum of these agents is indexed by \(\mathcal{J} := [0,1]\) and scaled to represent a total population of $\nev$,  grouped by their OD pairs. All EVs within the same OD category are treated as identical. For the \(k\)-th OD pair, the \emph{OD demand} is denoted by \(\rho_k > 0\), representing the fraction of EVs traveling from \(o_k\) to $d_k$, with the sum satisfying \(\sum_{k=1}^K \rho_k = 1\).
These agents form a continuum, grouped by their OD pairs, with each individual driver being infinitesimal. All EVs within the same OD category are treated as identical. For the \(k\)-th OD pair, the \emph{OD demand} is denoted by \(\rho_k > 0\), representing the mass of EVs traveling from \(o_k\) to \(d_k\).

\subsubsection{FL decisions}
For the FL system representing the  EV population, the critical internal decisions are route and charging station choices, captured by the traffic flow vector \(\Fr \in \mathbb{R}_+^{N_\mathrm{R}}\). Here, entry \(\flr\) denotes the mass of EVs selecting the \(r\)-th route \(R_r\) and charging at the charging station associated with the route.
We define the feasible set of traffic flows as
\begin{equation}
\mathcal{X} \!:=\! \big\{ \Fr\! \in\!\mathbb{R}_+^{N_\mathrm{R}} \!\mid\!
% \sum\nolimits_{R_r \in \mathcal{R}_k^\mathrm{tr}} \flr \!= \!\rho_k {\nev}, \ \forall k \in \{1,\!\dots,\!K\} \big\}.
\sum\nolimits_{R_r\! \in\! \mathcal{R}_k^\mathrm{tr}}\! \flr \!= \!\rho_k, \, \forall k \in \{1,\!\dots,\!K\} \big\}.
\label{eq:route-rate-b}%
\end{equation}
% That is, for each OD pair \(k\), the aggregate flow over all feasible routes in \(\mathcal{R}_k^\mathrm{tr}\) must equal the travel demand \(\rho_k{\nev}\), and flows are nonnegative on all routes.
That is, for each OD pair \(k\), the aggregate flow over all feasible routes in \(\mathcal{R}_k^\mathrm{tr}\) must equal the travel demand \(\rho_k\), and flows are nonnegative on all routes.

{For simplicity, each EV is assumed to charge a fixed amount of energy $q > 0$.  }
Given the traffic flow \({\Fr}\), the aggregate charging load at each power system bus satisfies~\eqref{eq:linear-map}, where matrix $\mathbf{A}^\mathrm{FL} $ is given by $\mathbf{A}^\mathrm{FL} = { q} \cdot (\mathbf{A}^\mathrm{CB})^\top \mathbf{A}^\mathrm{CR}$.
{\begin{remark}[Heterogeneous Charging Energy Requirements] 
% If EVs have heterogeneous charging energy requirements, partition them into $\Upsilon$ categories with per-EV energy $q_\upsilon$. Let $\mathbf{x}^{(\upsilon)}\in\mathbb{R}_+^{N_\mathrm{R}}$ be the route traffic flow vector for category $\upsilon$ and stack $\mathbf{x}:=[(\mathbf{x}^{(1)})^\top,\ldots,(\mathbf{x}^{(\Upsilon)})^\top]^\top$.  Then Assumption~\ref{ass:linear-map} still holds with matrix $\mathbf{A}^\mathrm{FL} $ given by
% previous (2026-09-07): \blue{If EVs have heterogeneous charging energy requirements, we assume they are partitioned into finitely many categories indexed by $\upsilon\in\{1,\ldots,\Upsilon\}$, where every EV in category $\upsilon$ charges a fixed amount of energy $q_\upsilon>0$. Each category has its own route traffic flow vector $\mathbf{x}^{(\upsilon)}\in\mathbb{R}_+^{N_\mathrm{R}}$, and we stack them as $\mathbf{x}:=[(\mathbf{x}^{(1)})^\top,\ldots,(\mathbf{x}^{(\Upsilon)})^\top]^\top$. Then Assumption~\ref{ass:linear-map} still holds with matrix $\mathbf{A}^\mathrm{FL}$ given by}
If EVs have heterogeneous charging energy requirements, we assume they fall into finitely many categories, where all EVs in the same category charge the same fixed amount of energy. Denoting by $q_1,\ldots,q_Q>0$ the charge amounts of the $Q$ categories and by $\mathbf{x}^{(1)},\ldots,\mathbf{x}^{(Q)}\in\mathbb{R}_+^{N_\mathrm{R}}$ their corresponding route traffic flow vectors, and stacking $\mathbf{x}:=[(\mathbf{x}^{(1)})^\top,\ldots,(\mathbf{x}^{(Q)})^\top]^\top$, Assumption~\ref{ass:linear-map} still holds with matrix $\mathbf{A}^\mathrm{FL}$ given by
\(
\mathbf{A}^\mathrm{FL}=[q_1(\mathbf{A}^\mathrm{CB})^\top\mathbf{A}^\mathrm{CR}\ \cdots\ q_Q(\mathbf{A}^\mathrm{CB})^\top\mathbf{A}^\mathrm{CR}].
\)
\end{remark}}

\subsubsection{Disutility functions}
Each edge \(e \in \mathcal{E}^\mathrm{tr}\) in the transportation network is associated with a flow-dependent travel cost function \(c_e: \mathbb{R}^+ \mapsto \mathbb{R}^+\), representing the monetary value of the travel time on edge \(e\).  
The flow on edge \(e\), denoted by \(\fer\), equals the sum of flow from all routes that include \(e\):
\begin{equation}
\fer = \sum\nolimits_{R_r \in \mathcal{R}^\mathrm{tr}: e \in R_r} \flr, \quad \forall e \in \mathcal{E}^\mathrm{tr}.
\label{eq:edge-route}
\end{equation}
Thus, edge flows are linear functions of the route flow vector \(\Fr\).  
Since every driver suffers $c_e(y_e(\Fr))$ while traveling through edge $e$ with the mass of such drivers being $y_e(\Fr)$, the total FL disutility \(\Phi^\mathrm{SW}(\Fr)\) is then calculated as:
% the sum of travel costs across all edges:
\begin{equation}\label{eq:phisw:ev}
    \Phi^\mathrm{SW}(\Fr) = \sum\nolimits_{e \in \mathcal{E}^\mathrm{tr}}\! c_e\left(\fer\right)\fer.
\end{equation}

\subsubsection{FL best response model}
We are interested in characterizing how the LMPs affect EVs' route and charging location choices. Given the LMPs, 
%In the nonatomic game setting, 
each EV driver selects a route and charging station aiming to minimize their individual cost.  
%This cost combines two components: the travel time along the chosen route and the charging cost determined by the electricity price at its associated charging station.  
The collective outcome of these selfish individual decisions is characterized by the Nash equilibrium of the {non-atomic} game, commonly referred to as the user equilibrium (UE) in the transportation literature \cite{sheffi1985urban}. 
%a User Equilibrium (UE) \cite{roughgarden2002bad}, a type of Nash Equilibrium where no driver can reduce their individual cost by unilaterally changing their route. 
% Compared to the classical transportation model where costs depend solely on travel time, the UE in the coupled system is extended to also account for charging costs.  
Specifically, given the LMPs \(\boldsymbol{\lambda}\), the cost associated with a route \(R_r\) is decomposed into travel and charging components:
\begin{subequations}\label{eq:route-costs}
    \begin{align}
    &\text{Travel cost:} &&
    \pitravel_r(\Fr) := \sum\nolimits_{e \in R_r} c_e(\fer), \label{eq:travel-cost}\\
    &\text{Charging cost:} &&
    \picharge_r(\boldsymbol{\lambda}) := \bigl((\mathbf{A}^\mathrm{FL})^\top \boldsymbol{\lambda}\bigr)_r, \label{eq:charge-cost}\\
    &\text{Route cost:} &&
    \piaug_r(\Fr, \boldsymbol{\lambda}) = \pitravel_r(\Fr) + \picharge_r(\boldsymbol{\lambda}), \label{eq:total-cost}
    \end{align}
\end{subequations}
%\jq{should charging cost include a $q$ factor?} \nanre{It is included in $\mathbf{A}^\mathrm{FL}$}
where the subscript \((\cdot)_r\) extracts the vector's $r$-th entry.  
Formally, the UE route flow under prices \(\boldsymbol{\lambda}\), denoted \(\Frs(\boldsymbol{\lambda})\), is defined by the condition that for any OD pair \(k\), every route \(R_r \in \mathcal{R}_k^\mathrm{tr}\) with positive flow \(\flr^\mathrm{UE} > 0\) satisfies
\begin{equation}
\piaug_r(\Frs, \boldsymbol{\lambda}) \leq \piaug_{r'}(\Frs, \boldsymbol{\lambda}), 
\quad \forall R_{r'} \in \mathcal{R}_k^\mathrm{tr}.
\label{def:UE}
\end{equation}
In other words, every EV in any used route does not have a unilateral incentive to pick a different route. 
The aggregate charging load $\mathbf s$ induced at the UE is captured by the price-responsive mapping
\(
\mathbf s = \boldsymbol{\sigma}(\boldsymbol{\lambda}) = \mathbf{A}^\mathrm{FL}\, \Frs(\boldsymbol{\lambda}).
\)

\subsection{\jq{Structural} Results for Coupled FL and Power System}\label{sec:resultsEV}
Although the UE characterization of the FL best response to LMPs $\boldsymbol{\lambda}$ \nan{in~\eqref{def:UE}} does not directly appear in the form of Assumption~\ref{ass:SFL-specific cost}, \nan{ it can be equivalently expressed as a} convex optimization formulation by properly defining \( \Phi(\Fr)\): 
\begin{lemma}[Equivalence of UE and Best-Response Model]
    \label{lem:UE-equivalence}
    For a fixed vector of LMPs \(\boldsymbol{\lambda}\), a feasible traffic flow \(\Fr\) is a UE if and only if it solves
    \begin{equation}
        \min_{\Fr \in \mathcal{X}} \ \Phi(\Fr) + \boldsymbol{\lambda}^\top \mathbf{A}^\mathrm{FL}\mathbf{x} ,
        \label{opt:UE}%
    \end{equation}
    %\sum\nolimits_{R_r \in \mathcal{R}^\mathrm{tr}} \pi_r(\boldsymbol{\lambda}) \flr,
    with the FL preference function formulated as follows:
    \begin{equation}\label{eq:phi:ev}
        \Phi(\Fr) %= \sum\nolimits_{e \in \mathcal{E}^\mathrm{tr}} h_e(\fer)
         := \sum\nolimits_{e \in \mathcal{E}^\mathrm{tr}} \int_0^{\fer} c_e(\xi) \, \mathrm{d}\xi.
    \end{equation} 
    % where \(h_e(x) := \int_0^x c_e(\xi) \, \mathrm{d}\xi\) is the integrated travel cost function for edge \(e\).
    % Problem \eqref{opt:UE} then matches the general FL best-response formulation~\eqref{eq:sfl-response}
    Moreover, \(\Phi(\Fr)\) is strictly convex if for every edge \(e \!\in\! \mathcal{E}^\mathrm{tr}\), the cost function $c_e(\cdot)$ is non-negative, {continuous} and increasing, and the mapping \(\Fr\!\mapsto\!(\fer)_{e\in\mathcal{E}^\mathrm{tr}}\) is injective on \(\mathcal{X}\). % integrable,
\end{lemma}
Note that the condition in Lemma \ref{lem:UE-equivalence}  on edge-level travel cost will probably hold in practice, since the travel time increases with the traffic flow on the edge due to potential traffic congestion.  
% Under , if the condition on edge-level travel cost holds, then
\jq{It follows that  Assumption \ref{ass:SFL-specific cost}  holds for electrified transportation systems under the described conditions for $c_e(\cdot)$.}
\jq{With \(\Phi(\Fr)\) defined in Lemma~\ref{lem:UE-equivalence}}, the associated value function \(J(\mathbf{s})\) follows directly from~\eqref{j-phi-re}, and \jq{the structural characterization of GCE in} Theorem \ref{thm:GCE-Structural} \jq{can be directly applied for coupled power and transportation systems.}

%  and is sufficient to ensure UE uniqueness \cite{smith1979existence}, though not strictly necessary (see Section~\ref{sec:conexp-trans}).

% The UE condition in~\eqref{def:UE} admits an equivalent convex optimization formulation.

% \begin{equation}
%     J(\mathbf{s}) :=
%     \min_{\Fr \in \mathcal{X}}
%     \left\{\Phi(\Fr)
%     \
%     \mathrm{s.t.}\ \eqref{eq:edge-route},\;\eqref{eq:linear-map}.\right\}
%     \label{def:UE-ev}
% \end{equation}%.
% \begin{assumption}[Edge-level Cost Function Properties]
%     for each edge \(e \in \mathcal{E}^\mathrm{tr}\), the travel cost function \(c_e(\cdot)\) is non-negative, continuously differentiable, and increasing.
% \label{ass:route}
% \end{assumption}

\subsubsection{Existence and uniqueness of GCE} 

% Since \(\Phi(\Fr)\)  is strictly convex, the resulting \(J(\mathbf{s})\) is strictly convex. 
With \(\Phi(\Fr)\) defined in \eqref{eq:phi:ev}, the price-responsive aggregate load mapping \(\boldsymbol{\sigma}(\boldsymbol{\lambda})\) can be equivalently expressed in \eqref{value:sfl},
which coincides with the UE-based response model \(
    \boldsymbol{\sigma}(\boldsymbol{\lambda}) = \mathbf{A}^\mathrm{FL} \, \Frs(\boldsymbol{\lambda}).
\)
\jq{By} Corollary \ref{thm:gce_existence}, the GCE exists and is unique. 
{This implies that, once  \(J(\mathbf{s})\) is available, the GCE of the coupled system can be computed directly from the system-level optimization problem~\eqref{prob:P_GCE} in Theorem~\ref{thm:GCE-Structural}, without requiring detailed information on OD demands, routes, and flows. This abstraction substantially reduces the informational burden and computational complexity to characterize the GCE, \jq{avoiding the need to work} with the high-dimensional FL decision space.}

\subsubsection{Social welfare property}
With \(\Phi^\mathrm{SW}(\Fr)\) defined in \eqref{eq:phisw:ev}, the SWM of the coupled system along with the value function \(J^{\mathrm{SW}}(\mathbf{s})\) is the solution to~\eqref{prob:P_SW} defined in Lemma~\ref{thm:GCE-SWM}.
With the above characterization, \(J(\mathbf{s})\) and \(J^{\mathrm{SW}}(\mathbf{s})\) have different forms, which indicates that the GCE generally fails to achieve the social optimum, and the gap between their solutions quantifies the efficiency loss due to selfish user behavior. 
A detailed comparison between GCE and SWM outcomes is illustrated via an example in Section~\ref{sec:conexp-trans}.
% \jq{your discussions are very short and leave many natural question unanswered for the reader.  is JUE and JSW same or different? Shall we expect GCE optimizes social welfare given structural similarity between (11) and (12)?}
%\jq{For example, for thm 1, why h is defined in that way?}
%-----------------------------------------------------------
% \begin{equation}
%     \min_{\Fr} \ \sum_{e \in \mathcal{E}^\mathrm{tr}} c_e(\fer)\fer \quad
%     \mathrm{s.t.} \ ~\eqref{eq:feas-fl}; ~\eqref{eq:ftos}.
% \end{equation}
% In the next subsection, we present a concrete example to compare the GCE solution with this welfare-optimal benchmark and highlight the potential inefficiencies arising from self-interested routing behavior.

\subsection{Illustrative Example}\label{sec:ev:eg}
\label{sec:conexp-trans}

\begin{figure}[t]
    \centering
    \includegraphics[width=0.95\linewidth]{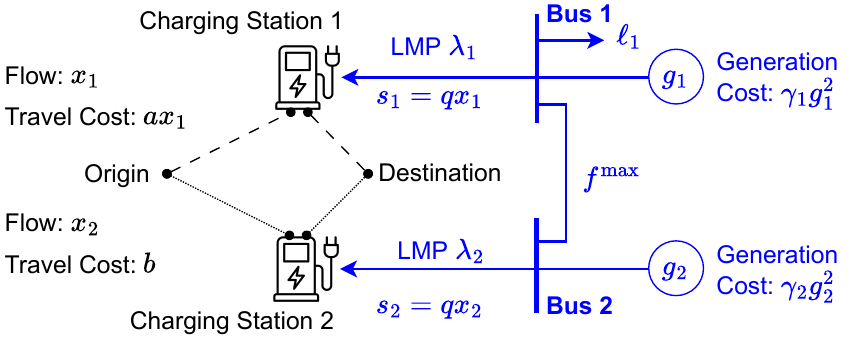}
    \caption{Coupled transportation and 2-bus power system.}
    \label{fig:trans}
     \vspace{-1em}
\end{figure}

% \jq{shall we also include gen cost on this figure? shall we make everything on the power side blue... so it's some form of color coding otherwise it might be unclear why the LMPs are blue}
As \jq{depicted} in Fig. \ref{fig:trans}, we consider a two-route transportation network with a common OD pair with unit travel demand. Each route includes a charging station node, which is connected to a distinct bus in a two-bus power system.
The travel cost for each EV user is $\pi^\mathrm{travel}_{1}(\Fr) = a \Fone$ on route 1 and a constant $\pi^\mathrm{travel}_2(\Fr)  = b$ on route 2, with \(a, b>0\). %\jq{consistency on the notation and terminology}
Given the LMPs, the total cost of traveling and charging along route 1 is \( a \Fone + \lambda_1 q \), while the corresponding cost for route 2 is \( b + \lambda_2 q \).
The aggregate EV charging loads at buses 1 and 2 are given by $s_1 = \Fone q$ and $s_2 = \Ftwo q$, respectively.
At each bus there is a generator, each with a cost function: \(\gamma_1g_1^2\) for generator 1 and \(\gamma_2g_2^2\) for generator 2. There is also a stationary demand \( \ell_1\) at bus 1.
The transmission line capacity is $f^{\max}$.
% \jq{consistency: you write Route 1, Bus 1, but generator 1. I wouldn't cap any of them in my writing though.}

The FL preference function
satisfies\footnote{Although route 2 has a constant cost, for this simple network we can show the FL preference function $\Phi(\mathbf{x})$ is strictly convex and Assumption \ref{ass:SFL-specific cost} holds.} $\Phi(\Fone, \Ftwo)= \tfrac{1}{2}a\Fone^2 + b\Ftwo$.  
{Following the definition~\eqref{j-phi-re}, \jq{we have \(J(s_1, s_2)=(a\,s_1^2)/(2q^2) + (b\,s_2)/q\)} in this example.}
% \begin{equation}
%     J(s_1, s_2) = \frac{a}{2q^2}s_1^2 + \frac{b}{q}s_2.
% \end{equation}
% is the optimal value to the following problem given a feasible \(s_1, s_2\)
% \begin{subequations}
%     \begin{align}
%    \min_{\Fone \ge 0,\ \Ftwo\ge 0} \quad & \Phi(\Fone, \Ftwo) \\
%      \mathrm{s.t.}\quad \quad \ & \Fone + \Ftwo = 1, \label{ev-each}\\
%     %  & (s_1, s_2)^\top = q \,(\Fone, \Ftwo)^\top.
%      & s_i = q x_i, \quad i= 1,2. \label{ev-0}
%   \end{align}
%   \label{eq:kkt-trans}%
% \end{subequations}%
\nan{The optimization-based characterization for the GCE is as follows:}
%\jq{we need to be always very clear about the difference between different concepts, as the concepts themselves are a part of our contributions. FL is not the same as FL system. GCE is not the same as our optimization-based characterization for GCE. }
\begin{subequations}
    \label{eq:num-GCE}
  \begin{align}
   \min_{\substack{s_1, s_2,g_1, g_2, p_1}} \ & J(s_1, s_2) + \gamma_1g_1^2 + \gamma_2g_2^2\\
  \mathrm{s.t.}\quad \quad
   & \lambda_1: p_1 = g_1 - \ell_1 - s_1, \label{ev-1'}\\& \lambda_2: -p_1 =  g_2 - s_2, \label{ev-1}\\
  & -{f}^{\max} \le p_1 \le {f}^{\max}, \label{ev-2}\\
%  & s_1 + s_2 = q {\nev}, \label{ev-4}\\
  & s_1 + s_2 = q, \label{ev-4}\\ 
  & s_i \ge 0, \quad  i= 1,2, \label{ev-5}\\
  & 0\le g_i \le g_i^{\max} , \quad  i= 1,2. \label{ev-3}
%   &  \mathbf{0} \le (g_1, g_2)^\top \le (g_1^{\max}, g_2^{\max})^\top. \label{ev-3}
\end{align}  %
\end{subequations}
Meanwhile, the total FL disutility equals $\Phi^\mathrm{SW}(\Fone, \Ftwo)= a\Fone^2 + b\Ftwo$, and \jq{so
\(J^\mathrm{SW}(s_1, s_2)= (a\,s_1^2)/(q^2) + (b\,s_2)/q\)} in this example.
% \begin{equation}
%     J^\mathrm{SW}(s_1, s_2) = \frac{a}{q^2}s_1^2 + \frac{b}{q}s_2.
% \end{equation}
% \begin{equation}
%  \min_{\Fone\ge 0,\ \Ftwo\ge 0} \  \Phi^\mathrm{SW}(\Fone, \Ftwo) \quad
%    \mathrm{s.t.} \ ~\eqref{ev-each},\eqref{ev-0}.
% \end{equation}%
Therefore, the welfare maximization problem $(\mathrm{P}_{\mathrm{SWM}})$ takes the following form:
\begin{subequations}
    \label{eq:num-SWM}
  \begin{align}
   \min_{\substack{s_1, s_2, g_1, g_2, p_1}} \ & J^\mathrm{SW}(s_1, s_2) + \gamma_1g_1^2 + \gamma_2g_2^2 \\
   \mathrm{s.t.}\quad \quad & \eqref{ev-1'}, \eqref{ev-1}, \eqref{ev-2}, \eqref{ev-3},  \eqref{ev-4}, \eqref{ev-5}.
  \end{align}
\end{subequations}

% We also assume that parameters satisfy $a > b$, $\gamma_2 \ll c_1$, and \(
% g_2^{\max} < c_1(\ell_1 + q)/(c_1+\gamma_2),
% \) indicating the generator 2 has limited capacity but better cost efficiency than generator 1. 
\jq{As $J$ and $J^\mathrm{SW}$ do not coincide, }we investigate whether \jq{the GCE} supports social welfare by comparing  solutions of \eqref{eq:num-GCE} and \eqref{eq:num-SWM} under different power system congestion regimes. \footnote{The power system is \emph{congested} if at least one transmission line flow limit binds at the dispatch solution, i.e., \(|[\mathbf{H}\mathbf{p}]_m| = f_m^{\max}\) for some line \(m\).}
\begin{lemma}[Comparison of $(\mathrm{P}_{\mathrm{GCE}})$ and $(\mathrm{P}_{\mathrm{SWM}})$]
    \label{lem:num-trans}
%    The links between solutions of \eqref{eq:num-GCE} and \eqref{eq:num-SWM} is as follows:
Let \(\mathbf{x}^{\mathrm{GCE}}\) and \(\mathbf{x}^{\mathrm{SWM}}\) denote the traffic flows at the GCE solved via~\eqref{eq:num-GCE} and at the SWM optimum in~\eqref{eq:num-SWM}, respectively. Define the \emph{price of anarchy} (PoA) as follows:
\[
\mathrm{PoA}:=\frac{\Phi^\mathrm{SW}(\mathbf{x}^{\mathrm{GCE}})+C(\mathbf{g}^{\mathrm{GCE}})}{\Phi^\mathrm{SW}(\mathbf{x}^{\mathrm{SWM}})+C(\mathbf{g}^{\mathrm{SWM}})}.
\]%
The following statements hold.
    \begin{enumerate}[label=\roman*), leftmargin = 0.5cm]
    % \item There exist parameter values \(q, a, b,\gamma_1, \gamma_2, f^{\max}, g_1^{\max}\) and \(g_2^{\max}\), such that GCE does not support social welfare.
    \item The PoA satisfies \(1\le\mathrm{PoA}\le\tfrac{4}{3}\).
%    \item Furthermore, \jq{if} \(a>b\), \(\gamma_2\ll\gamma_1\), \(g_2^{\max}< \gamma_1(\ell_1+q)/(\gamma_1+\gamma_2)\), and \(g_1^{\max} > q + \ell_1 \), then there exist two thresholds,
    \item Furthermore, \jq{if} \(a>b\), \(g_2^{\max}\in\bigl(q(1-b/(2a)),q\bigr]\), \(g_1^{\max}>\ell_1+q\), and \(\gamma_1(\ell_1+q-g_2^{\max})-\gamma_2g_2^{\max}>b/(2q)\), then there exist two thresholds, $\bar{f}^{\max} > \underline{f}^{\max}>0$, such that
    \begin{enumerate}[leftmargin = 0.3cm]
        \item If \(f^{\max} > \bar{f}^{\max} \), the power system is uncongested in   $(\mathrm{P}_{\mathrm{GCE}})$ and $(\mathrm{P}_{\mathrm{SWM}})$, and \(\mathrm{PoA}>1\), approaching \(4/3\) as \(b\) approaches \(a\) and the generation cost becomes negligible relative to the FL disutility.
        \item If \( \underline{f}^{\max}\!<\! f^{\max} \!\le\! \bar{f}^{\max} \), the power system congests only in $(\mathrm{P}_{\mathrm{GCE}})$, and the PoA decreases continuously from its value in a) to \(1\) as \(f^{\max}\) decreases from \(\bar{f}^{\max}\) to \(\underline{f}^{\max}\).
        \item If \(f^{\max} \le \underline{f}^{\max} \), the power system is congested in  $(\mathrm{P}_{\mathrm{GCE}})$ and $(\mathrm{P}_{\mathrm{SWM}})$, and \(\mathrm{PoA}=1\). 
    \end{enumerate}
    \end{enumerate}
\end{lemma}
The PoA measures the efficiency loss in the GCE, and its bound for the coupled system in Lemma~\ref{lem:num-trans}-i) coincides with the classical PoA bound for selfish routing with affine travel costs in an isolated transportation system \cite{roughgarden2002bad}.
% Lemma \ref{lem:num-trans} shows that, although
It also shows that, under certain parameter settings, %such as ranges of transmission capacity, generator limits, and route costs, 
the GCE can nonetheless support social welfare even when $J(\mathbf{s}) \neq J^{\mathrm{SW}}(\mathbf{s})$.

% ORIGINAL (replaced 2026-09-08): In the parameter regime of Lemma~\ref{lem:num-trans}-ii), route~1 is less costly than route~2 when its flow is below $b/a$, but becomes more expensive when this threshold is exceeded.
% \blue{The conditions on \(\gamma_1,\gamma_2, g_1^{\max}\) and \(g_2^{\max}\) keep the cheaper generator~2 at its capacity and generator~1 unconstrained.} The lemma further partitions outcomes into three transmission-capacity regimes, \emph{each corresponding to a distinct power system congestion pattern} and a \blue{different PoA outcome within \([1,4/3]\)}:
The parameter regime of Lemma~\ref{lem:num-trans}-ii) creates a setting where route~1 is less costly than route~2 at low flow but more expensive once its flow exceeds $b/a$, and where the cheaper generator~2 is at its capacity and generator~1 is unconstrained. Under it, the lemma further partitions outcomes into three transmission-capacity regimes, \emph{each corresponding to a distinct power system congestion pattern} and a different PoA outcome within \([1,4/3]\):
When $f^{\max}$ is sufficiently large, the uncongested system has uniform charging costs, and user choices depend solely on travel costs. In $(\mathrm{P}_{\mathrm{GCE}})$, users selfishly over-utilize the flexible route 1, leading to higher load at bus~1 \jq{(with the more expensive generator)} compared to the more balanced profile in $(\mathrm{P}_{\mathrm{SWM}})$. {In fact, it can be shown that} \(s_1^{\mathrm{GCE}} = 2s_1^{\mathrm{SWM}}>0\) in Lemma~\ref{lem:num-trans}-ii)-a). 
% \jq{when does this hold? just for ii-a? or more generally?}
% estabilishing Lemma~\ref{lem:num-trans}-i) and  Lemma~\ref{lem:num-trans}-ii)-a).
As a result, $(\mathrm{P}_{\mathrm{GCE}})$ relies heavily on the transmission line to import cheaper power from bus 2, making it more prone to line congestion. Thus, for an intermediate range of $f^{\max}$, {the} power system {is congested} only in $(\mathrm{P}_{\mathrm{GCE}})$. %, explaining Lemma \ref{lem:num-trans}-ii)-b). 
Finally, when $f^{\max}$ is small enough \jq{so the transmission constraint is binding} for both models, \nan{the higher LMP at bus 1 discourages drivers from choosing route 1. The charging load at bus~1 can be shown to be identical for both models, satisfying \(s_1 = q - g_2^{\max} + f^{\max}\) in Lemma~\ref{lem:num-trans}-ii)-c).} 
\vspace{-0.18cm}
\section{Coupled Power and Datacenter Networks}
\label{sec:datacenter}
% {This section presents a stylized model of datacenter workloads as FLs that link the power and computing infrastructures. While abstracted for clarity, the model retains the essential dynamics between workload distribution and electricity pricing. We aim to characterize the system constraints, formalize the price-response mapping \(\boldsymbol{\sigma}\), and analyze the resulting GCE and SWM in this coupled setting.}
This section adapts the general framework from Section~\ref{sec:unifying_framework} to datacenter networks, another important class of FL systems. 
The spatial flexibility  stems from how computing workloads can be geographically distributed among multiple datacenters.
Section~\ref{sec:FLelement:DC} introduces the elements to model datacenter operations as an FL system, which then enables us to apply the general results to examine the existence, uniqueness, and efficiency of the GCE in the coupled power-computing systems in Section~\ref{sec:resultsDC}. The section concludes with a stylized example highlighting how GCE and SWM outcomes diverge under varying system and market parameters.

\vspace{-0.18cm}
\subsection{Elements of the FL System}
\label{sec:FLelement:DC}
Suppose there are \( N_\mathrm{D} \) datacenters in the system, and each datacenter is connected to a power grid bus.
We define the datacenter-bus incidence matrix as \(\mathbf{A}^\mathrm{DB} \in \mathbb{R}^{N_\mathrm{D} \times N}\), where entry \((j, i)\) is 1 if the \(j\)-th datacenter is connected to bus \(i\) in the power grid. 
Each unit of computing workload consumes a \jq{fixed amount of electricity} \(q > 0\) \cite{Desislavov_2023}.
\subsubsection{Collective decision model}
We model the setting as a \emph{finite-player game}, where the agents are \( K \) AI companies, denoted by set \(\mathcal{K}\), each allocating computing workloads spatially across \( N_\mathrm{D} \) datacenters. 
{When $K=1$, our model reduces to the case that one company balances workloads across its geographically dispersed datacenters. In this case, the decision model for the FL system is a centralized optimization.}
\subsubsection{FL decisions}
% original: The workload allocation strategy of company \(k\) is represented by a vector \(\mathbf{x}^{(k)}\), where \( x_{j}^{(k)} \) denotes the portion of workload assigned to datacenter \(j\).
The workload allocation decision of company \(k\) is represented by a vector \(\mathbf{x}^{(k)}\), where \( x_{j}^{(k)} \) denotes the amount of computing service company \(k\) obtains at datacenter \(j\).\footnote{Computing service can be modeled as continuously divisible for elastic workloads such as inference, and for training that can be distributed across sites; a job that must be executed entirely at a single site induces a discrete placement decision and is outside this model.} Stacking all companies' decisions yields the FL decision vector \(\mathbf{x} \in \mathbb{R}^{K N_\mathrm{D}}\):
\(
\mathbf{x} = \bigl( \mathbf{x}^{(1)\top}, \dots, \mathbf{x}^{(K)\top} \bigr)^\top.
\)
We define the block summation matrix \(\mathbf{A}^\mathrm{sum} \in \mathbb{R}^{N_\mathrm{D} \times K N_\mathrm{D}}\) as
\(\mathbf{A}^\mathrm{sum} = \left( \mathbf{I}_{N_\mathrm{D}}, \dots, \mathbf{I}_{N_\mathrm{D}} \right)\), which aggregates all companies' allocations to each datacenter via
$\mathbf{A}^\mathrm{sum}\mathbf{x}$.
We denote \(\mathcal{X}^{(k)}\) as the feasible set of company \(k\)'s workload allocation, which characterizes the constraints on computing workload across datacenters, and define  \(\mathcal{X} := \prod_{k \in \mathcal{K}} \mathcal{X}^{(k)} \) as the aggregate feasible set for all companies.
% The electricity consumption at datacenter \(j\) is
% \(q \sum_{k=1}^K x_j^{(k)}.\) 
Hence, the aggregate load profile across datacenters satisfies~\eqref{eq:linear-map},  with the matrix $\mathbf{A}^\mathrm{FL}$ given by $\mathbf{A}^\mathrm{FL} := q {(\mathbf{A}^\mathrm{DB})}^\top \mathbf{A}^\mathrm{sum} $.
% \begin{equation}
% \mathbf{s} = q \cdot {(\mathbf{A}^\mathrm{DB})}^\top \mathbf{A}^\mathrm{sum} \mathbf{x}.
% \label{eq:dc-load}
% \end{equation}
\subsubsection{Disutility functions}
The utility of company \(k\) from allocating workloads is denoted by a function \(u_k(\mathbf{x}^{(k)}; \mathbf{x}^{(-k)})\), where \(\mathbf{x}^{(-k)}\) represents the collective decisions of all other companies. This notation captures both the intrinsic value of company $k$'s allocation and the externalities induced by the actions of competitors, \jq{e.g., due to increased delays when shared communication/computing infrastructure at the datacenters is congested}.  
The total FL disutility function $\Phi^\mathrm{SW}(\mathbf{x})$ is then the negative sum of the utility functions from computing workloads across all AI companies, which satisfies
\begin{equation}
    \Phi^\mathrm{SW}(\mathbf{x}) = - \sum\nolimits_{k=1}^{K}u_k\big(\mathbf{x}^{(k)}; \mathbf{x}^{(-k)}\big).
    \label{eq:swm-dc}
\end{equation}

\subsubsection{FL best response model}
In our model, each AI company directly pays for the electricity consumed by its workloads in datacenters, an arrangement consistent with self-operated or colocation datacenters, and serving as a stylized abstraction of electricity bill pass-through in cloud computing environments.
The net payoff of company \(k\) is then the utility from computing workloads minus the cost of electricity. In response to  the LMPs \(\boldsymbol{\lambda}\), each company distributes its computing workload across datacenters to maximize its net payoff:
\begin{equation}
    \max\nolimits_{\mathbf{x}^{(k)}\in \mathcal{X}^{(k)}} \ u_k\big(\mathbf{x}^{(k)}; \mathbf{x}^{(-k)}\big) - q{\big(\mathbf{A}^\mathrm{DB}\boldsymbol{\lambda}\big)}^\top\mathbf{x}^{(k)}.
    % \mathrm{s.t.} \ & \mathbf{x}^{(k)} \geq 0; \quad \mathbf{1}^\top \mathbf{x}^{(k)} \le \rho_k,
    \label{eq:dc-NE}
\end{equation}
% where \(\boldsymbol{\lambda}^\mathrm{D} \in \mathbb{R}^{N_\mathrm{D}}\) denotes the LMPs at datacenter buses, given by \(\boldsymbol{\lambda}^\mathrm{D} = \mathbf{A}^\mathrm{DB}\boldsymbol{\lambda}\).
% \jq{let's get rid of $\lambda^\mathrm{D}$?}
A Nash equilibrium (NE) arises when no company can improve its payoff by unilaterally changing its decision. Denote an NE decision as \(\mathbf{x}^\mathrm{NE}(\boldsymbol{\lambda})\). The aggregate price-responsive load $\mathbf{s}$ induced at equilibrium is then characterized by the mapping:
\(
\boldsymbol{\sigma}(\boldsymbol{\lambda}) = \mathbf{A}^\mathrm{FL}\, \mathbf{x}^\mathrm{NE}(\boldsymbol{\lambda}).
\)

\vspace{-0.27cm}
\subsection{\jq{Structural} Results for Coupled FL and Power System}
\label{sec:resultsDC}
This subsection establishes the equivalence between the NE characterization of the FL best response to the LMPs $\bm{\lambda}$ and Assumption~\ref{ass:SFL-specific cost}.
\nan{Let $\mathbf{F}\!:\!\mathcal{X}\!\to\!\mathbb{R}^{K N_\mathrm{D}}$ stack the gradients of each company's utility $u_k$ with respect to its own decision variable $\mathbf{x}^{(k)}$, i.e.,
    \(
       \jq{ \mathbf{F}(\mathbf{x}) \!=\! \bigl(\nabla_{\mathbf{x}^{(1)}} u_1(\mathbf{x})^\top,\dots,\nabla_{\mathbf{x}^{(K)}} u_K(\mathbf{x})^\top\bigr)^\top. }
    \)%
    }
    %     \bigl[\nabla_{\mathbf{x}^{(k)}}u_k\!\bigl(\mathbf{x}^{(k)};\mathbf{x}^{(-k)}\bigr)\bigr]_{k=1}^{K}.
With this notation, we impose the following conditions under which the preference function $\Phi(\mathbf{x})$ can be defined.
\begin{lemma}[Equivalence of NE and Best Response Model]
    \label{lemma:equiva_NE}
    Suppose each utility function \(u_k(\mathbf{x}^{(k)};\mathbf{x}^{(-k)})\) is continuously differentiable. The following statements hold.
    \begin{enumerate}[label=\roman*), leftmargin = 0.5cm]
    \item For any fixed LMP vector \(\boldsymbol{\lambda}\), an NE \(\mathbf{x}^\mathrm{NE} \in \mathcal{X}\) is a solution to the variational inequality (VI):
    \begin{equation}
        \left\langle\!  - \nan{\mathbf{F}(\mathbf{x})}\! +\! q {(\mathbf{A}^\mathrm{sum})}^\top\!  \mathbf{A}^\mathrm{DB}\boldsymbol{\lambda}, \mathbf{x} \! -\!  \mathbf{x}^\mathrm{NE}\!  \right\rangle \! \ge \!  0, \ \forall\, \mathbf{x} \in \mathcal{X}.
    \label{eq:vi-dc}
    \end{equation}
    % where \nan{ $\nabla_{\mathbf{x}}\mathbf{u}(\mathbf{x}^\mathrm{NE})=
    % \bigl(\nabla_{\mathbf{x}^{(1)}} u_1(\mathbf{x}),\dots,\nabla_{\mathbf{x}^{(K)}} u_K(\mathbf{x})\bigr)^\top\!\in\!\mathbb{R}^{K N_\mathrm{D}}$ stacks all individual gradients of the players' utilities for all \(k\in\mathcal K\).}
    % \jq{what is the dimension of $\nabla_{\mathbf{x}} \mathbf{u}(\mathbf{x}^\mathrm{NE})$?}
    \item If the symmetry condition holds:
    \begin{equation}
        \frac{\partial}{\partial \mathbf{x}^{(\tilde{k})}} \left( \nabla_{\mathbf{x}^{(k)}} u_k \right)
        = \frac{\partial}{\partial \mathbf{x}^{(k)}} \left( \nabla_{\mathbf{x}^{(\tilde{k})}} u_{\tilde{k}} \right), \ \forall\, k,\tilde{k} \in \mathcal K,
        \label{ass:dc_potential}
    \end{equation}
    % \jq{change $i$, $j$ to some notion of $k$. }
    then there exists a function \(\Phi(\mathbf{x})\) such that \(\nabla_{\mathbf{x}^{(k)}} \Phi(\mathbf{x}) = -\nabla_{\mathbf{x}^{(k)}} u_k(\mathbf{x}^{(k)};\mathbf{x}^{(-k)})\) for all \(k\), and the VI and thus the NE is equivalent to \nan{\eqref{eq:sfl-response} with $\Phi(\mathbf{x})$ defined this way.}
    % \begin{equation}
    %     \min_{\mathbf{x} \in \mathcal{X}}~ \Phi(\mathbf{x}) + \boldsymbol{\lambda}^\top \mathbf{A}^\mathrm{FL}\mathbf{x}.
    % \end{equation}
    % \jq{shall we just say (3) with $\Phi$ defined this way?}
    \item If, in addition, each \(u_k\) is strictly concave and each feasible set \(\mathcal{X}^{(k)}\) is nonempty, convex, and compact, then  the optimization \eqref{eq:sfl-response} is convex and has a unique solution.
    \end{enumerate}
\end{lemma}

The term $\tfrac{\partial}{\partial \mathbf{x}^{(\tilde{k})}}\!\left(\nabla_{\mathbf{x}^{(k)}} u_k\right)$ represents the Jacobian matrix of the gradient of company $k$'s utility with respect to another company's decision, i.e., the cross-partial derivatives $\tfrac{\partial^2 u_k}{\partial \mathbf{x}^{(\tilde{k})}\partial \mathbf{x}^{(k)}}$.
The symmetry cross-partial-derivative condition~\eqref{ass:dc_potential} ensures that the game is a \emph{potential game} \cite{monderer1996potential}, permitting the existence of a potential function \(\Phi(\mathbf{x})\).
This condition frequently holds in datacenter environments, as companies sharing computational resources commonly experience reciprocal performance effects, such as shared queueing delays, leading to symmetric interactions in their utility functions.
Concave utility functions \(u_k\) are also typical in practice, as the marginal value of computing resources generally diminishes with increasing workload; under the strict concavity in Lemma~\ref{lemma:equiva_NE}-iii), the resulting \(\Phi(\mathbf{x})\) is strictly convex.
% ORIGINAL (shortened 2026-09-08):
% Given the conditions in Lemma~\ref{lemma:equiva_NE}, Assumption \ref{ass:SFL-specific cost} holds for datacenter networks.
% Given the definition of \(\Phi(\mathbf{x})\),
% the corresponding value function \(J(\mathbf{s})\) follows from~\eqref{j-phi-re}, and {the optimization-based characterization for the GCE in Theorem~\ref{thm:GCE-Structural} then applies to the setting here.}
Given the conditions in Lemma~\ref{lemma:equiva_NE}, Assumption \ref{ass:SFL-specific cost} holds for datacenter networks, the value function \(J(\mathbf{s})\) follows from~\eqref{j-phi-re}, and the optimization-based characterization for the GCE in Theorem~\ref{thm:GCE-Structural} applies here.
If the symmetry condition~\eqref{ass:dc_potential} is not satisfied, a merit-function VI approach is used; see Appendix~\ref{app:non-potential}.
% and Section~\ref{dc:concrete_example} for details
% \jq{NE replaced with $\Phi$ and optimization formulation. it sounds a bit weird to say ``GCE formulation proceeds as Theorem 1''. perhaps change to the optimization-based characterization for GCE in Theorem 1 then applies to the setting here. Again, we are trying to highlight the results apply, and ``formulation'' usually is not a result. }

\subsubsection{Existence and uniqueness of GCE}
Under the defined preference function \(\Phi(\mathbf{x})\), the aggregate price-responsive load mapping \(\boldsymbol{\sigma}(\boldsymbol{\lambda}) = \mathbf{A}^\mathrm{FL}\, \mathbf{x}^\mathrm{NE}(\boldsymbol{\lambda})\) can be equivalently characterized by the solution to~\eqref{value:sfl}.
When the conditions in Lemma~\ref{lemma:equiva_NE} are satisfied, the GCE of the coupled power-datacenter system exists and is unique according to Corollary \ref{thm:gce_existence}.
Consequently, once the value function \(J(\mathbf{s})\) is specified, the GCE can be directly computed from~\eqref{prob:P_GCE} in Theorem~\ref{thm:GCE-Structural} without the need to track detailed workload allocations at the company level, simplifying the analysis of aggregate system behavior.

\subsubsection{Social welfare property}

%-----------------------------------------------------------
% We formulate the following SWM problem, which is measured as the total users' utility from computing services, offset by the cost of power generation:

% \begin{corollary}[{SWM of the Power-Datacenter System}]\label{prop:sw_dc}
%     The coupled systems satisfy the following properties:
%     \begin{enumerate}[label=\roman*), leftmargin=*, ref=\thetheorem.\roman*]
%         \item 
%         The total FL distulity is defined as the negative sum of the utility functions,
%         \(
%             \Phi^\mathrm{SW}(\mathbf{x}) = - \sum\nolimits_{k=1}^{K}u_k\left(\mathbf{x}^{(k)}; \mathbf{x}^{(-k)}\right).
%         \)
%         \item 
%         Value function \(J^{\mathrm{SW}}(\mathbf{s})\) follows \eqref{j-phi-re} with the new definition of  $\Phi^\mathrm{SW}(\mathbf{x})$ and the SWM of the coupled system is the solution to~\eqref{prob:P_SW}.
%     \end{enumerate}
%     \end{corollary}
Value function \(J^{\mathrm{SW}}(\mathbf{s})\) follows {the definition in Lemma~\ref{thm:GCE-SWM}} with the new form of $\Phi^\mathrm{SW}(\mathbf{x})$ in~\eqref{eq:swm-dc} and the SWM of the coupled system is the solution to~\eqref{prob:P_SW} defined in Lemma~\ref{thm:GCE-SWM}.
% \jq{hard to find this definition. good to give it an eqn number and refer to the number here. }
Function \(J(\mathbf{s})\) corresponds to decentralized equilibria under self-interested company behavior, whereas \(J^{\mathrm{SW}}(\mathbf{s})\) characterizes centralized coordination that maximizes aggregate utility. An example of their explicit forms and the inefficiencies due to misaligned incentives is detailed in Section~\ref{dc:concrete_example}. 
% \jq{the last sentence sounds like the next section gives ``explicit forms'' for general setting, rather than just for an example.}

\vspace{-0.18cm}
\subsection{Illustrative Example}
\label{dc:concrete_example}
\begin{figure}[t]
    \centering
    \includegraphics[width=0.98\linewidth]{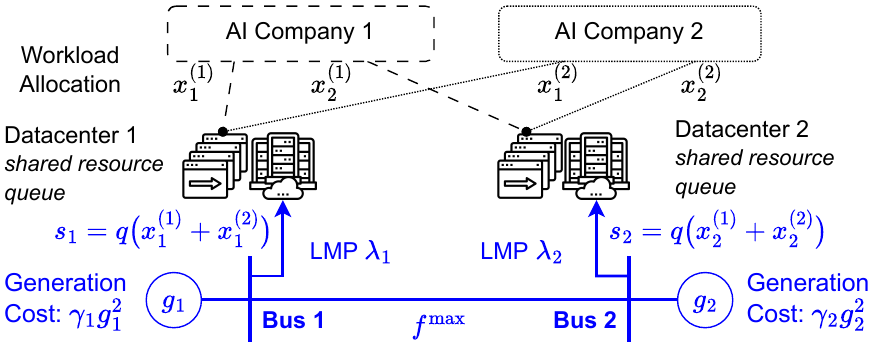}
    \caption{Coupled datacenter and 2-bus power system.}
    % \jq{what does interdependent task queue mean?} \nanre{Shared resource queue?} 
    \label{fig:datacenter}
     \vspace{-1em}
\end{figure}
As illustrated in Fig. \ref{fig:datacenter}, the system consists of two AI companies that allocate their computing workloads across two datacenters to optimize their utility, i.e., $K=2$. 
%\jq{write out $K=2$ etc so it's easy for people to find. }
In this setup, the two datacenters are connected to power system buses 1 and 2, respectively.
We do not enforce capacity limits on the datacenter loads, allowing them to be flexible in their load-shifting decisions.
There is no stationary demand at the buses; all load comes from the datacenters.
Consequently, the total demand at bus 1 is $s_1 = q\big(x_1^{(1)} + x_1^{(2)}\big)$, and the total demand at bus 2 is $s_2 = q\big(x_2^{(1)} + x_2^{(2)}\big)$.
% \jq{these and some other large parentheses really stood out. Could you get away with smaller parentheses for at least some of them?}
Additionally, two generators at bus 1 and bus 2 provide power outputs $g_1$ and $g_2$, respectively, with quadratic cost function $\gamma_i g_i^2$, $i =1, 2$. 
%cost functions as defined in Section \ref{sec:conexp-trans}.

In this example, we set the utility function from computing workload for company $k$, \(u_k\left(\mathbf{x}^{(k)}; \mathbf{x}^{(-k)}\right)\), to be quadratic:
% \begin{equation}
% \boldsymbol{\eta}_k^\top \mathbf{x}^{(k)} -\frac{1}{2} \left( \mathbf{x}^{(k)} \right)^\top \mathbf{M}_{k,k} \mathbf{x}^{(k)} - \left( \mathbf{M}_{k,-k}\mathbf{x}^{(-k)} \right)^\top  \mathbf{x}^{(k)},
% \label{eq:quadratic-utility}
% \end{equation}
% \jq{you may avoid left and right commands for parentheses to make them smaller, and potentially less prominent, like example below.}
\begin{equation}
\boldsymbol{\eta}_k^\top \mathbf{x}^{(k)} -\frac{1}{2}  {\mathbf{x}^{(k)}}^\top \mathbf{M}_{k,k} \mathbf{x}^{(k)} - \big( \mathbf{M}_{k,-k}\mathbf{x}^{(-k)} \big)^\top  \mathbf{x}^{(k)},
\label{eq:quadratic-utility}
\end{equation}
where \(\boldsymbol{\eta}_k \in \mathbb{R}^{N_\mathrm{D}}_{\ge 0}\) denotes the baseline utility, reflecting factors such as locational preference and latency penalties associated with geographical distance between datacenters and companies. The symmetric matrix \(\mathbf{M}_{k,k} \in \mathbb{R}^{N_\mathrm{D} \times N_\mathrm{D}}\) captures diminishing returns to scale in resource utilization across datacenters, while \(\mathbf{M}_{k,-k} \in \mathbb{R}^{N_\mathrm{D} \times (K-1) N_\mathrm{D}}\) represents the influence of other companies' workloads on company \(k\)'s utility, including queueing delays at each datacenter \cite{tsiligkaridis2025distributed}.
% where \( \boldsymbol{\eta}_k \in \mathbb{R}^{N_\mathrm{D}} \) reflects the baseline utility, incorporating factors such as data locality and latency penalties due to geographical distance between datacenters and company locations;
%  the symmetric matrix \( \mathbf{M}_{k,k} \in \mathbb{R}^{N_\mathrm{D} \times N_\mathrm{D}} \) captures diminishing returns to scale in resource utilization across datacenters;  and \( \mathbf{M}_{k,-k} \in \mathbb{R}^{ N_\mathrm{D} \times (K-1) N_\mathrm{D}} \) encodes the impact of other companies' workloads on company \(k\)'s utility, such as queueing delays due to shared infrastructure \cite{tsiligkaridis2025distributed}.

Given a vector of LMPs \(\boldsymbol{\lambda}\), the NE can be expressed as a VI problem that finds a vector \(\mathbf{x}^\star \in \mathcal{X}\) such that: 
\begin{equation}
\left\langle \mathbf{M}^\mathrm{NE} \mathbf{x}^\star - \boldsymbol{\eta} + {(\mathbf{A}^\mathrm{FL})}^\top \boldsymbol{\lambda} , \ \mathbf{x} - \mathbf{x}^\star \right\rangle \ge 0, \ \forall \mathbf{x} \in \mathcal{X},
\label{exa:vi-dc}
\end{equation}
% \begin{equation}
% \left\langle \mathbf{M}^\mathrm{NE} \mathbf{x}^\star - \boldsymbol{\eta} + q {(\mathbf{A}^\mathrm{sum})}^\top {\mathbf{A}^\mathrm{DB}} \boldsymbol{\lambda} , \ \mathbf{x} - \mathbf{x}^\star \right\rangle \ge 0, \ \forall \mathbf{x} \in \mathcal{X},
% \label{exa:vi-dc}
% \end{equation}
where $\boldsymbol{\eta} = (\boldsymbol{\eta}_1^\top, \boldsymbol{\eta}_2^\top)^\top$ stacks the baseline utility vectors, and \( \mathbf{M}^\mathrm{NE}  \in \mathbb{R}^{K N_\mathrm{D} \times K N_\mathrm{D}}\) captures the inter-company coupling:
\begin{equation}
\mathbf{M}^\mathrm{NE} = \begin{pmatrix}
\mathbf{M}_{1,1} & \mathbf{M}_{1,2} \\
\mathbf{M}_{2,1}  & \mathbf{M}_{2,2} 
\end{pmatrix}.
\end{equation}
Under the {symmetry} condition~\eqref{ass:dc_potential}, $\mathbf{M}^{\mathrm{NE}}$ is required to be symmetric, i.e., \(\mathbf{M}^{\mathrm{NE}}=(\mathbf{M}^{\mathrm{NE}})^\top\). According to Lemma~\ref{lemma:equiva_NE}, the corresponding FL preference function \( \Phi(\mathbf{x}) \) is given by
\begin{equation}
    \Phi(\mathbf{x})\! = 
        \frac{1}{2}\mathbf{x}^\top\mathbf{M}^\mathrm{NE} \mathbf{x} - \boldsymbol{\eta}^\top \mathbf{x}.
\end{equation}
Following definition~\eqref{j-phi-re}, we have
\(
    J(\mathbf{s})\! = 
    \min\nolimits_{\mathbf{x}\in \mathcal{X}} \{\Phi(\mathbf{x}):\eqref{eq:linear-map}\}. 
\)
   {Then,} $(\mathrm{P}_{\mathrm{GCE}})$ takes the form of~\eqref{eq:num-GCE} (without~\eqref{ev-4}) with this instantiated \(J(\mathbf{s})\).
If $\mathbf{M}^{\mathrm{NE}}$ is positive definite, the solution $(\mathbf{g}^{\star}, \mathbf{s}^{\star})$ is unique.  

Regarding the SWM benchmark, 
the detailed characterization of~\eqref{eq:swm-dc} for \(\Phi^\mathrm{SW}(\mathbf{x})\) in this example is given by:
\begin{equation}
    \Phi^\mathrm{SW}(\mathbf{x}) = 
    \frac{1}{2}\mathbf{x}^\top\mathbf{M}^\mathrm{SW} \mathbf{x} - \boldsymbol{\eta}^\top \mathbf{x},
\end{equation}
where 
% \jq{why is this sentence here?}
matrix \(\mathbf{M}^\mathrm{SW} \in \mathbb{R}^{K N_\mathrm{D} \times K N_\mathrm{D}}\) is symmetric positive definite, which satisfies:
\begin{equation}
\mathbf{M}^\mathrm{SW} := \mathbf{M}^\mathrm{NE} + \left( \mathbf{M}^\mathrm{NE} \right)^\top - \operatorname{blkdiag} \left( \mathbf{M}_{1,1}, \mathbf{M}_{2,2}\right),
\end{equation}
where \(\operatorname{blkdiag}(\cdot)\) denotes the block diagonal matrix.
Following the definition in Lemma~\ref{thm:GCE-SWM}, we have
\(
J^\mathrm{SW}(\mathbf{s}) =  
\min\nolimits_{\mathbf{x}\in \mathcal{X}} \{\Phi^\mathrm{SW}(\mathbf{x}):~\eqref{eq:linear-map}\}.
\)
For this example, $(\mathrm{P}_{\mathrm{SWM}})$ takes the same form as~\eqref{eq:num-SWM}, with \(J^\mathrm{SW}(\mathbf{s})\) instantiated.

% \jq{why NE$\to$SW $\to$ NE?}

The following lemma establishes whether the \jq{GCE outcome} maximizes social welfare. 
\begin{lemma}[Characterization of $(\mathrm{P}_{\mathrm{GCE}})$]
    \label{lem:cha-dc}
    % The properties of the game $(\mathrm{P}_{\mathrm{GCE}})$ are determined by :
    \nan{Assume that both $\mathbf{M}^{\mathrm{NE}}$ and $\mathbf{M}^{\mathrm{SW}}$ are  positive definite\jq{.}}
    Denote by \(\mathbf{x}^{\mathrm{GCE}}\) and \(\mathbf{x}^{\mathrm{SWM}}\) the workload allocations at the GCE and at the SWM optimum, and let \(\Delta\mathbf{M}:=\mathbf{M}^{\mathrm{SW}}-\mathbf{M}^{\mathrm{NE}}\) collect the cross-company interaction blocks.
            % \jq{W}hether the GCE supports social welfare depends on the cross-company externalities, i.e., the structure of the interaction matrix $\mathbf{M}^{\mathrm{NE}}$:
            \jq{W}hether the GCE supports social welfare depends on \(\Delta\mathbf{M}\), shown as follows:
            \begin{enumerate}[label=\roman*)]
                % \item If there exists \emph{no cross-company externalities}, i.e., \(\bigl(\mathbf{M}_{1,2}=\mathbf{M}_{2,1}=\mathbf{0}\bigr)\), \jq{the GCE outcome} coincides with the social-welfare benchmark.
        % \item For \emph{generic cross-terms}, i.e., \(\bigl(\mathbf{M}_{1,2}\!=\!\mathbf{M}_{2,1}\!\neq \!\mathbf{0}\bigr)\), there exist parameter values \(q, \boldsymbol{\eta}, \mathbf{M}^{\mathrm{NE}},\gamma_1, \gamma_2, f^{\max}, g_1^{\max}\) and \(g_2^{\max}\) where \jq{the GCE} does not support social welfare.
        \item The welfare gap between the GCE and the SWM, i.e., \(\Phi^{\mathrm{SW}}(\mathbf{x}^{\mathrm{GCE}})+C(\mathbf{g}^{\mathrm{GCE}})-\Phi^{\mathrm{SW}}(\mathbf{x}^{\mathrm{SWM}})-C(\mathbf{g}^{\mathrm{SWM}})\), is nonnegative and bounded above by
        
        \(
        \tfrac{1}{2}\bigl(\Delta\mathbf{M}\,\mathbf{x}^{\mathrm{GCE}}\bigr)^{\!\top}\bigl(\mathbf{M}^{\mathrm{SW}}\bigr)^{-1}\bigl(\Delta\mathbf{M}\,\mathbf{x}^{\mathrm{GCE}}\bigr)
        \).
        When \(\Delta\mathbf{M}\!=\!\mathbf{0}\), the GCE outcome coincides with the SWM. Moreover, there exist \(\Delta\mathbf{M}\) and parameter values for which the welfare gap is strictly positive.
        \item 
        % \nan{If the baseline-utility vectors are identical, $\boldsymbol{\eta}_1 = \boldsymbol{\eta}_2 = (1,1)^\top$; the per-unit energy coefficients are normalized to $q_1 = q_2 = 1$; the generators have sufficient capacity; and the transmission line capacity $f^{\max}$ is sufficiently large such that both GCE and SWM operate without congestion in the power network. Under the condition
        % $\mathbf{1}^\top (\mathbf{M}^\mathrm{NE})^{-1} \mathbf{1} \ge \mathbf{1}^\top (\mathbf{M}^\mathrm{SW})^{-1} \mathbf{1},$
        % the LMPs in GCE are higher than those in SWM, i.e.,
        % $\lambda_1^{\mathrm{GCE}} = \lambda_2^{\mathrm{GCE}} > \lambda_1^{\mathrm{SWM}} = \lambda_2^{\mathrm{SWM}}.$}
	       {Suppose grid constraints do not bind, i.e., $f^{\max} = g_i^{\max} = \infty$, $i=1,2$, and the AI companies have identical baseline-utility, i.e., $\bm \eta_1 = \bm \eta_2 \propto \mathbf 1$. If $(\mathbf{M}^{\mathrm{NE}})^{-1}\mathbf{1}>(\mathbf{M}^{\mathrm{SW}})^{-1}\mathbf{1}>\mathbf{0}$ componentwise,
        then  LMPs in GCE are higher than those in SWM, i.e.,
        $\lambda_i^{\mathrm{GCE}} > \lambda_i^{\mathrm{SWM}} ,$ $i=1,2$.  } 
        % \jq{[isn't $q$ a scalar? why did we need that condition?]}
\end{enumerate}
\end{lemma}
The lemma shows that in a competitive setting, AI companies' self-interested decisions may generate negative externalities, as each company accounts for shared infrastructure only through its own utility and electricity payments, leading to inefficient spatial allocation of computation, which is quantified by Lemma~\ref{lem:cha-dc}-i). 
In contrast, a social planner would internalize these externalities and balance workloads across locations to minimize total system cost.
In particular, under the condition in Lemma~\ref{lem:cha-dc}-ii), AI companies tend to induce excessive demand, resulting in higher aggregate demand and consequently higher LMPs.
The LMP ordering in Lemma~\ref{lem:cha-dc}-ii) is driven by the negative externality of the competitive allocation (see Lemma~\ref{lem:ces-positive-externality} in Appendix~\ref{app:ces-positive-externality} for an example of a positive externality from shared-service scaling).

% \input{3-concrete_examples}
% \input{4-general-results}
% \input{2-equilbrium}
% needed in second column of first page if using \IEEEpubid
%\IEEEpubidadjcol

\vspace{-0.18cm}
\section{Case Study}
\label{sec:case_study}
\begin{figure*}[t]
    \centering
    \includegraphics[width=0.9\textwidth]{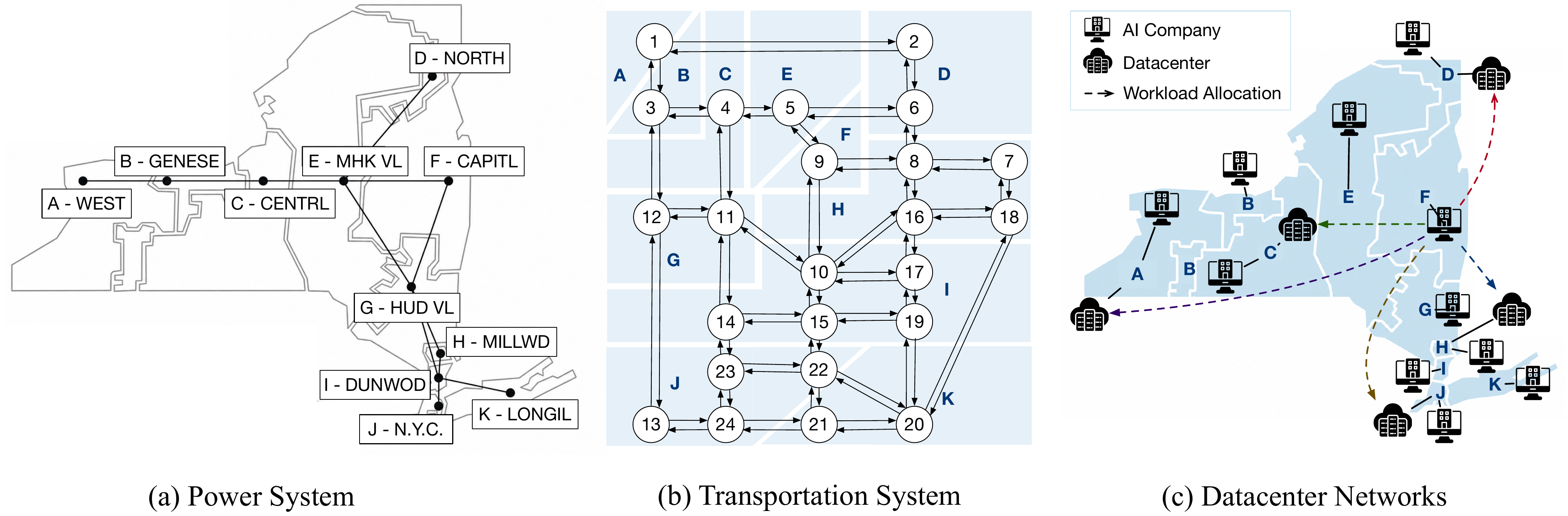}
    \caption{The interconnected (a) power system, (b) transportation system, and (c) datacenter networks.}
    \label{fig:coupling_sys}
    \vspace{-1em}
\end{figure*}
\label{sec:para-case}

We study an integrated system comprising the power grid, transportation network, and datacenter network. 
% This case study illustrates how the framework developed in Section~\ref{sec:unifying_framework} can be applied to real-world settings, and examines how the resulting GCE captures the interactions among these networks and affects welfare outcomes.
% We evaluate the GCE in discrete one-hour market-clearing periods, treating all energy-use activities within each hour as simultaneous.
We evaluate the GCE in a steady-state setting in which the travel demand, the stationary loads, and the generation costs are time-invariant, thereby abstracting from intertemporal constraints. We characterize the state of the coupled infrastructure systems over a one-hour horizon, on which all reported quantities, including the generation cost, the EV travel cost, and the computing utility, are evaluated.

\vspace{-0.36cm}
\subsection{Configurations of the Power and Multi-FL Systems}
We form a stylized test system, in which the transportation and datacenter networks are electrically connected to the same power system, with configuration details illustrated below.

\subsubsection{\jq{Power system}}
The power system configuration \nan{is adapted from} the setting in~\cite{10636017}, based on an 11-zone aggregation of the NYISO network that includes generation data, network parameters, baseline load profiles, and inter-zone distance information. 
The 11 zones, indexed by A-K, are shown in Fig.~\ref{fig:coupling_sys}(a), where each zone represents a major operating region (e.g., WEST, CENTRL, N.Y.C), and the transmission lines connecting them indicate the network topology of the regional grid. 
% {The transportation and datacenter networks are electrically connected to the same power system, with configuration details illustrated below.}
% MOVED (R1-9) to the opening of Section VI-A above, reworded so that the subsection opens with one
% sentence before the first subsubsection. ORIGINAL: {The transportation and datacenter networks are electrically connected to the same power system\blue{, together forming a composite stylized test system of coupled infrastructures}, with configuration details illustrated below.}
To increase the sensitivity of LMPs to  FL load-profile variations, we scale the quadratic coefficient in the generation cost by a factor of ten relative to~\cite{10636017}, which steepens marginal generation costs.
\vspace{2pt}
% The coefficient of the quadratic term in the generation cost is set to be ten times larger than that in \cite{10636017}, reflecting a steeper marginal generation cost profile.
% \nan{The configuration details can be found in Appendix~\ref{sec:para-case}.}
\subsubsection{\jq{Electrified transportation system}}
The transportation network, shown in Fig.~\ref{fig:coupling_sys}(b), is adapted from the Sioux Falls road network~\cite{SiouxFalls}, consisting of 24 nodes and 76 edges. 
Each node is equipped with an EV charging station, while also representing the origins and destinations of EV trips.
% These nodes each connect electrically to one of the 11 NYISO zones in Fig.~\ref{fig:coupling_sys}(a); the shaded blue areas indicate this geographic correspondence and determine~$\mathbf{A}^\mathrm{CB}$.
% R3-7 (2026-09-08): "electrical assignment" replaced by a plain description of what the shaded areas encode.
% These nodes each connect electrically to one of the 11 NYISO zones in Fig.~\ref{fig:coupling_sys}(a); the shaded blue areas indicate this \blue{electrical assignment} and determine~$\mathbf{A}^\mathrm{CB}$.
These 24 nodes are grouped into the 11 NYISO zones by the shaded blue areas in Fig.~\ref{fig:coupling_sys}(a); this grouping determines~$\mathbf{A}^\mathrm{CB}$, and charging stations in different zones face different LMPs.
Arrowed lines indicate the bidirectional links between nodes, representing the road connections that form the transportation network.

% The EV-population data and OD-demand data are also adapted from~\cite{SiouxFalls}, yielding 182 OD pairs with OD demand~$\boldsymbol{\rho}$ and a total of {{158{,}080}} EV \emph{agents}.
The EV-population data and OD-demand data are also adapted from~\cite{SiouxFalls}, yielding 182 OD pairs with OD demand~$\boldsymbol{\rho}$, where $\mathbf{1}^\top\boldsymbol{\rho} = 158{,}080$.
To better capture real-world congestion effects, we rescale the OD demands.
For each OD pair, we first identify the two routes with the lowest free-flow travel times.
% ORIGINAL (R1-9, the first sentence below is now the footnote above): This construction yields a compact route--charging choice set that keeps the decision space low dimensional; allowing charging at intermediate locations would simply introduce additional routes without altering the modeling framework. The resulting route--charging configuration determines $\mathbf{A}^\mathrm{CR}$.
Given this fixed route set, charging is assigned to either the origin or the destination of each route, so that an EV's decision trades off travel time against nodal charging prices.\footnote{This construction yields a compact route--charging choice set; allowing charging at intermediate locations would simply introduce additional routes without altering the modeling framework.}
The resulting route--charging configuration determines $\mathbf{A}^\mathrm{CR}$.
The travel cost function on edge $e$ is modeled as \( c_e(\fer) = c_e^0 \big(1 + \kappa_e\fer\big) \), where $c_e^0$ represents the base cost, computed as the free-flow travel time multiplied by a value of time of \$17.09/hour \cite{van2025political}. The coefficient $\kappa_e$ reflects the marginal effect of traffic volume and is obtained by dividing the delay parameter by the edge capacity.
Following~\cite{SiouxFalls}, we triple the free-flow travel times and scale the delay parameters by a factor of ten to increase the sensitivity of per-driver travel cost to flow relocation.
{Each EV requires an identical charging amount $q^\mathrm{EV} \in [1,20]$ kWh, corresponding to the energy delivered during one hour of Level 1-2 charging~\cite{EVchargingspeed}.}

\subsubsection{\jq{Datacenter network}}
The datacenter network consists of 11 AI companies and 5 datacenters.
% Figure~\ref{fig:coupling_sys}(c) marks the geographic locations of all companies and datacenters, and links each of them to a power-zone index (A--K) in Fig.~\ref{fig:coupling_sys}(a).
% We place the five datacenters in the WEST, CENTRL, NORTH, MILLWD, and N.Y.C zones following~\cite{10636017}; this placement defines the datacenter-to-bus coupling matrix~$\mathbf{A}^\mathrm{DB}$.
% Each AI company can dispatch workload across all five datacenters; dashed lines in Fig.~\ref{fig:coupling_sys}(c) illustrate this connectivity for one representative company, with different colors corresponding to different datacenters.
Figure~\ref{fig:coupling_sys}(c) marks the locations of all companies and datacenters and links each to a power-zone index (A--K) in Fig.~\ref{fig:coupling_sys}(a).
Following~\cite{10636017}, we place the five datacenters in the WEST, CENTRL, NORTH, MILLWD, and N.Y.C zones, which defines the datacenter-to-bus coupling matrix~$\mathbf{A}^\mathrm{DB}$. 
Each AI company can dispatch workload across all five datacenters; dashed lines in Fig.~\ref{fig:coupling_sys}(c) illustrate this for one representative company, with colors distinguishing the datacenters.
% MOVED to app.tex Appendix~\ref{app:case-utility} for R1-9: Table~\ref{tab:case_study_parameters}
% and the parameter interpretation / block structure of $\mathbf{M}$ that followed this sentence.
% ORIGINAL: The utility function of each AI company $k$ takes the quadratic form~\eqref{eq:quadratic-utility}, where the parameters are specified as follows:
The utility function of each AI company $k$ takes the quadratic form~\eqref{eq:quadratic-utility}. Its parameters, their interpretation, and the resulting block structure of $\mathbf{M}$ are specified in Appendix~\ref{app:case-utility}.
% The specifications of parameters in the utility function are documented in Appendix~\ref{sec:para-case}.
% ORIGINAL: {Each standardized \emph{workload} (e.g., an AI inference/training batch) has an energy requirement of $q^\mathrm{DC} \in [0,10]$ kWh.
{Each standardized \emph{workload} has an energy requirement of $q^\mathrm{DC} \in [0,10]$ kWh.
}
{Each AI company optimizes the allocation of its fixed total workload across datacenters} 
\jq{to maximize its net payoff,}
{and the total amount of workloads is 364,803.}

\begin{figure}[t]
    \centering
    \begin{subfigure}[b]{0.46\linewidth}
        \centering
        \includegraphics[width=1\linewidth]{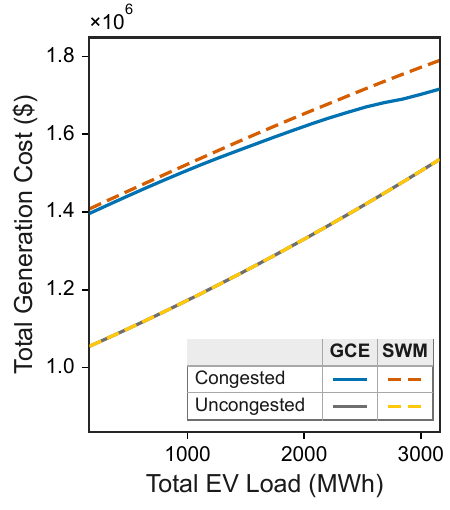}
        \caption{ Cost sensitivity to $q^\mathrm{EV}$\\ with $q^\mathrm{DC} = 10$ kWh.}
        \label{fig:genCost_EV}
    \end{subfigure}
    \begin{subfigure}[b]{0.46\linewidth}
        \centering
        \includegraphics[width=1\linewidth]{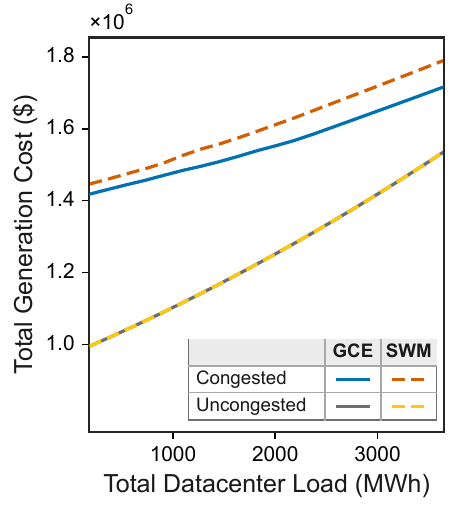}
        \caption{ Cost sensitivity to $q^\mathrm{DC}$ with\\ $q^\mathrm{EV} = 20$ kWh.}
        \label{fig:genCost_DC}
    \end{subfigure}
    \caption{{Impact of FL parameters on generation cost.}}
    \label{fig:genCost}
     \vspace{-1em}
\end{figure}
\begin{figure}[t]
    \centering
    \begin{subfigure}[b]{0.46\linewidth}
        \centering
        \includegraphics[width=1\linewidth]{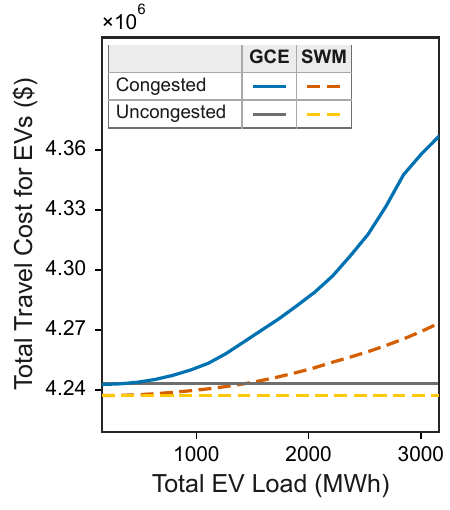}
        \caption{Internal-system effect \\ with $q^\mathrm{DC} = 10$ kWh.}
        \label{fig:self}
    \end{subfigure}
    \begin{subfigure}[b]{0.46\linewidth}
        \centering
        \includegraphics[width=1\linewidth]{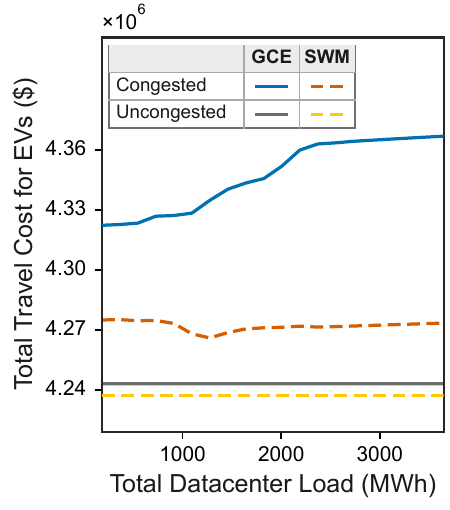}
        \caption{Cross-system impact \\with $q^\mathrm{EV} = 20$ kWh.}
        \label{fig:mutual}
    \end{subfigure}
    \caption{{Travel cost under increasing EV and datacenter loads.}}
    \label{fig:net benefit}
     \vspace{-1em}
\end{figure}

\vspace{-1em}
\subsection{GCE-Based Interaction and Welfare Analysis}
Using the GCE and SWM formulations developed in Section~\ref{sec:general_results}, we compute the FL decisions, generator dispatch, and LMPs $(\mathbf{g}^\star,\mathbf{s}^\star,\boldsymbol{\lambda}^\star)$ under GCE 
% framework \jq{what does the word ``framework'' mean here?} 
and compare them with the corresponding \jq{social welfare }optimal outcomes.
We then vary key parameters, namely, the transmission line capacities, the EV charging amount $q^\mathrm{EV}$, and the energy demand per workload $q^\mathrm{DC}$, to examine their impacts on the performance of each subsystem.
{For comparison, we consider two experimental settings: the \emph{congested} case, based on the original dataset where limited line capacity causes congestion even without interconnection of the FL systems, and the \emph{uncongested} case, a synthetic scenario in which all line-capacity constraints are relaxed to ensure an uncongested power system.}
% \jq{perhaps mention which line is congested and why it matters} \nanre{Actually there are three congested lines}

% \subsubsection{Generation cost of the power system}
% 

Fig.~\ref{fig:genCost} illustrates the impact of FLs' electricity consumption on total generation cost under GCE and SWM, where network congestion results in higher costs for both models.
% Under a congested power system, the GCE yields a lower generation cost than the SWM, indicating that self-interested FL behaviors can, counterintuitively, better utilize available generation by exploiting LMP signals.
Under a congested power system, the GCE yields a lower generation cost than the SWM (but a higher total social cost), as the two models spatially allocate the same total FL demand differently. Compared with the SWM allocation, self-interested FL agents under the GCE shift more consumption toward lower-LMP buses with cheaper marginal generators, which lowers the generation cost. The SWM instead accepts a higher generation cost in exchange for a lower total FL disutility, thereby attaining a higher social welfare than the GCE. In the uncongested case, all LMPs are equal, so FL agents have no incentive to shift consumption toward lower-price buses, and the generation costs under the GCE and the SWM coincide.
As the FL load increases, the generation cost gap between GCE and SWM widens, reaching 4.30\% \nan{at the rightmost points of both Fig. \ref{fig:genCost_EV} and Fig. \ref{fig:genCost_DC}}, where total EV and datacenter loads account for 8.52\% and 9.83\% of the total load, respectively.

{Fig. \ref{fig:net benefit} illustrates how the total EV travel cost varies with increasing EV and datacenter loads. When the power system is uncongested, travel cost is insensitive to load growth; when it is congested, the travel cost gap between the GCE and SWM widens with EV charging demand in Fig.~\ref{fig:self}, reflecting higher traffic latency caused by EVs' self-interested routing.
In Fig.~\ref{fig:mutual}, under the GCE, rising datacenter load raises the {total EV travel cost} by increasing LMPs and shifting their locational pattern, which redistributes traffic and heightens latency, showing a negative cross-system externality between the datacenter and transportation networks.}
The changes in generation cost, travel cost, and computing utility partially offset, and the GCE's social welfare loss relative to the SWM reaches $1.33\%$ in a congested power system at $q^\mathrm{EV}=20$~kWh and $q^\mathrm{DC}=10$~kWh, an estimated \$0.59 billion per year.

\begin{figure}[t]
    \centering
    \begin{subfigure}[b]{0.32\linewidth}
        \centering
        \includegraphics[width=1\linewidth]{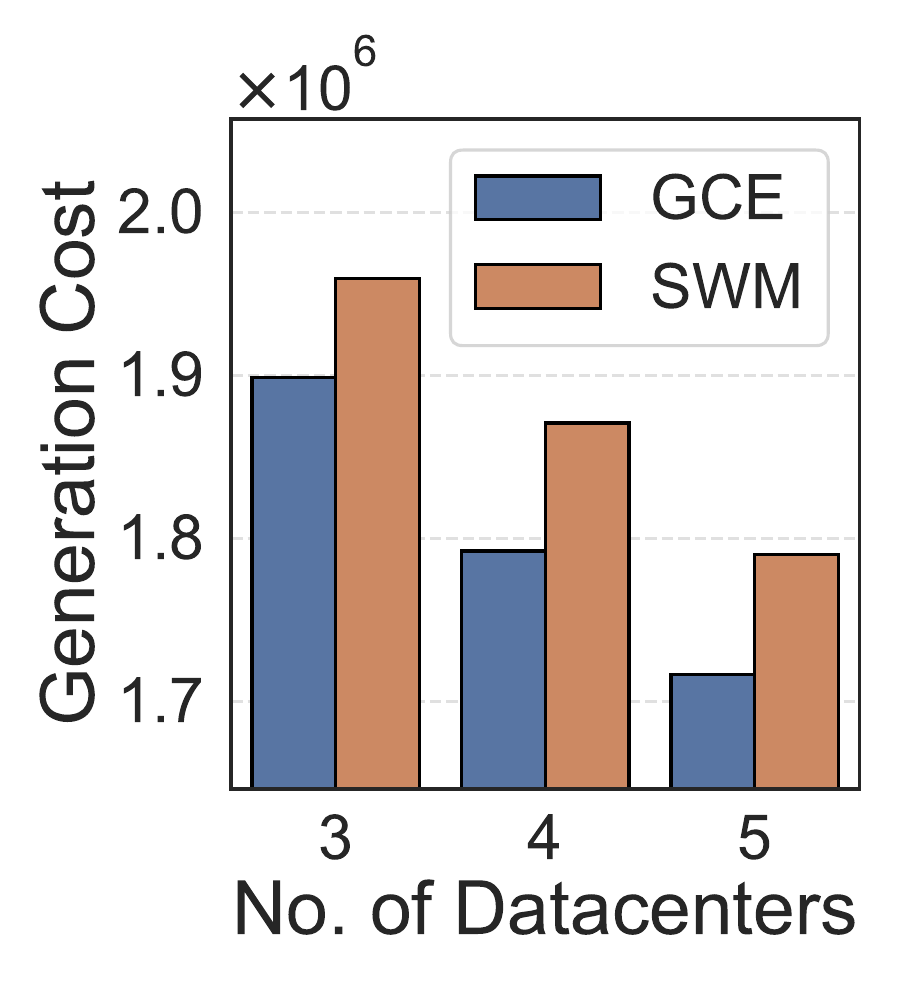}
        \caption{Generation Cost.}
        \label{fig:fl-gen}
    \end{subfigure}
    \begin{subfigure}[b]{0.32\linewidth}
        \centering
        \includegraphics[width=1\linewidth]{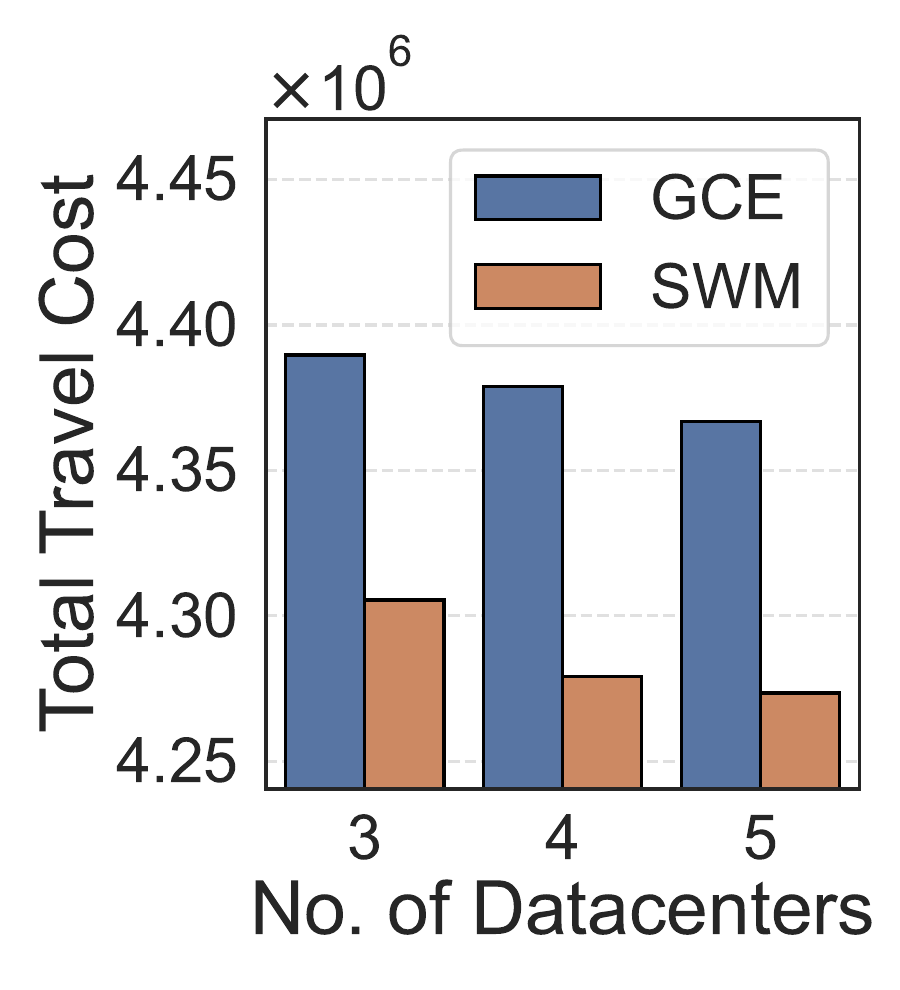}
        \caption{Travel Cost.}
        \label{fig:fl-ev}
    \end{subfigure}
    \begin{subfigure}[b]{0.32\linewidth}
        \centering
        \includegraphics[width=1\linewidth]{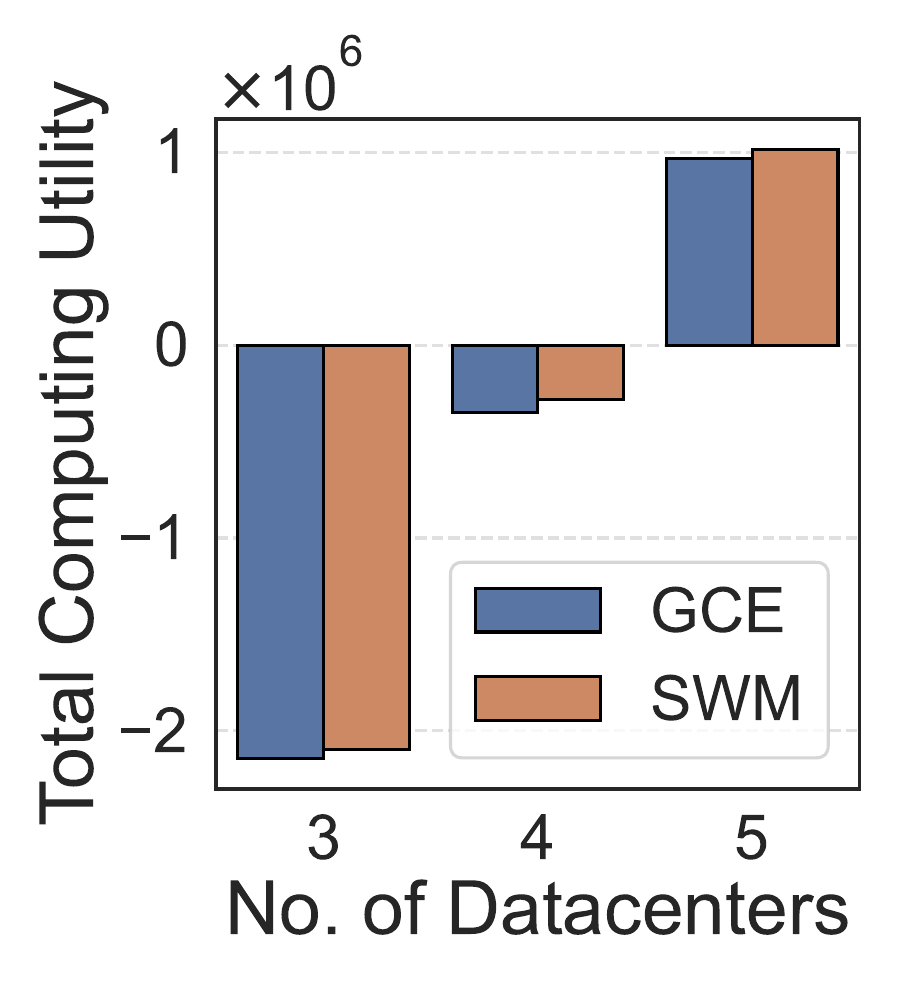}
        \caption{Computing Utility.}
        \label{fig:fl-dc}
    \end{subfigure}
    \caption{Impact of spatial flexibility with $q^\mathrm{EV} = 20$ kWh and $q^\mathrm{DC} = 10$ kWh in the congested power system.}
    \label{fig:impact_FL}
    \vspace{-1em}
\end{figure}

% We \jq{examine the impact of  spatial flexibility} of the datacenter network by varying the number of available datacenters from 1 to 5 in Fig.~\ref{fig:impact_FL} \jq{with the total workload from the AI companies fixed}, where fewer datacenters indicate less flexibility for AI companies to allocate their workloads. 
We examine the impact of datacenter spatial flexibility by varying the number of datacenters that each AI company can access, from one to five, in Fig.~\ref{fig:impact_FL}, while keeping the total workload for each company fixed.
Fewer available datacenters correspond to a more constrained workload-allocation decision space and less spatial flexibility.
Concretely, we construct a nested sequence of access sets by progressively enabling connections from each company to additional datacenters: starting with the site in Zone~J, then adding Zones~H, D, C, and A.
Thus, each step from 1 to 5 introduces one more siting option for workload placement, isolating the value of additional spatial flexibility.
The one- and two-datacenter cases are infeasible because both sites lie within the power system's most congested zone.
In Fig.~\ref{fig:fl-gen}, increasing the number of datacenters decreases generation cost, as greater spatial flexibility enables workload redistribution toward more efficient and less congested regions of the grid.
Spatial flexibility of the datacenter network likewise reduces EVs' total travel cost under both GCE and SWM frameworks in Fig.~\ref{fig:fl-ev} by flattening LMP disparities, thus mitigating  traffic congestion in the transportation network \jq{induced by drivers chasing lower charging prices}.
As shown in Fig.~\ref{fig:fl-dc}, computing utility improves with spatial flexibility due to reduced queuing delays. When datacenter locations are restricted, workloads concentrate in a few sites, resulting in negative utility from excessive waiting.
Overall, Fig. \ref{fig:impact_FL}  demonstrates that expanding spatial flexibility benefits all coupled infrastructures even in the GCE model.

\vspace{-0.24cm}
\section{Conclusion}
\label{sec:conclusion}
This paper introduces a unified framework for electricity markets that explicitly captures the bidirectional interaction between FLs and LMPs through the development and analysis of a GCE. 
We establish the key structural conditions under which this GCE is guaranteed to exist, be unique, and align with the social welfare optimum. 
A significant strength of this framework is that these desirable properties hold without requiring the system operator to have explicit knowledge of the FL agents' individual decision-making processes.
Through case studies integrating the NYISO grid with the Sioux Falls transportation and datacenter networks, we validate our theoretical findings. These simulations confirm the GCE's applicability and provide quantitative evidence of the significant mutual influence these coupled systems exert on one another, reinforcing the need for such an integrated equilibrium model.
Future work will relax the time-invariant setting by extending the GCE framework to multi-period models with intertemporal constraints across coupled infrastructure systems.

\vspace{-0.24cm}
\bibliographystyle{IEEEtran}
% argument is your BibTeX string definitions and bibliography database(s)
\bibliography{myabbrv, ref}

@STRING{IEEE_J_CNS        = "{IEEE} Trans. Control of Network Systems"}

@STRING{IEEE_J_PDS        = "{IEEE} Trans. Parallel Distrib. Syst."}

@STRING{IEEE_J_PWRAS      = "{IEEE} Trans. Power App. Syst."}

@STRING{IEEE_J_PWRS       = "{IEEE} Trans. Power Syst."}

@STRING{IEEE_J_SG 	      = "{IEEE} Trans. Smart Grid"}

@STRING{IEEE_J_STE 	      = "{IEEE} Trans. Sustain. Energy"}

@book{nocedal2006numerical,
  title={Numerical optimization},
  author={Nocedal, Jorge and Wright, Stephen J},
  year={2006},
  publisher={Springer}
}

@article{faechinei2003finite,
  title={Finite-dimensional variational inequalities and complementarity problems},
  author={Faechinei, F and Pang, JS},
  journal={New York: Springer-Verlag},
  year={2003}
}

@article{roughgarden2002bad,
  title={How bad is selfish routing?},
  author={Roughgarden, Tim and Tardos, {\'E}va},
  journal={Journal of the ACM},
  volume={49},
  number={2},
  pages={236--259},
  year={2002},
  publisher={ACM New York, NY, USA}
}

@article{alizadeh2016optimal,
  title={Optimal pricing to manage electric vehicles in coupled power and transportation networks},
  author={Alizadeh, Mahnoosh and Wai, Hoi-To and Chowdhury, Mainak and Goldsmith, Andrea and Scaglione, Anna and Javidi, Tara},
  journal=IEEE_J_CNS,
  volume={4},
  number={4},
  pages={863--875},
  year={2016},
  publisher={IEEE}
}

@Misc{ev-increase,
note = {\url{https://www.eia.gov/todayinenergy/detail.php?id=62083}},
title = {
{U.S}. electricity consumption by light-duty vehicles likely surpassed rail in 2023
},
author={{U.S.} {Energy} {Information} Administration}
}

@Misc{dc-increase,
note = {\url{https://www.iea.org/reports/world-energy-outlook-2024}},
title = {
World Energy Outlook 2024
},
author={IEA}
}

@ARTICLE{10636017,
  author={Dvorkin, Vladimir},
  journal=IEEE_J_PWRS, 
  title={Agent Coordination via Contextual Regression {(AgentCONCUR)} for Data Center Flexibility}, 
  year={2025},
  volume={40},
  number={2},
  pages={1832-1842},
  doi={10.1109/TPWRS.2024.3442954}}

@article{ZHANG2020106723,
title = {Flexibility from networks of data centers: A market clearing formulation with virtual links},
journal = {Electric Power Systems Research},
volume = {189},
pages = {106723},
year = {2020},
issn = {0378-7796},
doi = {https://doi.org/10.1016/j.epsr.2020.106723},
author = {Weiqi Zhang and Line A. Roald and Andrew A. Chien and John R. Birge and Victor M. Zavala}
}

@ARTICLE{7967870,
  author={Wei, Wei and Wu, Lei and Wang, Jianhui and Mei, Shengwei},
  journal=IEEE_J_SG, 
  title={Network Equilibrium of Coupled Transportation and Power Distribution Systems}, 
  year={2018},
  volume={9},
  number={6},
  pages={6764-6779},
  doi={10.1109/TSG.2017.2723016}}

@ARTICLE{7572952,
  author={Wei, Wei and Mei, Shengwei and Wu, Lei and Shahidehpour, Mohammad and Fang, Yujuan},
  journal=IEEE_J_SG, 
  title={Optimal Traffic-Power Flow in Urban Electrified Transportation Networks}, 
  year={2017},
  volume={8},
  number={1},
  pages={84-95},
  doi={10.1109/TSG.2016.2612239}}

@ARTICLE{9891825,
  author={Shao, Chengcheng and Li, Ke and Qian, Tao and Shahidehpour, Mohammad and Wang, Xifan},
  journal=IEEE_J_SG, 
  title={Generalized User Equilibrium for Coordination of Coupled Power-Transportation Network}, 
  year={2023},
  volume={14},
  number={3},
  pages={2140-2151},
  doi={10.1109/TSG.2022.3206511}}

@ARTICLE{8737720,
  author={Rossi, Federico and Iglesias, Ramon and Alizadeh, Mahnoosh and Pavone, Marco},
  journal=IEEE_J_CNS, 
  title={On the Interaction Between Autonomous Mobility-on-Demand Systems and the Power Network: Models and Coordination Algorithms}, 
  year={2020},
  volume={7},
  number={1},
  pages={384-397},
  doi={10.1109/TCNS.2019.2923384}}

@article{tsiligkaridis2025distributed,
  title={Distributed Economic Dispatch in Power Networks Incorporating Data Center Flexibility},
  author={Tsiligkaridis, Athanasios and Andrianesis, Panagiotis and Coskun, Ayse K and Caramanis, Michael C and Paschalidis, Ioannis Ch},
  journal={IEEE Trans. Sustain. Comput.},
  year={2025},
  publisher={IEEE}
}

@Misc{SiouxFalls,
note = {\url{https://github.com/bstabler/TransportationNetworks/}},
title = {
Sioux Falls Network
},
author={{Transportation} {Networks} for Research Core Team}
}

@Misc{EVchargingspeed,
note = {\url{https://www.ev.guide/lesson-articles/ev-charging-speed}},
title = {
{EV} Charging Speed
},
author={EV. Guide}
}

@article{hall2024carbon,
  title={Carbon-Aware Computing for Data Centers with Probabilistic Performance Guarantees},
  author={Hall, Sophie and Micheli, Francesco and Belgioioso, Giuseppe and Radovanovi{\'c}, Ana and D{\"o}rfler, Florian},
  journal={arXiv preprint arXiv:2410.21510},
  year={2024}
}

@inproceedings{li2024towards,
  title={Towards environmentally equitable {AI} via geographical load balancing},
  author={Li, Pengfei and Yang, Jianyi and Wierman, Adam and Ren, Shaolei},
  booktitle={Proceedings of the 15th ACM International Conference on Future and Sustainable Energy Systems},
  pages={291--307},
  year={2024}
}

@INPROCEEDINGS{6322266,
  author={Lin, Minghong and Liu, Zhenhua and Wierman, Adam and Andrew, Lachlan L. H.},
  booktitle={2012 IGCC}, 
  title={Online algorithms for geographical load balancing}, 
  year={2012},
  volume={},
  number={},
  pages={1-10},
  doi={10.1109/IGCC.2012.6322266}}

@ARTICLE{8540414,
  author={Zhou, Zhi and Liu, Fangming and Chen, Shutong and Li, Zongpeng},
  journal=IEEE_J_PDS, 
  title={A Truthful and Efficient Incentive Mechanism for Demand Response in Green Datacenters}, 
  year={2020},
  volume={31},
  number={1},
  pages={1-15},
  doi={10.1109/TPDS.2018.2882174}}

@ARTICLE{9963634,
  author={Zhang, Weiqi and Roald, Line and Zavala, Victor},
  journal={IEEE Trans. Sustain. Comput.}, 
  title={Exploring the Impacts of Power Grid Signals on Data Center Operations Using a Receding-Horizon Scheduling Model}, 
  year={2023},
  volume={8},
  number={2},
  pages={245-256},
  doi={10.1109/TSUSC.2022.3224668}}

@inproceedings{werner2021pricing,
  title={Pricing flexibility of shiftable demand in electricity markets},
  author={Werner, Lucien and Wierman, Adam and Low, Steven H},
  booktitle={Proceedings of the Twelfth ACM International Conference on Future Energy Systems},
  pages={1--14},
  year={2021}
}

@ARTICLE{7517380,
  author={Kim, Kibaek and Yang, Fan and Zavala, Victor M. and Chien, Andrew A.},
  journal=IEEE_J_STE, 
  title={Data Centers as Dispatchable Loads to Harness Stranded Power}, 
  year={2017},
  volume={8},
  number={1},
  pages={208-218},
  doi={10.1109/TSTE.2016.2593607}}

@inproceedings{wierman2014opportunities,
  title={Opportunities and challenges for data center demand response},
  author={Wierman, Adam and Liu, Zhenhua and Liu, Iris and Mohsenian-Rad, Hamed},
  booktitle={International Green Computing Conference},
  pages={1--10},
  year={2014},
  organization={IEEE}
}

@ARTICLE{9770383,
  author={Radovanović, Ana and Koningstein, Ross and Schneider, Ian and Chen, Bokan and Duarte, Alexandre and Roy, Binz and Xiao, Diyue and Haridasan, Maya and Hung, Patrick and Care, Nick and Talukdar, Saurav and Mullen, Eric and Smith, Kendal and Cottman, MariEllen and Cirne, Walfredo},
  journal=IEEE_J_PWRS, 
  title={Carbon-Aware Computing for Datacenters}, 
  year={2023},
  volume={38},
  number={2},
  pages={1270-1280},
  doi={10.1109/TPWRS.2022.3173250}}

@article{fridgen2017shifting,
  title={Shifting load through space--The economics of spatial demand side management using distributed data centers},
  author={Fridgen, Gilbert and Keller, Robert and Thimmel, Markus and Wederhake, Lars},
  journal={Energy Policy},
  volume={109},
  pages={400--413},
  year={2017},
  publisher={Elsevier}
}

@article{zhang2022remunerating,
  title={Remunerating space--time, load-shifting flexibility from data centers in electricity markets},
  author={Zhang, Weiqi and Zavala, Victor M},
  journal={Applied Energy},
  volume={326},
  pages={119930},
  year={2022},
  publisher={Elsevier}
}

@ARTICLE{9330803,
  author={Cui, Yan and Hu, Zechun and Duan, Xiaoyu},
  journal=IEEE_J_SG, 
  title={Optimal Pricing of Public Electric Vehicle Charging Stations Considering Operations of Coupled Transportation and Power Systems}, 
  year={2021},
  volume={12},
  number={4},
  pages={3278-3288},
  doi={10.1109/TSG.2021.3053026}}

@ARTICLE{8445618,
  author={Wang, Xu and Shahidehpour, Mohammad and Jiang, Chuanwen and Li, Zhiyi},
  journal=IEEE_J_PWRS, 
  title={Coordinated Planning Strategy for Electric Vehicle Charging Stations and Coupled Traffic-Electric Networks}, 
  year={2019},
  volume={34},
  number={1},
  pages={268-279},
  doi={10.1109/TPWRS.2018.2867176}}

@ARTICLE{9748107,
  author={Xie, Shiwei and Wu, Qiuwei and Hatziargyriou, Nikos D. and Zhang, Menglin and Zhang, Yachao and Xu, Yinliang},
  journal=IEEE_J_PWRS, 
  title={Collaborative Pricing in a Power-Transportation Coupled Network: A Variational Inequality Approach}, 
  year={2023},
  volume={38},
  number={1},
  pages={783-795},
  doi={10.1109/TPWRS.2022.3162861}}

@article{sheng2021coordinated,
  title={Coordinated pricing of coupled urban power-traffic networks: The value of information sharing},
  author={Sheng, Yujie and Guo, Qinglai and Chen, Feng and Xu, Luo and Zhang, Yang},
  journal={Applied Energy},
  volume={301},
  pages={117428},
  year={2021},
  publisher={Elsevier}
}

@ARTICLE{8304623,
  author={Wang, Shuoyao and Bi, Suzhi and Zhang, Ying-Jun Angela and Huang, Jianwei},
  journal=IEEE_J_STE, 
  title={Electrical Vehicle Charging Station Profit Maximization: Admission, Pricing, and Online Scheduling}, 
  year={2018},
  volume={9},
  number={4},
  pages={1722-1731},
  doi={10.1109/TSTE.2018.2810274}}

@article{zhang2020power,
  title={Power and transport nexus: Routing electric vehicles to promote renewable power integration},
  author={Zhang, Hongcai and Hu, Zechun and Song, Yonghua},
  journal=IEEE_J_SG,
  volume={11},
  number={4},
  pages={3291--3301},
  year={2020},
  publisher={IEEE}
}

@article{tan2015real,
  title={Real-time charging navigation of electric vehicles to fast charging stations: A hierarchical game approach},
  author={Tan, Jun and Wang, Lingfeng},
  journal=IEEE_J_SG,
  volume={8},
  number={2},
  pages={846--856},
  year={2015},
  publisher={IEEE}
}

@article{guo2014rapid,
  title={Rapid-charging navigation of electric vehicles based on real-time power systems and traffic data},
  author={Guo, Qinglai and Xin, Shujun and Sun, Hongbin and Li, Zhengshuo and Zhang, Boming},
  journal=IEEE_J_SG,
  volume={5},
  number={4},
  pages={1969--1979},
  year={2014},
  publisher={IEEE}
}

@article{hsu1997introduction,
  title={An introduction to the pricing of electric power transmission},
  author={Hsu, Michael},
  journal={Utilities Policy},
  volume={6},
  number={3},
  pages={257--270},
  year={1997},
  publisher={Elsevier}
}

@ARTICLE{4111733,
  author={Caramanis, M. C. and Bohn, R. E. and Schweppe, F. C.},
  journal=IEEE_J_PWRAS, 
  title={Optimal Spot Pricing: Practice and Theory}, 
  year={1982},
  volume={PAS-101},
  number={9},
  pages={3234-3245},
  doi={10.1109/TPAS.1982.317507}}

@book{sheffi1985urban,
  title={Urban transportation networks},
  author={Sheffi, Yosef},
  volume={6},
  year={1985},
  publisher={Prentice-Hall, Englewood Cliffs, NJ}
}

@book{urruty1993convex,
  title={Convex analysis and minimization algorithms},
  author={Urruty, Jean-Baptiste Hiriart and Lemar{\'e}chal, Claude},
  year={1993},
  publisher={Springer-Verlag}
}

@article{monderer1996potential,
  title={Potential games},
  author={Monderer, Dov and Shapley, Lloyd S},
  journal={Games and economic behavior},
  volume={14},
  number={1},
  pages={124--143},
  year={1996},
  publisher={Elsevier}
}

@article{van2025political,
  title={The Political Economy of Congestion Pricing},
  author={Van Doren, Peter and Weisman, Dennis L},
  journal={Regulation},
  volume={48},
  pages={14},
  year={2025},
  publisher={HeinOnline}
}

@INPROCEEDINGS{10644866,
  author={Mou, Minghao and Qian, Sean and Qin, Junjie},
  booktitle={2024 American Control Conference}, 
  title={Nexus Cognizant Pricing of Workplace Electric Vehicle Charging}, 
  year={2024},
  volume={},
  number={},
  pages={4275-4282},
  doi={10.23919/ACC60939.2024.10644866}}

@article{Desislavov_2023,
   title={Trends in {AI} inference energy consumption: Beyond the performance-vs-parameter laws of deep learning},
   volume={38},
   ISSN={2210-5379},
   url={http://dx.doi.org/10.1016/j.suscom.2023.100857},
   DOI={10.1016/j.suscom.2023.100857},
   journal={Sustainable Computing: Informatics and Systems},
   publisher={Elsevier BV},
   author={Desislavov, Radosvet and Martínez-Plumed, Fernando and Hernández-Orallo, José},
   year={2023},
   month=apr, pages={100857} }

@ARTICLE{4956966,
  author={Stott, Brian and Jardim, Jorge and Alsac, Ongun},
  journal={IEEE Transactions on Power Systems}, 
  title={{DC} Power Flow Revisited}, 
  year={2009},
  volume={24},
  number={3},
  pages={1290-1300},
  doi={10.1109/TPWRS.2009.2021235}}

@misc{brenner2026strategicdatacenterload,
      title={Strategic Data Center Load Shifting: Implications for Market Efficiency and Transmission Value}, 
      author={Aron Brenner and Line Roald and Saurabh Amin},
      year={2026},
      eprint={2510.20805},
      archivePrefix={arXiv},
      primaryClass={eess.SY},
      url={https://arxiv.org/abs/2510.20805}, 
}
% \newpage
\appendices

\section{Non-Potential Game Case for Lemma \ref{lemma:equiva_NE}}
\label{app:non-potential}
When condition \eqref{ass:dc_potential} does not hold, the FL
system's best response to LMP can no longer be written as a convex program in the form of~\eqref{eq:sfl-response}. The definitions of $\Phi(\mathbf{x})$ and $J(s)$ in Section~\ref{sec:unifying_framework} do not apply here.
Nevertheless, we can still characterize the price-load mapping and the associated GCE by introducing the
\emph{merit function} of the VI:
% \begin{assumption}
%     Assume $\mathbf{F}$ is strictly {monotone} on the compact, convex set $\mathcal{X}$.
%     \label{ass:nonpotential}
% \end{assumption}
\begin{theorem}[GCE for a Non-Potential Game]
    \label{thm:nonpotential_UE_sigma}
        % \item
        %  {Merit function for NE.}
% Denote \(\mathbf{A}^\mathrm{FL} = q\bigl(\mathbf{A}^\mathrm{DB}\bigr)^{\!\top}\mathbf{A}^\mathrm{sum}\).
% \begin{enumerate}[label=\textnormal{(\roman*)},leftmargin=*]
%     \item  and
%     \item the aggregation matrix $\mathbf{A}\!\in\!\mathbb{R}^{N\times KN^\mathrm{D}}$ has full row-rank.
% \end{enumerate}
Suppose each utility function \(u_k(\mathbf{x}^{(k)};\mathbf{x}^{(-k)})\) is continuously differentiable and let $\mathbf{F}(\mathbf{x})$ follow the definition in Section~\ref{sec:resultsDC}. Let $\nabla \mathbf{C}(\mathbf{g})$ denote the gradient of the total generation cost function $C(\mathbf{g})$ with respect to the generator output vector $\mathbf{g}$. Assume that for every $\tilde{\mathbf{x}}\in\mathcal{X}$ and $\tilde{\mathbf{g}}\in\mathcal{G}$, the functions $\mathbf{x}\mapsto-\mathbf{F}(\mathbf{x})^{\top}(\mathbf{x}-\tilde{\mathbf{x}})$ and $\mathbf{g}\mapsto\nabla \mathbf{C}(\mathbf{g})^{\top}(\mathbf{g}-\tilde{\mathbf{g}})$ are convex on $\mathcal{X}$ and $\mathcal{G}$, respectively. The following statements hold.
    % % \nan{Let $\mathbf{F}\!:\!\mathcal{X}\!\to\!\mathbb{R}^{K N_\mathrm{D}}$ stack the gradients of each company's utility $u_k$ with respect to their own decision variable $\mathbf{x}^{(k)}$, i.e.,
    % \(
    %    \jq{ \mathbf{F}(\mathbf{x}) \!=\! \bigl(\nabla_{\mathbf{x}^{(1)}} u_1(\mathbf{x})^\top,\dots,\nabla_{\mathbf{x}^{(K)}} u_K(\mathbf{x})^\top\bigr)^\top. }
    % \)%
    % }
\begin{enumerate}[label=\roman*), leftmargin = 0.5cm]
\item For any fixed LMP vector $\boldsymbol{\lambda}\!\in\!\mathbb{R}^{N}$, an NE \(\mathbf{x}^\mathrm{NE}(\boldsymbol{\lambda})\) satisfies~\eqref{eq:vi-dc}, which can also be expressed as the minimizer of the following merit-function optimization:
\begin{equation}
\min_{\mathbf{x}\in\mathcal{X}}
\sup_{\tilde{\mathbf{x}}\in\mathcal{X}}
-\mathbf{F}(\mathbf{x})^{\!\top}(\mathbf{x}-\tilde{\mathbf{x}})
\!+\!
\boldsymbol{\lambda}^{\!\top}\!\bigl(\mathbf{A}^\mathrm{FL}\mathbf{x}-\mathbf{A}^\mathrm{FL}\tilde{\mathbf{x}}\bigr).
\label{eq:nonpotential_UE}
\end{equation}
Assuming $-\mathbf{F}$ is strictly {monotone} on the compact, convex set $\mathcal{X}$, we have that \(\mathbf{x}^\mathrm{NE}(\boldsymbol{\lambda})\) is unique.

\item  % {Price-Load Mapping.}
It follows that the \emph{price-load mapping}
$\mathbf{s}^\star = \boldsymbol{\sigma}(\boldsymbol{\lambda})$
can be expressed as the minimizer of the following optimization:
\begin{equation}
\min_{\mathbf{s}\in\mathcal{S}}
\;
\sup_{\tilde{\mathbf{s}}\in\mathcal{S}}
\Bigl[
J(\mathbf{s},\tilde{\mathbf{s}})
+\boldsymbol{\lambda}^{\top}(\mathbf{s}-\tilde{\mathbf{s}})
\Bigr],
\label{non-p:price-load}
\end{equation}
where $J(\mathbf{s},\tilde{\mathbf{s}})$ maps each given $(\mathbf{s},\tilde{\mathbf{s}})$ to the optimal value of the following problem:
\[
\min_{\mathbf{x}\in\mathcal{X}}
\left\{
  \sup_{\tilde{\mathbf{x}}\in\mathcal{X}}
  -\mathbf{F}(\mathbf{x})^{\top}(\mathbf{x}-\tilde{\mathbf{x}})
  \ \mathrm{s.t.}\ 
  \mathbf{s} = \mathbf{A}^{\mathrm{FL}}\mathbf{x},\
  \tilde{\mathbf{s}} = \mathbf{A}^{\mathrm{FL}}\tilde{\mathbf{x}}
\right\}.
\]
\item % {GCE Characterization.}
%  Let $J$ be as defined in Theorem~\ref{thm:nonpotential_UE_sigma}, and let $\mathbf{C}(\cdot)$ denote the convex generation-cost function.
A triple $(\mathbf{g}^{\star},\!\mathbf{s}^{\star},\boldsymbol{\lambda}^{\star})$ constitutes a GCE if and only if \((\mathbf{s}^{\star},\!\mathbf{g}^{\star})\) is the minimizer to the following problem
\begin{equation}
    \begin{aligned}
        \min_{\substack{\mathbf{s}\in\mathcal{S},\,\mathbf{g}\in\mathcal{G}}}\;
        \sup_{\substack{\tilde{\mathbf{s}}\in\mathcal{S},\,\tilde{\mathbf{g}}\in\mathcal{G}}}
        & \Bigl[
            J(\mathbf{s},\tilde{\mathbf{s}})
            + \nabla \mathbf{C}(\mathbf{g})^{\top}(\mathbf{g}-\tilde{\mathbf{g}})
        \Bigr]\\
\mathrm{s.t.}   \quad
&\mathbf{p} = \mathbf{g} - \boldsymbol{\ell} - \mathbf{s} \in \mathcal{P},
\
\tilde{\mathbf{p}} = \tilde{\mathbf{g}} - \boldsymbol{\ell} - \tilde{\mathbf{s}} \in \mathcal{P},
    \end{aligned}
    \label{eq:GCE-minmax}
\end{equation}
and the corresponding LMP vector $\boldsymbol{\lambda}^{\star}$ is the Lagrange multiplier associated with the nodal constraint in the economic dispatch problem~\eqref{unify:economic dispatch} given $\mathbf{s}^{\star}$.
\end{enumerate}

\end{theorem}

\section{Parameters of the AI Companies' Utility Functions in the Case Study}
\label{app:case-utility}
In~\eqref{eq:quadratic-utility}, each component $\eta_{k,j}$ of the baseline utility vector represents company $k$'s willingness to pay (or preference) for datacenter $j$, denoted by $\tau^{(k)}_j$, minus a communication cost proportional to the distance $\gamma_k d^{(k)}_j$ between them.
The quadratic cost term captures two effects.
First, datacenter queueing~\cite{tsiligkaridis2025distributed} implies that the per-job waiting cost at datacenter $j$ increases with the aggregate workload $\sum_{\tilde{k}} x_j^{(\tilde{k})}$, scaled by a queueing coefficient $\alpha_j\ge 0$.
As a result, company $k$ incurs a total waiting cost $\sum_j \alpha_j x_j^{(k)}\sum_{\tilde{k}} x_j^{(\tilde{k})}$, which couples the decisions of competing companies.
Second, diminishing marginal returns for company $k$ are captured by an internal quadratic cost $\tfrac{1}{2}\omega_k\sum_j (x_j^{(k)})^2$ with $\omega_k>0$.
These two effects yield the block structure of~$\mathbf{M}$: for each $\tilde{k}\ne k$, the off-diagonal blocks satisfy $\mathbf{M}_{k,\tilde{k}}=\operatorname{diag}(\alpha_1,\ldots,\alpha_{N^\mathrm{D}})$, and the row block $\mathbf{M}_{k,-k}$ concatenates $\{\mathbf{M}_{k,\tilde{k}}\}_{\tilde{k}\ne k}$.
The diagonal blocks satisfy $\mathbf{M}_{k,k}=\operatorname{diag}(2\alpha_1+\omega_k,\ldots,2\alpha_{N^\mathrm{D}}+\omega_k)$, ensuring concavity of company $k$'s utility.
\begin{table}[t]
    \centering
    \caption{Parameters in AI Companies' Utility Functions~\eqref{eq:quadratic-utility}.}
    \label{tab:case_study_parameters}
    \begin{tabular}{c|l|l}
        \toprule
        Param & Description & Value \\ \midrule
        \( \tau_{j}^{(k)} \) & Company $k$'s willingness to pay for datacenter $j$ & $10^{1} \cdot\mathcal{U}(1, 2)$ \\ 
        \( \gamma_k \) & Company $k$'s sensitivity to distance & $10^{-1} \cdot \mathcal{U}(2, 4)$ \\ 
        \( d_{j}^{(k)} \)  & Distance from company $k$ to datacenter $j$ & based on \cite{10636017} \\ 
        \( \omega_k \)& Diminishing returns coefficient &  $ 10^{-4} \cdot \mathcal{U}(1, 5) $ \\ 
        \( \alpha_{j} \) & Queueing coefficient of datacenter $j$ & $10^{-4} \cdot \mathcal{U}(1, 2)$   \\ 
        $\zeta_k$ & Total workload of user $k$ & $10^{4} \cdot  \mathcal{U}(1, 5)$ \\
        \bottomrule
\end{tabular}
\end{table}
% Therefore, the off-diagonal blocks as $\mathbf{M}_{k,\tilde{k}} = \operatorname{diag}(\alpha_1,\ldots,\alpha_{N^\mathrm{D}})$ for every $\tilde{k} \ne k$, and the row block matrix $\mathbf{M}_{k,-k}$ horizontally concatenates all the interaction matrices $\mathbf{M}_{k,\tilde{k}}$.
% Then the diagonal blocks satisfy $\mathbf{M}_{k,k} = \operatorname{diag}(2\alpha_1+\omega_k,\ldots,2\alpha_{N^\mathrm{D}}+\omega_k)$. This formulation guarantees the utility function's concavity, as the requirement $\mathbf{M}_{k,k}\succeq 0$ is always satisfied.
\nan{
    These parameters of the utility functions are selected according to the aforementioned criteria and summarized in Table~\ref{tab:case_study_parameters}, where $\mathcal{U}(\cdot)$ denotes the uniform distribution. Once \jq{the random components of the parameters in the table are} determined, these parameters remain fixed throughout all experiments for consistency. }

\section{Proofs and Supplementary Results in Section \ref{sec:general_results}}
\label{app:sec3-proofs}
% \section{Proof for Theorem \ref{thm:GCE-Structural}}
\begin{proof}[Proof of Theorem \ref{thm:GCE-Structural}]
    The primal problem~\eqref{prob:P_GCE} is convex, since $J(\mathbf{s})$ and $C(\mathbf{g})$ are convex and the coupling constraint $\mathbf{p} = \mathbf{g} - \boldsymbol{\ell} - \mathbf{s}$ is affine. 
    Define the Lagrangian:
    \begin{equation}
        \mathcal{L}^\mathrm{GCE}(\mathbf{g}, \mathbf{s}, \mathbf{p}, \boldsymbol{\lambda})
        = J(\mathbf{s}) + C(\mathbf{g})
        + \boldsymbol{\lambda}^\top(\mathbf{p} - \mathbf{g} + \boldsymbol{\ell} + \mathbf{s}).
    \end{equation}
    By convex analysis, the KKT conditions are both necessary and sufficient for optimality. Moreover, the KKT system for the primal-dual optimum 
$(\mathbf{g}^\star,\mathbf{s}^\star,\mathbf{p}^\star,\boldsymbol{\lambda}^\star)$
can be written as the following stationarity and primal feasibility conditions:
\begin{subequations}\label{kkt-con-gce}
\begin{align}
    - \partial  C(\mathbf{g}^\star) + \boldsymbol{\lambda}^\star   &\in  N_{\mathcal{G}}(\mathbf{g}^\star), \label{kkt:g}\\
    -\partial J(\mathbf{s}^\star) - \boldsymbol{\lambda}^\star  &\in  N_{\mathcal{S}}(\mathbf{s}^\star), \label{kkt:s}\\
    - \boldsymbol{\lambda}^\star &\in  N_{\mathcal{P}}(\mathbf{p}^\star), 
              \label{kkt:p}\\
\mathbf{0} &= \mathbf{p}^\star - \mathbf{g}^\star + \boldsymbol{\ell} + \mathbf{s}^\star. 
              \label{kkt:feas}
\end{align}
\end{subequations}
Here, $N_{\mathcal{G}}(\cdot)$, $N_{\mathcal{S}}(\cdot)$, and $N_{\mathcal{P}}(\cdot)$ denote the normal cones to the corresponding feasible sets, where $N_{\mathcal{X}}(x)$ is defined as $N_{\mathcal{X}}(x)=\{\,\nu\mid \nu^\top(y-x)\le0,\,\forall\,y\in\mathcal{X}\,\}$.

    Each condition is equivalent to the optimality condition of a separate subproblem:  
    \begin{subequations}
    \begin{align}
        \eqref{kkt:g} \Leftrightarrow \quad &
        \mathbf{g}^\star 
        = \argmin_{\mathbf{g}\in\mathcal{G}}
          \{\,C(\mathbf{g}) - \boldsymbol{\lambda}^{\star\top}\mathbf{g}\,\},\label{g-kkt}\\ 
          \eqref{kkt:s} \Leftrightarrow\quad & \mathbf{s}^\star \in \argmin_{\mathbf{s} \in \mathcal{S}} \left\{ J(\mathbf{s}) + \boldsymbol{\lambda}^{\star\top}\mathbf{s} \right\}, \label{s-kkt}\\
          \eqref{kkt:p} \Leftrightarrow \quad & \mathbf{p}^\star \in \argmin_{\mathbf{p} \in \mathcal{P}} \left\{ \boldsymbol{\lambda}^{\star\top}\mathbf{p} \right\}, \label{p-kkt}\\
          \eqref{kkt:feas} \Leftrightarrow \quad & \mathbf{p}^\star = \mathbf{g}^\star - \boldsymbol{\ell} - \mathbf{s}^\star \label{feas-kkt}.
    \end{align}
    \end{subequations}
    We now show that these KKT conditions are a restatement of the GCE definition.
    \begin{enumerate}[label=\roman*), leftmargin=*]
        \item Classical CE Condition: 
        A pair $(\mathbf{g}^\star, \boldsymbol{\lambda}^\star)$ forms a classical CE given $\mathbf{s}^\star$ if and only if:
        \begin{itemize}[leftmargin = *]
            \item    Generator individual rationality holds, which is identical to \eqref{g-kkt};
            \item Feasibility holds, which is identical to~\eqref{feas-kkt};
            \item    Price optimality holds; the detailed reason is as follows: Given $\mathbf{s}$,  the Lagrangian of the economic dispatch problem~\eqref{unify:economic dispatch} is
            \(
               \mathcal{L}_{\mathrm{ED}}(\mathbf{g},\mathbf{p};\boldsymbol{\lambda})
               = C(\mathbf{g}) + \boldsymbol{\lambda}^\top\bigl(\mathbf{p}-\mathbf{g}+\boldsymbol{\ell}+\mathbf{s}\bigr).
            \)
            The dual function decomposes additively:
            \[
               Q(\boldsymbol{\lambda})
               = {\inf_{\mathbf{g}\in\mathcal{G}}\bigl\{C(\mathbf{g})-\boldsymbol{\lambda}^\top\mathbf{g}\bigr\}}
                 + {\inf_{\mathbf{p}\in\mathcal{P}}\boldsymbol{\lambda}^\top\mathbf{p}}
                 + \boldsymbol{\lambda}^\top(\boldsymbol{\ell}+\mathbf{s}).
            \]
            Hence the dual problem is $\max_{\boldsymbol{\lambda}} Q(\boldsymbol{\lambda})$. 
            At a dual maximizer $\boldsymbol{\lambda}^\star$, the corresponding primal minimizer is decomposed into two components:
            \[
               \mathbf{g}^\star \in \argmin_{\mathbf{g}\in\mathcal{G}}\{C(\mathbf{g})-\boldsymbol{\lambda}^{\star\top}\mathbf{g}\}, 
               \
               \mathbf{p}^\star \in \argmin_{\mathbf{p}\in\mathcal{P}}\{\boldsymbol{\lambda}^{\star\top}\mathbf{p}\},
            \]
            which are exactly~\eqref{g-kkt} and~\eqref{p-kkt}, respectively.
            % together with primal feasibility $\mathbf{p}^\star=\mathbf{g}^\star-\boldsymbol{\ell}-\mathbf{s}$. 
            % Equivalently, the KKT condition for the $\mathbf{p}$-subproblem is exactly~\eqref{p-kkt}.
        \end{itemize}
   \item FL Best Response Condition:  The FL power consumption $\mathbf{s}^\star$ is the aggregate best response, 
   $\mathbf{s}^\star = \boldsymbol{\sigma}(\boldsymbol{\lambda}^\star)$, 
   if and only if it solves $\min_{\mathbf{s} \in \mathcal{S}} \{ J(\mathbf{s}) + \boldsymbol{\lambda}^{\star\top}\mathbf{s} \}$, according to the definition in Section \ref{mode:sfl}.
   This is identical to~\eqref{s-kkt}.
    \end{enumerate}
Since each component of the GCE definition maps bi-directionally to~\eqref{kkt-con-gce}, a triple $(\mathbf{g}, \mathbf{s}, \boldsymbol{\lambda})$ is a GCE if and only if it is a primal-dual optimal solution to ($\mathrm{P}_{\mathrm{GCE}}$).
\end{proof}

% \section{Proof for Corollary~\ref{thm:gce_existence}}
\begin{proof}[Proof of Corollary~\ref{thm:gce_existence}]
    \emph{Existence.} We have:
    \begin{enumerate}[label=(C\arabic*),leftmargin=*]
    \item $\mathcal G,\mathcal S$ and the associated $\mathcal F$ are nonempty, convex, and compact;
    \item $J$ and $C$ are continuous, and convex on $\mathcal S$ and $\mathcal G$, respectively.
    \end{enumerate}
    Then the objective \(J(\mathbf s)+C(\mathbf g)\) is continuous on the compact feasible set $\mathcal F$. By Weierstrass' theorem, a minimizer \((\mathbf g^\star,\mathbf s^\star,\mathbf p^\star)\) exists. Since the constraint $\mathbf{p} = \mathbf{g} - \boldsymbol{\ell} - \mathbf{s}$ is affine, KKT ensures the existence of a multiplier \(\boldsymbol{\lambda}^\star\) satisfying the system \eqref{kkt-con-gce}. Hence a GCE exists.
    
    \emph{Uniqueness.} Since  $J$ is strictly convex on $\mathcal S$ and $C$ is strictly convex on $\mathcal G$,
 over $\mathcal F$ the objective reduces to the strictly convex function \(J(\mathbf s)+C(\mathbf g)\) in the variables $(\mathbf g,\mathbf s)$ with a convex feasible set $\mathcal F$. A strictly convex function has at most one minimizer; hence $(\mathbf g^\star,\mathbf s^\star)$ is unique, and so is $\mathbf p^\star=\mathbf g^\star-\boldsymbol{\ell}-\mathbf s^\star$.   
    The uniqueness of the KKT multiplier \(\boldsymbol{\lambda}^\star\) is ensured if linear independence constraint qualification holds for the active constraints at  \((\mathbf g^\star,\mathbf s^\star,\mathbf p^\star)\). 
\end{proof}

% \section{Proof for Lemma \ref{thm:GCE-SWM}}

\begin{proof}[Proof of Lemma \ref{thm:GCE-SWM}]
    \label{app:wel-ev}
Substituting the definition of \(J^{\mathrm{SW}}\) into \eqref{prob:P_SW} merges the outer and inner minimizations, yielding exactly the social-welfare program~\eqref{eq:swm_ori} since $\mathbf{s} = \mathbf{A}^\mathrm{FL}\mathbf{x}$.
Hence, an optimal solution \((\mathbf{s}^\star,\mathbf{g}^\star,\mathbf{p}^\star)\) admits an \(\mathbf{x}^\star\) attaining \(J^\mathrm{SW}(\mathbf{s}^\star)\) with \(\mathbf{A}^\mathrm{FL}\mathbf{x}^\star=\mathbf{s}^\star\).
Conversely, an \((\mathbf{x}^\star,\mathbf{g}^\star,\mathbf{p}^\star)\) solves \eqref{eq:swm_ori}, establishing the equivalence.
\end{proof}

% \section{Proof for Corollary~\ref{thm:gce_efficiency}}
\begin{proof}[Proof of Corollary~\ref{thm:gce_efficiency}]
    % Let $\varphi(\mathbf{s},\mathbf{g}) := J^{\mathrm{SW}}(\mathbf{s}) + C(\mathbf{g})$. By Lemma~\ref{thm:GCE-SWM}, SWM is equivalent to
    % \[
    % \min_{(\mathbf{s},\mathbf{g})\in \mathcal{F}} \ \varphi(\mathbf{s},\mathbf{g}).
    % \]
    Since $J^{\mathrm{SW}}$ and $C$ are convex and $\mathcal{F}$ is closed and convex, the standard subgradient optimality condition for minimizing a convex function over a convex set (see \cite{urruty1993convex}) implies \((\mathbf{s}^\star,\mathbf{g}^\star)\) is optimal for \(\min_{(\mathbf{s},\mathbf{g})\in\mathcal{F}} J^{\mathrm{SW}}(\mathbf{s}) + C(\mathbf{g})\) if and only if there exist \(\bm{\mu}\in\partial J^{\mathrm{SW}}(\mathbf{s}^\star),\ \bm{\nu}\in\partial C(\mathbf{g}^\star)\) such that
    \begin{equation}
\langle \bm{\mu},\, \mathbf{s}-\mathbf{s}^\star\rangle+\langle \bm{\nu},\, \mathbf{g}-\mathbf{g}^\star\rangle \ge 0,\ \forall\,(\mathbf{s},\mathbf{g})\in\mathcal{F}.
\end{equation}
    % Taking $(\mathbf{s}^\star,\mathbf{g}^\star)=(\mathbf{s}^{\mathrm{GCE}},\mathbf{g}^{\mathrm{GCE}})$ gives exactly condition~\eqref{eq:vi-eff}. 
    Consequently, $(\mathbf{s}^{\mathrm{GCE}},\mathbf{g}^{\mathrm{GCE}})$ solves~\eqref{prob:P_SW} if and only if~\eqref{eq:vi-eff} holds, i.e., the GCE is efficient if and only if~\eqref{eq:vi-eff} holds.
    \end{proof}
% Finally, if \(J = J^{\mathrm{SW}}\), then problems \eqref{prob:P_GCE} and \eqref{prob:P_SW} coincide in both objective and constraints, proving Corollary~\ref{thm:gce_efficiency}.

% \section{Proof for Theorem~\ref{thm:GCE-multi}}
\begin{proof}[Proof of Theorem~\ref{thm:GCE-multi}]
The generalization follows closely from the proof of Theorem~\ref{thm:GCE-Structural}, with only notational extensions to accommodate multiple FL systems.
Under Definition~\ref{def:stacked-FL}, the optimization problem~\eqref{prob:P_GCE}
% \[
% \min_{\mathbf{s}\in\mathcal{S},\,\mathbf{g}\in\mathcal{G},\,\mathbf{p}\in\mathcal{P}}
% \bigl\{ J(\mathbf{s}) + C(\mathbf{g}) : \mathbf{p} = \mathbf{g} - \boldsymbol{\ell} - \mathbf{S}\mathbf{1} \bigr\}
% \]
is convex, since $J(\mathbf{s})=\sum_w J_w(\mathbf{S}_w)$ and $C(\mathbf{g})$ are convex and the coupling constraint is affine.
The Lagrangian is
\begin{equation}
\mathcal{L}^\mathrm{multi}(\mathbf{g},\mathbf{s},\mathbf{p},\boldsymbol{\lambda})
= J(\mathbf{s}) + C(\mathbf{g})
  + \boldsymbol{\lambda}^\top(\mathbf{p} - \mathbf{g} + \boldsymbol{\ell} + \sum_w\mathbf{E}_w\mathbf{s}).
\end{equation}
% By convex analysis, the KKT conditions are necessary and sufficient for optimality.
For a primal-dual optimum $(\mathbf{g}^\star,\mathbf{s}^\star,\mathbf{p}^\star,\boldsymbol{\lambda}^\star)$, all KKT conditions are identical to~\eqref{kkt-con-gce}, except the $\mathbf{s}$-stationarity~\eqref{kkt:s}, which is:
\begin{equation}
    - \partial J(\mathbf{s}^\star) - \mathbf{1}\!\otimes\!\boldsymbol{\lambda}^\star \in N_{\mathcal{S}}(\mathbf{s}^\star),
    \label{kkt-multi}
\end{equation}
% \begin{subequations}
% \begin{align}
% - \partial C(\mathbf{g}^\star) + \boldsymbol{\lambda}^\star &\in N_{\mathcal{G}}(\mathbf{g}^\star),\\
% - \partial J(\mathbf{s}^\star) - \mathbf{1}\!\otimes\!\boldsymbol{\lambda}^\star &\in N_{\mathcal{S}}(\mathbf{s}^\star),\\
% - \boldsymbol{\lambda}^\star &\in N_{\mathcal{P}}(\mathbf{p}^\star),\\
% \mathbf{0} &= \mathbf{p}^\star - \mathbf{g}^\star + \boldsymbol{\ell} + \mathbf{S}^\star\mathbf{1}.
% \end{align}
% \label{kkt-multi}
% \end{subequations}
where the Kronecker product $\mathbf{1}\!\otimes\!\boldsymbol{\lambda}^\star \in \mathbb{R}^{WN}$ replicates the same LMP vector $\boldsymbol{\lambda}^\star$ for each subsystem.
 
Because $J(\mathbf{s})$ and $\mathcal{S}$ are separable across systems, \eqref{kkt-multi} implies that each subsystem $w$ satisfies
\begin{equation}
\mathbf{S}_w^\star
\in \argmin_{\mathbf{S}_w \in \mathcal{S}_w}
\bigl\{ J_w(\mathbf{S}_w) + \boldsymbol{\lambda}^{\star\top}\mathbf{S}_w \bigr\},
\end{equation}
which recovers each FL agent's best response. 
The generator and network subproblems remain identical to~\eqref{g-kkt}-\eqref{p-kkt} in the single-system case.

For each element of the GCE definition, we have:
\begin{itemize}[leftmargin=*]
\item FL best response: Each FL subsystem $w$ minimizes its own objective $J_w(\mathbf{S}_w)+\boldsymbol{\lambda}^{\top}\mathbf{S}_w$ given $\boldsymbol{\lambda}$.
\item Classical CE condition: The generator and network decisions $(\mathbf{g},\mathbf{p})$ solve the economic-dispatch and network-feasibility problems given $\boldsymbol{\lambda}$;
with feasibility constraints replaced by $\mathbf{p} = \mathbf{g} - \boldsymbol{\ell} - \sum_w\mathbf{E}_w\mathbf{s}$, which ensures balance.
\end{itemize}
Hence, $(\mathbf{g}^\star,\mathbf{s}^\star,\boldsymbol{\lambda}^\star)$ satisfies the definition of a GCE for multiple independent FL systems.

Because the stacked problem is convex and separable, all arguments for existence, uniqueness, and efficiency carry over verbatim from Corollaries~\ref{thm:gce_existence} and~\ref{thm:gce_efficiency} by substituting $J(\mathbf{s})$ and $J^{\mathrm{SW}}(\mathbf{s})$ with their separable forms in~\eqref{eq:new-def}.
% Therefore, the triple $(\mathbf{g},\mathbf{s},\boldsymbol{\lambda})$ is a GCE of the coupled multi-FL system if and only if it is a primal-dual optimal solution to~\eqref{prob:P_GCE} under Definition~\ref{def:stacked-FL}.
\end{proof}

\noindent \emph{Remark and Supplementary Information to Theorem~\ref{thm:GCE-Structural}:} Theorem~\ref{thm:GCE-Structural} computes the GCE outcome through \eqref{prob:P_GCE}, where the input value function $J$ comes from an estimation of the FL system's aggregate response. Note that $J$ is not required by the SO, which solves the economic dispatch \eqref{unify:economic dispatch} with the demand as input; instead, $J$ matters to those planning and operating the coupled infrastructure systems, such as an urban planner, and can be reused when the generation fleet is updated or new FL systems come online. Since a value function $\hat{J}$ learned from historical data or reported by the FL system may deviate from the true $J$, the following corollary supplements Theorem~\ref{thm:GCE-Structural} by quantifying how the approximation error in $\hat{J}$ propagates to the outcome computed from it. Conditions i) and ii) show that the computed triple is a $2\varepsilon$-approximate GCE: grid consistency holds exactly, while the FL best-response objective at the computed LMPs $\hat{\boldsymbol{\lambda}}$ is within $2\varepsilon$ of its minimum. Conditions iii) and iv) bound the GCE-objective error and the load-profile error relative to the exact GCE, respectively. The result extends to the multiple FL systems of Theorem~\ref{thm:GCE-multi}, with $\varepsilon$ equal to the sum of the per-system estimation errors.

\begin{corollary}[Approximate GCE under an Estimated Value Function $\hat J(\mathbf{s})$]
\label{cor:approx-gce}
Let Assumptions~\ref{ass:linear-map} and~\ref{ass:SFL-specific cost} hold and suppose $\mathcal{F}\neq\emptyset$. Let $\hat{J}\colon\mathcal{S}\to\mathbb{R}$ be continuous and convex, with $\sup_{\mathbf{s}\in\mathcal{S}}\lvert \hat{J}(\mathbf{s})-J(\mathbf{s})\rvert\le\varepsilon$ for some $\varepsilon\ge0$. Let $(\hat{\mathbf{s}},\hat{\mathbf{g}},\hat{\mathbf{p}},\hat{\boldsymbol{\lambda}})$ be a primal-dual optimal solution of \eqref{prob:P_GCE} with $J$ replaced by $\hat{J}$, which exists by the same argument as Corollary~\ref{thm:gce_existence}, and let $(\mathbf{s}^{\mathrm{GCE}},\mathbf{g}^{\mathrm{GCE}},\mathbf{p}^{\mathrm{GCE}})$ be a primal optimal solution of the exact problem. The following statements hold.
\begin{enumerate}[label=\roman*), leftmargin=*]
    \item The pair $(\hat{\mathbf{g}},\hat{\boldsymbol{\lambda}})$ is a classical CE given $\hat{\mathbf{s}}$.
    \item For every $\mathbf{s}\in\mathcal{S}$,
    $J(\hat{\mathbf{s}})+\hat{\boldsymbol{\lambda}}^{\top}\hat{\mathbf{s}} \le J(\mathbf{s})+\hat{\boldsymbol{\lambda}}^{\top}\mathbf{s}+2\varepsilon$.
    \item The objectives at $(\hat{\mathbf{s}},\hat{\mathbf{g}})$ and at the exact GCE satisfy
    \[
    0\le J(\hat{\mathbf{s}})+C(\hat{\mathbf{g}})
    -J(\mathbf{s}^{\mathrm{GCE}})-C(\mathbf{g}^{\mathrm{GCE}})
    \le2\varepsilon.
    \]
    \item If $J$ is $\varrho$-strongly convex on $\mathcal{S}$ for some $\varrho>0$,\footnote{$J$ is $\varrho$-strongly convex on $\mathcal{S}$ if $J(\mathbf{s})-\frac{\varrho}{2}\|\mathbf{s}\|_2^2$ is convex on $\mathcal{S}$.} then
    $\|\hat{\mathbf{s}}-\mathbf{s}^{\mathrm{GCE}}\|_{2}\le 2\sqrt{\varepsilon/\varrho}$.
\end{enumerate}
\end{corollary}

\begin{proof}[Proof of Corollary~\ref{cor:approx-gce}]
Primal-dual optimality of the estimated problem gives the KKT system \eqref{kkt-con-gce} with $J$ replaced by $\hat J$. Conditions \eqref{kkt:g}, \eqref{kkt:p}, and \eqref{kkt:feas} are unchanged and establish i), while \eqref{kkt:s} gives $\hat{\mathbf{s}}\in\argmin_{\mathbf{s}\in\mathcal{S}}\{\hat{J}(\mathbf{s})+\hat{\boldsymbol{\lambda}}^{\top}\mathbf{s}\}$.

For any $\mathbf{s}\in\mathcal{S}$,
\begin{align*}
J(\hat{\mathbf{s}})+\hat{\boldsymbol{\lambda}}^{\top}\hat{\mathbf{s}}
&\le \hat{J}(\hat{\mathbf{s}})+\hat{\boldsymbol{\lambda}}^{\top}\hat{\mathbf{s}}+\varepsilon
\le \hat{J}(\mathbf{s})+\hat{\boldsymbol{\lambda}}^{\top}\mathbf{s}+\varepsilon\\
&\le J(\mathbf{s})+\hat{\boldsymbol{\lambda}}^{\top}\mathbf{s}+2\varepsilon,
\end{align*}
which proves ii).

Exact optimality gives the lower bound in iii). For the upper bound,
\begin{align*}
J(\hat{\mathbf{s}})+C(\hat{\mathbf{g}})
&\le \hat{J}(\hat{\mathbf{s}})+C(\hat{\mathbf{g}})+\varepsilon
\le \hat{J}(\mathbf{s}^{\mathrm{GCE}})+C(\mathbf{g}^{\mathrm{GCE}})+\varepsilon\\
&\le J(\mathbf{s}^{\mathrm{GCE}})+C(\mathbf{g}^{\mathrm{GCE}})+2\varepsilon.
\end{align*}

Finally, $\varrho$-strong convexity of $J$, convexity of $C$, and exact optimality imply the quadratic-growth bound
\[
J(\hat{\mathbf{s}})+C(\hat{\mathbf{g}})
\ge J(\mathbf{s}^{\mathrm{GCE}})+C(\mathbf{g}^{\mathrm{GCE}})
+\tfrac{\varrho}{2}\,\|\hat{\mathbf{s}}-\mathbf{s}^{\mathrm{GCE}}\|_{2}^{2}.
\]
Combining this inequality with iii) proves iv).
\end{proof}

\section{Proofs in Section \ref{sec:transportation}}
% \section{Proof for Lemma \ref{lem:UE-equivalence}}
\begin{proof}[Proof of Lemma \ref{lem:UE-equivalence}]
    \label{app-gce-ev}
    Since $c_e(\cdot)$ is integrable, define
    $h_e(y):=\int_0^{y} c_e(\xi)\,\mathrm{d}\xi$.
    For any $y_1\neq y_2$ and $\theta\in(0,1)$, let $y_\theta := \theta y_1 + (1-\theta) y_2$. Without loss of generality suppose $y_1<y_2$, hence $y_1<y_\theta<y_2$. Then if $c_e$ is increasing and continuous,
    \begin{align*}
    &\theta h_e(y_1) + (1-\theta) h_e(y_2) - h_e(y_\theta)\\
    =& \theta\!\int_{0}^{y_1}\! c_e(\xi)\,d\xi + (1-\theta)\!\int_{0}^{y_2}\! c_e(\xi)\,d\xi - \int_{0}^{y_\theta}\! c_e(\xi)\,d\xi\\
    =&-\theta\!\int_{y_1}^{y_\theta}\! c_e(\xi)\,d\xi + (1-\theta)\!\int_{y_\theta}^{y_2}\! c_e(\xi)\,d\xi \\
    > & -\theta (y_\theta-y_1)\, c_e(y_\theta) + (1-\theta)(y_2-y_\theta)\, c_e(y_\theta) = 0,
    \end{align*}
    where the strict inequality is due to $c_e(\xi)<c_e(y_\theta)$ for $\xi\in(y_1,y_\theta)$ and $c_e(\xi)>c_e(y_\theta)$ for $\xi\in(y_\theta,y_2)$. Therefore, $h_e(y_\theta) < \theta h_e(y_1) + (1-\theta) h_e(y_2)$, proving that $h_e$ is strictly convex.
    % Moreover, if every $c_e(\cdot)$ is continuous and nonnegative, $h_e$ is continuous by the fundamental theorem of calculus.
    Since each $y_e(\mathbf{x})$ is affine in $\mathbf{x}$, the mapping $\mathbf{x}\mapsto(y_e(\mathbf{x}))_e$ is injective on $\mathcal{X}$, and $\Phi(\mathbf{x})=\sum_e h_e\big(y_e(\mathbf{x})\big)$, it follows that $\Phi(\mathbf{x})$ is continuous and strictly convex in $\mathbf{x}$. 

    Consider the convex optimization problem~\eqref{opt:UE}. Because $\Phi(\mathbf{x})$ is continuous and strictly convex, the KKT condition is necessary and sufficient. Let $\mu_k$ denote the multiplier for the equality constraint $\sum_{R_r \in \mathcal{R}_k^{\mathrm{tr}}} x_r = \rho_k$, and $\nu_r \ge 0$ the multiplier for $x_r \ge 0$. Then for each $k$ and $r\in\mathcal{R}_k^{\mathrm{tr}}$,
    \begin{equation}
        \label{UE:stationarity}
        0 \in \partial_{x_r}\Phi(\mathbf{x}^\star) + ((\mathbf{A}^{\mathrm{FL}})^\top \boldsymbol{\lambda})_r + \mu_k - \nu_r.
    \end{equation}
    Note that
\(  \partial_{x_r}\Phi(\mathbf{x}) = \{\pi^{\mathrm{travel}}_r(\mathbf{x})\}.\)
    Since the charging term is linear, \(((\mathbf{A}^{\mathrm{FL}})^\top \boldsymbol{\lambda})_r=\pi_r^{\mathrm{charge}}(\boldsymbol{\lambda})\). 
    % Substituting into~\eqref{UE:stationarity} gives
    % \[
    % 0\;=\; g_r\; +\; \pi_r^{\mathrm{charge}}(\boldsymbol{\lambda})\; +\; \mu_k\; -\; \nu_r.
    % \]
    Applying complementary slackness $\nu_r x_r^\star = 0$ yields two cases:
    \begin{itemize}[leftmargin= *]
    \item If $x_r^\star > 0$, then $\nu_r = 0$ and $\pi_r^{\mathrm{travel}}(\mathbf{x}^\star) + \pi_r^{\mathrm{charge}}(\boldsymbol{\lambda}) = -\mu_k$, meaning all used routes for OD pair $k$ have equal minimal route cost.
    \item If $x_r^\star = 0$, then $\nu_r \ge 0$ and $\pi_r^{\mathrm{travel}}(\mathbf{x}^\star) + \pi_r^{\mathrm{charge}}(\boldsymbol{\lambda}) \ge -\mu_k$, so unused routes have no lower route cost.
    \end{itemize}
    % Here $\pi_r^{\mathrm{travel}}(\mathbf{x}^\star)$ denotes a valid selection within the subgradient interval $[\pi_r^{\mathrm{travel}}(\mathbf{x}^\star)^-,\,\pi_r^{\mathrm{travel}}(\mathbf{x}^\star)^+]$, chosen consistently across routes.
    Hence, for each OD pair $k$,
    \[
    \pi_r(\mathbf{x}^\star, \boldsymbol{\lambda}) \le \pi_{r'}(\mathbf{x}^\star, \boldsymbol{\lambda}), \quad \forall R_{r'} \in \mathcal{R}_k^{\mathrm{tr}},\; \text{with } x_r^\star > 0,
    \]
    which is exactly the UE condition~\eqref{def:UE}. Conversely, any $\mathbf{x}^\star$ satisfying~\eqref{def:UE} admits such multipliers $\mu_k, \nu_r$ and therefore satisfies the KKT conditions, proving necessity and sufficiency. This establishes the equivalence between the optimization problem~\eqref{opt:UE} and the UE condition~\eqref{def:UE}. A similar derivation appears in the classical transportation equilibrium theory~\cite{sheffi1985urban}, where the charging component is absent.
\end{proof}

% \section{Proof for Lemma~\ref{lem:num-trans}}
\begin{proof}[Proof of Lemma~\ref{lem:num-trans}]
\emph{Proof of i).} Write \(\boldsymbol{\lambda}:=\boldsymbol{\lambda}^{\mathrm{GCE}}\), let \(\mathbf{x}^{\mathrm{GCE}}\) and \(\mathbf{x}^{\mathrm{SWM}}\) denote the route flows underlying \(\mathbf{s}^{\mathrm{GCE}}\) and \(\mathbf{s}^{\mathrm{SWM}}\), and set \(y_e^{\mathrm{GCE}}:=y_e(\mathbf{x}^{\mathrm{GCE}})\) and \(y_e^{\mathrm{SWM}}:=y_e(\mathbf{x}^{\mathrm{SWM}})\). Since the GCE outcome is feasible for \eqref{prob:P_SW}, \(\mathrm{PoA}\ge1\).

By Theorem~\ref{thm:GCE-Structural}, the GCE is primal-dual optimal for \eqref{prob:P_GCE}. Evaluating \eqref{kkt:g} at \(\mathbf{g}^{\mathrm{SWM}}\in\mathcal{G}\) and \eqref{kkt:p} at \(\mathbf{p}^{\mathrm{SWM}}\in\mathcal{P}\) gives \(C(\mathbf{g}^{\mathrm{SWM}})-\boldsymbol{\lambda}^{\top}\mathbf{g}^{\mathrm{SWM}}\ge C(\mathbf{g}^{\mathrm{GCE}})-\boldsymbol{\lambda}^{\top}\mathbf{g}^{\mathrm{GCE}}\) and \(\boldsymbol{\lambda}^{\top}\mathbf{p}^{\mathrm{SWM}}\ge\boldsymbol{\lambda}^{\top}\mathbf{p}^{\mathrm{GCE}}\). Adding them and substituting \(\mathbf{p}=\mathbf{g}-\boldsymbol{\ell}-\mathbf{s}\) at both solutions yields
\begin{equation}\label{eq:poa-gen}
C(\mathbf{g}^{\mathrm{SWM}})-C(\mathbf{g}^{\mathrm{GCE}})\ \ge\ \boldsymbol{\lambda}^{\top}(\mathbf{s}^{\mathrm{SWM}}-\mathbf{s}^{\mathrm{GCE}}).
\end{equation}%
By Lemma~\ref{lem:UE-equivalence}, \(\mathbf{x}^{\mathrm{GCE}}\) minimizes the convex function \(\Phi(\mathbf{x})+\boldsymbol{\lambda}^{\top}\mathbf{A}^\mathrm{FL}\mathbf{x}\) over \(\mathcal{X}\), with \(\partial\Phi(\mathbf{x})/\partial x_r=\pi_r^{\mathrm{travel}}(\mathbf{x})\). First-order optimality in the feasible direction toward \(\mathbf{x}^{\mathrm{SWM}}\in\mathcal{X}\), together with the linearity of the edge flows~\eqref{eq:edge-route} in \(\mathbf{x}\), gives
\begin{equation}\label{eq:poa-route}
\sum\nolimits_{e}c_e(y_e^{\mathrm{GCE}})\,(y_e^{\mathrm{SWM}}-y_e^{\mathrm{GCE}})+\boldsymbol{\lambda}^{\top}(\mathbf{s}^{\mathrm{SWM}}-\mathbf{s}^{\mathrm{GCE}})\ \ge\ 0.
\end{equation}
Adding \eqref{eq:poa-route} to \eqref{eq:poa-gen} cancels the price terms, and \(\Phi^{\mathrm{SW}}(\mathbf{x}^{\mathrm{GCE}})=\sum_e c_e(y_e^{\mathrm{GCE}})y_e^{\mathrm{GCE}}\) from~\eqref{eq:phisw:ev} yields
\begin{equation}\label{eq:poa-comb}
\Phi^{\mathrm{SW}}(\mathbf{x}^{\mathrm{GCE}})+C(\mathbf{g}^{\mathrm{GCE}})\le\sum\nolimits_{e}c_e(y_e^{\mathrm{GCE}})\,y_e^{\mathrm{SWM}}+C(\mathbf{g}^{\mathrm{SWM}}).%
\end{equation}%
In this example \(\pi_1^{\mathrm{travel}}(\mathbf{x})=ax_1\) and \(\pi_2^{\mathrm{travel}}(\mathbf{x})=b\), so we have
\begin{equation}\label{eq:poa-affine}
\begin{split}
&\sum\nolimits_{e}c_e(y_e^{\mathrm{GCE}})\,y_e^{\mathrm{SWM}}-\Phi^{\mathrm{SW}}(\mathbf{x}^{\mathrm{SWM}})-\tfrac{1}{4}\Phi^{\mathrm{SW}}(\mathbf{x}^{\mathrm{GCE}})\\
&\qquad=-a\bigl(x_1^{\mathrm{SWM}}-\tfrac{1}{2}x_1^{\mathrm{GCE}}\bigr)^{2}-\tfrac{b}{4}\,x_2^{\mathrm{GCE}}\ \le\ 0 .
\end{split}
\end{equation}
Substituting \eqref{eq:poa-affine} into \eqref{eq:poa-comb} gives
\begin{align*}
&\Phi^{\mathrm{SW}}(\mathbf{x}^{\mathrm{GCE}})+C(\mathbf{g}^{\mathrm{GCE}})\\
\le & \Phi^{\mathrm{SW}}(\mathbf{x}^{\mathrm{SWM}})+C(\mathbf{g}^{\mathrm{SWM}})+\tfrac{1}{4}\Phi^{\mathrm{SW}}(\mathbf{x}^{\mathrm{GCE}})\\
\le &\Phi^{\mathrm{SW}}(\mathbf{x}^{\mathrm{SWM}})+C(\mathbf{g}^{\mathrm{SWM}})+\tfrac{1}{4} \left(\Phi^{\mathrm{SW}}(\mathbf{x}^{\mathrm{GCE}})+ C(\mathbf{g}^{\mathrm{GCE}}) \right),
    \end{align*} 
which gives \(\mathrm{PoA}\le4/3\) after rearranging.

\emph{Proof of ii).}%
\label{app-ev}
    \begin{table}[t]
        \centering
     \caption{Illustrative Example: Non-congested Case}
        \begin{tabular}{c|c|c|c|c}\toprule
        & $s_1^{*}$ & $s_2^{*}$  & $\lambda_1^*$ &  $\lambda_2^*$  \\\midrule
        GCE     &  $\frac{bq}{a}$ &  $\frac{(a-b)q}{a}$   & $2\gamma_1(\ell_1 + q- g_2^{\max})$  & $2\gamma_1(\ell_1 + q- g_2^{\max})$ \\ \hline
        SWM   & $\frac{bq}{2a}$ &  $\frac{(2a- b)q}{2a}$ & $2\gamma_1 (\ell_1 + q- g_2^{\max})$  & $2\gamma_1(\ell_1 + q- g_2^{\max})$ \\\bottomrule
        \end{tabular}
        \label{tab:scene_1}
    \end{table}
    Under the parameter regime in Lemma~\ref{lem:num-trans}-ii), both objectives of~\eqref{eq:num-GCE} and~\eqref{eq:num-SWM} are decreasing in \(g_2\) over \(0\le g_2\le g_2^{\max}\). 
Thus, in both GCE and SWM problems, \(g_2^\star=g_2^{\max}\) and \(g_1^\star=\ell_1+q-g_2^{\max}\). The two thresholds are
    \begin{align*}
    \bar f^{\max}
    =g_2^{\max}-\frac{(a-b)q}{a}, \quad
    \underline f^{\max}
    =g_2^{\max}-\frac{(2a-b)q}{2a}.
    \end{align*}
    The assumed range of \(g_2^{\max}\) ensures that the thresholds are positive and ordered.
Three mutually exclusive cases for \(f^{\max}\) illustrate Lemma \ref{lem:num-trans}:
\begin{enumerate}[label=\roman*), leftmargin = *]
    \item \textit{No congestion} \(\bigl(f^{\max}>\bar f^{\max}\bigr)\).  
      The line is not congested in either program.   Table~\ref{tab:scene_1} demonstrates the results under this parameter regime.
      We have 
      \(s_1^{\mathrm{GCE}}=2s_1^{\mathrm{SWM}}\) while the LMPs remain uniform,  
      \(\lambda_1^{\mathrm{GCE}}=\lambda_2^{\mathrm{GCE}}=\lambda_1^{\mathrm{SWM}}=\lambda_2^{\mathrm{SWM}}\).  
      Both problems dispatch \(g_2^\star=g_2^{\max}\) and \(g_1^\star=\ell_1+q-g_2^{\max}\), so the generation cost equals \(C_0:=\gamma_1(\ell_1+q-g_2^{\max})^2+\gamma_2(g_2^{\max})^2\) in both. With unit travel demand, the UE condition \(ax_1=b\) gives \(\Phi^{\mathrm{SW}}(\mathbf{x}^{\mathrm{GCE}})=a(b/a)^2+b(1-b/a)=b\), while minimizing \(ax_1^2+b(1-x_1)\) gives \(x_1^{\mathrm{SWM}}=b/(2a)\) and \(\Phi^{\mathrm{SW}}(\mathbf{x}^{\mathrm{SWM}})=b-b^2/(4a)\). This gives \(\mathrm{PoA}=\overline{\mathrm{PoA}}:=\bigl(b+C_0\bigr)\big/\bigl(b-b^2/(4a)+C_0\bigr)>1\), which approaches \(4/3\) as \(b\to a\) and \(C_0/b\to0\), showing that the bound in i) is tight.
% Even if the ISO selects electricity LMPs based on a demand profile that maximizes social welfare, spatially flexible loads may still act selfishly, deviating toward a routing or dispatch profile that is not socially optimal.

\item \textit{Congestion only in GCE} \(\bigl(\underline f^{\max}<f^{\max}\le\bar f^{\max}\bigr)\).
      The flexible load chosen under the GCE saturates the line, whereas the SWM solution remains uncongested.  
      Hence \(s_1^{\mathrm{GCE}} = q-g_2^{\max}+f^{\max} > s_1^{\mathrm{SWM}}\) and the LMPs split,
      \begin{subequations}
        \begin{align}
        \lambda_1^{\mathrm{GCE}} = & 2\gamma_1(\ell_1 + q- g_2^{\max}),\\
        \lambda_2^{\mathrm{GCE}} = & \lambda_1^{\mathrm{GCE}} - \frac{a}{q^2}{\left( g_2^{\max} - \frac{a-b}{a}q -f^{\max}\right)}.
        \end{align}
        \label{obs:price-ev}
      \end{subequations}
         Therefore,
      \(\lambda_2^{\mathrm{GCE}}<\lambda_1^{\mathrm{GCE}}=\lambda_1^{\mathrm{SWM}}=\lambda_2^{\mathrm{SWM}}\).  
      The generation cost is still \(C_0\) in both problems and the SWM remains uncongested, so the denominator of the PoA is unchanged. Writing \(\bar{x}:=(q-g_2^{\max}+f^{\max})/q\) for the GCE flow on route~1, which ranges over \((b/(2a),\,b/a]\) as \(f^{\max}\) ranges over \((\underline{f}^{\max},\,\bar{f}^{\max}]\), we obtain \(\mathrm{PoA}=\bigl(a\bar{x}^2+b(1-\bar{x})+C_0\bigr)\big/\bigl(b-b^2/(4a)+C_0\bigr)\). Since \(a\bar{x}^2+b(1-\bar{x})\) is minimized at \(\bar{x}=b/(2a)\) and is therefore increasing on \((b/(2a),\,b/a]\), the PoA decreases continuously from \(\overline{\mathrm{PoA}}\) to \(1\) as \(f^{\max}\) decreases to \(\underline{f}^{\max}\).
    %   The earlier onset of congestion in the GCE confirms Proposition~\ref{obs:ev1}.

\item \textit{Congestion in both problems} \(\bigl(f^{\max}\le\underline f^{\max}\bigr)\).   
      The transmission limit binds in both optimization models, and both attain
      \(
      s_1^{\mathrm{GCE}}=s_1^{\mathrm{SWM}}=q-g_2^{\max}+f^{\max}.
      \)
      Together with the common generation dispatch established above, this implies that both problems attain the same social cost, and hence \(\mathrm{PoA}=1\).
    %   This completes Proposition~\ref{prop-ev-all} for the congested regime.

    %  \item The line capacity satisfy $g_2^{\max} - {(2a-b)q}/{(2a)}  < f^{\max} <g_2^{\max} - {(a-b)q}/{a} $:
    %  Congestion presents in $\text{(P$_\mathrm{GCE}$)}$ and do not exist in $\text{(P$_{\mathrm{SWM}}$)}$.
    % Therefore, for $\text{(P$_{\mathrm{SWM}}$)}$, the results are the same, while for $\text{(P$_\mathrm{GCE}$)}$, we have:
    % $s_1^{*} = {q-g_2^{\max}+f^{\max}}-\ell_1$ ,  $s_2^{*} ={g_2^{\max}-f^{\max}}$   , 
    %  The price profiles under the two optimization frameworks manifest that the LMPs supporting equilibrium may differ from those calculated based on the demand profile for social welfare maximization.
    %  \item The line capacity satisfy $ f^{\max} \le g_2^{\max} - {(2a-b)q}/{(2a)} $: In both optimization problems, congestion arises. Under this scenario, the two optimizations are equivalent, sticking to the $\text{(P$_\mathrm{GCE}$)}$ derived in the last item. 
\end{enumerate}
\end{proof}

% \section{Proof for Lemma \ref{lemma:equiva_NE}}
\section{Proofs and Supplementary Results in Section \ref{sec:datacenter}}
\label{app:ces-positive-externality}
\begin{proof}[Proof of Lemma \ref{lemma:equiva_NE}]
    \label{app:gce-dc}
Since each utility \(u_k\) is differentiable and concave and each feasible set \(\mathcal X^{(k)}\) is convex and compact, the companies' optimization problem in \eqref{eq:dc-NE} is convex.  Its KKT conditions are therefore necessary and sufficient.  Collecting the \(K\) stationarity conditions yields exactly the VI in \eqref{eq:vi-dc}; conversely, any \(\mathbf{x}^\star\) solving \eqref{eq:vi-dc} satisfies every player's KKT system and is thus a Nash equilibrium.  This one-to-one correspondence follows from Proposition~1.3.4 of \cite{faechinei2003finite}.

Now suppose the cross-partial symmetry condition \eqref{ass:dc_potential} holds.
By standard results in potential game theory \cite{monderer1996potential}, there exists a continuously differentiable scalar function $\Phi(\mathbf{x})$ satisfying $\nabla_{\mathbf{x}^{(k)}}\Phi = - \nabla_{\mathbf{x}^{(k)}}u_k$ for all $k$.
Substituting this relation into \eqref{eq:vi-dc} gives
\[\big\langle
\nabla_{\mathbf{x}}\!\big(\Phi(\mathbf{x}^\mathrm{NE})
    +	\boldsymbol{\lambda}^\top\mathbf{A}^\mathrm{FL}\mathbf{x}^\mathrm{NE}\big),\,
\mathbf{x}-\mathbf{x}^\mathrm{NE}
\big\rangle
\ge 0,\quad\forall\,\mathbf{x}\in\mathcal{X},\]
since $\mathbf{A}^\mathrm{FL}=q(\mathbf{A}^\mathrm{DB})^\top\mathbf{A}^\mathrm{sum}.$
This is precisely the first-order optimality condition of the convex program
$\min_{\mathbf{x}\in\mathcal{X}}\;
\Phi(\mathbf{x}) + \boldsymbol{\lambda}^\top\mathbf{A}^\mathrm{FL}\mathbf{x},$
proving the equivalence between the VI and the optimization problem~\eqref{eq:sfl-response}.

Finally, if each $u_k$ is strictly concave and each feasible set $\mathcal{X}^{(k)}$ is nonempty, convex, and compact, then $-\mathbf{F}(\mathbf{x})$ is strictly monotone.
Thus, the negative gradient mapping $-\mathbf{F}(\mathbf{x}) =\nabla\Phi(\mathbf{x})$ is strictly monotone, implying that the potential function $\Phi(\mathbf{x})$ is strictly convex. Consequently, the optimization problem is convex.  
\end{proof}

% \section{Proof for Lemma \ref{lem:cha-dc}}
% \begin{proposition}[Congestion-matched scenario]\label{obs:congestion}
%     When the transmission line is congested under both the GCE and the social-welfare benchmark, the resulting prices satisfy
%     \(\lambda_1^{\mathrm{GCE}}>\lambda_1^{\mathrm{SWM}}\) and
%     \(\lambda_2^{\mathrm{GCE}}>\lambda_2^{\mathrm{SWM}}\).
%     \end{proposition}
% \begin{proposition}[Load-distribution difference]\label{obs:loadshift}
%     With asymmetric cross-company externalities \(\bigl(M_{1,2}\neq M_{2,1}\bigr)\), which implies \(m_s \ne m_u\) and \(m_t \ne m_v\).
%     The GCE allocations satisfy
%     \((1-m_u)x_{1}^{(1),\mathrm{GCE}}=(1-m_s)x_{1}^{(2),\mathrm{GCE}}\) and
%     \((1-m_v)x_{2}^{(1),\mathrm{GCE}}=(1-m_t)x_{2}^{(2),\mathrm{GCE}}\),
%     whereas the welfare-optimal solution equalizes workloads,
%     \(\mathbf{x}^{(1),\mathrm{SWM}}=\mathbf{x}^{(2),\mathrm{SWM}}\).
%     This discrepancy quantifies the unpriced externality ignored by self-interested firms.
%     \end{proposition}
% We elaborate on Lemma \ref{lem:cha-dc} by firstly deriving closed-form expressions for the .
\begin{proof}[Proof of Lemma \ref{lem:cha-dc}]
    \label{app-dc}
The only structural difference between optimization frameworks $(\mathrm{P}_{\mathrm{GCE}})$ and $(\mathrm{P}_{\mathrm{SWM}})$ in this example is the interaction matrix, i.e., $\mathbf{M}^\mathrm{NE}$ versus $\mathbf{M}^\mathrm{SW}$.
If there are no cross-company externalities, 
 i.e., $\mathbf{M}_{1,2}=\mathbf{M}_{2,1}=\mathbf{0}$, then $\mathbf{M}^\mathrm{NE}=\mathbf{M}^\mathrm{SW}$ and hence $J(\mathbf{s})\equiv J^{\mathrm{SW}}(\mathbf{s})$. The two optimization problems coincide, so the outcomes agree, establishing the no-externality case of Lemma~\ref{lem:cha-dc}-i).

We next prove the bound in Lemma~\ref{lem:cha-dc}-i). Write \(\boldsymbol{\lambda}:=\boldsymbol{\lambda}^{\mathrm{GCE}}\) and \(\Delta\mathbf{x}:=\mathbf{x}^{\mathrm{SWM}}-\mathbf{x}^{\mathrm{GCE}}\). The GCE outcome \((\mathbf{x}^{\mathrm{GCE}},\mathbf{g}^{\mathrm{GCE}},\mathbf{p}^{\mathrm{GCE}})\) is feasible for \eqref{eq:swm_ori}, which gives the lower bound. The same argument that gives \eqref{eq:poa-gen} in the proof of Lemma~\ref{lem:num-trans} yields \(C(\mathbf{g}^{\mathrm{SWM}})-C(\mathbf{g}^{\mathrm{GCE}})\ge\boldsymbol{\lambda}^{\top}\mathbf{A}^{\mathrm{FL}}\Delta\mathbf{x}\) here, while first-order optimality of \(\mathbf{x}^{\mathrm{GCE}}\) for the convex problem~\eqref{eq:sfl-response} gives \(\langle\nabla\Phi(\mathbf{x}^{\mathrm{GCE}}),\Delta\mathbf{x}\rangle+\boldsymbol{\lambda}^{\top}\mathbf{A}^{\mathrm{FL}}\Delta\mathbf{x}\ge0\). Adding the two eliminates the price terms and leaves \(C(\mathbf{g}^{\mathrm{GCE}})-C(\mathbf{g}^{\mathrm{SWM}})\le\langle\nabla\Phi(\mathbf{x}^{\mathrm{GCE}}),\Delta\mathbf{x}\rangle\), so the welfare gap is at most \(\Phi^{\mathrm{SW}}(\mathbf{x}^{\mathrm{GCE}})-\Phi^{\mathrm{SW}}(\mathbf{x}^{\mathrm{SWM}})+\langle\nabla\Phi(\mathbf{x}^{\mathrm{GCE}}),\Delta\mathbf{x}\rangle\).
With \(\nabla\Phi(\mathbf{x})=\mathbf{M}^{\mathrm{NE}}\mathbf{x}-\boldsymbol{\eta}\) and \(\nabla\Phi^{\mathrm{SW}}(\mathbf{x})=\mathbf{M}^{\mathrm{SW}}\mathbf{x}-\boldsymbol{\eta}\), the baseline utilities cancel and \(\nabla\Phi^{\mathrm{SW}}(\mathbf{x}^{\mathrm{GCE}})-\nabla\Phi(\mathbf{x}^{\mathrm{GCE}})=\Delta\mathbf{M}\,\mathbf{x}^{\mathrm{GCE}}\). Expanding the quadratic \(\Phi^{\mathrm{SW}}\) exactly around \(\mathbf{x}^{\mathrm{GCE}}\), we obtain
\begin{align*}
&\Phi^{\mathrm{SW}}(\mathbf{x}^{\mathrm{GCE}})+C(\mathbf{g}^{\mathrm{GCE}})
 -\Phi^{\mathrm{SW}}(\mathbf{x}^{\mathrm{SWM}})-C(\mathbf{g}^{\mathrm{SWM}})\\
&\qquad\le\ -\langle\Delta\mathbf{M}\,\mathbf{x}^{\mathrm{GCE}},\Delta\mathbf{x}\rangle-\tfrac{1}{2}\Delta\mathbf{x}^{\top}\mathbf{M}^{\mathrm{SW}}\Delta\mathbf{x}\\
&\qquad\le\ \sup\nolimits_{\mathbf{v}}\bigl\{-\langle\Delta\mathbf{M}\,\mathbf{x}^{\mathrm{GCE}},\mathbf{v}\rangle-\tfrac{1}{2}\mathbf{v}^{\top}\mathbf{M}^{\mathrm{SW}}\mathbf{v}\bigr\},
\end{align*}
where the supremum over \(\mathbf{v}\in\mathbb{R}^{KN_\mathrm{D}}\) is attained at \(\mathbf{v}=-(\mathbf{M}^{\mathrm{SW}})^{-1}\Delta\mathbf{M}\,\mathbf{x}^{\mathrm{GCE}}\) and equals the stated bound.
%  The same argument bounds the welfare gap by \(\|\nabla\Phi^{\mathrm{SW}}(\mathbf{x}^{\mathrm{GCE}})-\nabla\Phi(\mathbf{x}^{\mathrm{GCE}})\|_2^2/(2\varsigma)\) for any FL system whose \(\Phi^{\mathrm{SW}}\) is \(\varsigma\)-strongly convex, irrespective of the sign of the externality.

When there exist cross-company externalities, we write $\mathbf{M}\in\{\mathbf{M}^\mathrm{NE},\mathbf{M}^\mathrm{SW}\}$  for compactness. 
Note that for any fixed LMP vector $\boldsymbol{\lambda}$, the KKT conditions for the VI problem \eqref{exa:vi-dc} imply each company's optimal workload satisfies
\begin{equation}
\mathbf{x}^{\star}(\boldsymbol{\lambda}) 
= \mathbf{M}^{-1}\!\left( \boldsymbol{\eta} - \mathbf{A}^{\mathrm{FL}\top}\boldsymbol{\lambda} \right)
= \mathbf{M}^{-1}\!\left( \boldsymbol{\eta} - q 
\begin{pmatrix} \boldsymbol{\lambda}\\[1pt] \boldsymbol{\lambda}\end{pmatrix}\! \right).
\label{eq:xstar-compact}
\end{equation}

The strict positivity in Lemma \ref{lem:cha-dc}-i) can be demonstrated through an example 
in the parameter regime of Lemma \ref{lem:cha-dc}-ii).
Specifically, the generators are assumed to have sufficient capacity, such that their output limits are non-binding.
We then assess system equilibrium when the power system is not congested.
In this case, the LMPs are the same for both buses, i.e., $\lambda^\star= \lambda_1^\star = \lambda_2^\star=  \Gamma (g_1^\star + g_2^\star)$, where $\Gamma = \frac{2\gamma_1\gamma_2}{\gamma_1 + \gamma_2}$. The power balance gives  $g_1^\star + g_2^\star = q \mathbf{1}^\top \mathbf{x}^{\star}$.
Therefore, substituting~\eqref{eq:xstar-compact} leads to the closed-form derivation for LMPs, which is:
\begin{equation}
    \begin{aligned}
        \lambda^\star =& \frac{\Gamma  q \left(\mathbf{1}^\top (\mathbf{M})^{-1} \boldsymbol{\eta}\right)}{1 + \Gamma  q^2 \left(\mathbf{1}^\top (\mathbf{M})^{-1} \mathbf{1}\right)}
    \end{aligned}
    \label{eq:lmp-closed}
\end{equation}
If $\bm \eta_1 = \bm \eta_2 = \bar{\eta}\,\mathbf 1 \propto \mathbf 1$, where $\bar{\eta}$ is a constant,  $\lambda$ equals:
%  and $q_1=q_2=1$
\[
\lambda^\star
=\frac{\Gamma\,q\,\bar{\eta}\, \mathbf{1}^\top \mathbf{M}^{-1}\mathbf{1}}
      {1+\Gamma\,q^2\, \mathbf{1}^\top \mathbf{M}^{-1}\mathbf{1}}
= 
\left(\frac{q}{\bar{\eta}}+\frac{1}{\Gamma\,q\,\bar{\eta}\,\mathbf{1}^\top \mathbf{M}^{-1}\mathbf{1}}\right)^{-1}.
\]
Therefore, $\lambda^\star$ is increasing in $\mathbf{1}^\top \mathbf{M}^{-1}\mathbf{1}$.
	 Since the inequalities in Lemma~\ref{lem:cha-dc}-ii) hold componentwise, they imply
\[
\mathbf{1}^\top (\mathbf{M}^\mathrm{NE})^{-1}\mathbf{1} \;>\; \mathbf{1}^\top (\mathbf{M}^\mathrm{SW})^{-1}\mathbf{1},
\]
and therefore
\[
\lambda_1^{\mathrm{GCE}}=\lambda_2^{\mathrm{GCE}} \;>\; \lambda_1^{\mathrm{SWM}}=\lambda_2^{\mathrm{SWM}}.
\]
This proves Lemma~\ref{lem:cha-dc}-ii).

It remains to show that  there exist parameters for \(\mathbf{1}^\top (\mathbf{M}^\mathrm{NE})^{-1}\mathbf{1} \;\ge\; \mathbf{1}^\top (\mathbf{M}^\mathrm{SW})^{-1}\mathbf{1}\) to hold, so the parameter regime discussed above constitutes an example to prove  the strict positivity in Lemma~\ref{lem:cha-dc}-i). 
Particularly, consider the case when the diagonal blocks are the \(2\times2\) identity matrix, i.e., \(\mathbf{M}_{1,1}=\mathbf{M}_{2,2}=\mathbf{I}_2\). 
The cross-company interaction blocks are diagonal,  
\(\mathbf{M}_{1,2}=\mathbf{M}_{2,1} = \operatorname{diag}(\vartheta_1,\vartheta_2)\), with parameters satisfying \(0<\vartheta_1,\vartheta_2 < 1/2\).

Under the above parameter setting, we can verify that
\begin{equation}
    \begin{aligned}
        \mathbf{1}^\top(\mathbf{M}^\mathrm{NE})^{-1}\mathbf{1}
=&2\!\left(\frac{1}{1+\vartheta_1}+\frac{1}{1+\vartheta_2}\right), \\
\mathbf{1}^\top(\mathbf{M}^\mathrm{SW})^{-1}\mathbf{1} =  & 2\!\left(\frac{1}{1+2\vartheta_1}+\frac{1}{1+2\vartheta_2}\right) .
    \end{aligned}
\end{equation}
Hence, condition 
\(\mathbf{1}^\top(\mathbf{M}^\mathrm{NE})^{-1}\mathbf{1}
\ge \mathbf{1}^\top(\mathbf{M}^\mathrm{SW})^{-1}\mathbf{1}\)
is satisfied, leading to higher LMPs under GCE.
Since the LMPs differ and $\Gamma (g_1^\star + g_2^\star) = \lambda^\star$, the resulting workload profiles also differ, as $\mathbf{1}^\top \mathbf{x}^\star = (g_1^\star + g_2^\star)$ differs.
Therefore, \((\mathrm{P}_{\mathrm{GCE}})\) does not support social welfare, 
which establishes the strict positivity in Lemma~\ref{lem:cha-dc}-i).
\end{proof}

\noindent\emph{Remark to Lemma~\ref{lem:cha-dc}-ii):} The LMP ordering
can be reversed in the following example, in which the cross-company
externalities are positive, i.e., each company's activity increases the
benefits that the other companies obtain from shared datacenter resources or
services. Each company's utility is modeled as a constant elasticity of
substitution (CES) function, which illustrates that the utility may take a form
beyond the quadratic one in~\eqref{eq:quadratic-utility}, as long as it
satisfies the conditions of Lemma~\ref{lemma:equiva_NE}.

\begin{lemma}[CES Utility with a Positive Externality]
\label{lem:ces-positive-externality}
Consider the two-company, two-datacenter system in
Section~\ref{dc:concrete_example}, and suppose the grid constraints do not
bind, i.e., \(f^{\max}=g_i^{\max}=\infty\), \(i=1,2\). Let each company's
utility be
\begin{equation}
u_k\bigl(\mathbf{x}^{(k)};\mathbf{x}^{(-k)}\bigr)
=A\,v\bigl(\mathbf{x}^{(k)}\bigr)
+B\,v\Bigl(\sum\nolimits_{\tilde k=1}^{K}\mathbf{x}^{(\tilde k)}\Bigr),
\label{eq:ces-positive-utility}
\end{equation}
where \(A,B>0\) and the CES aggregator
\(
v(\mathbf{y}):=\bigl(\sum_{j=1}^{N_{\mathrm D}}(\delta+y_j)^{\chi}\bigr)^{\beta/\chi}
\)
has parameters \(\delta>0\), \(\chi<1\) nonzero, and \(\beta\in(0,1)\). Then
\(
\lambda_i^{\mathrm{SWM}}>\lambda_i^{\mathrm{GCE}},
\)
\(i=1,2\).
\end{lemma}

\begin{proof}[Proof of Lemma \ref{lem:ces-positive-externality}]

Condition~\eqref{ass:dc_potential} holds and Lemma~\ref{lemma:equiva_NE} leads to the following form for $\Phi(\mathbf{x})$ and $\Phi^{\mathrm{SW}}$:
\begin{align*}
\Phi(\mathbf{x})&=-A\sum\nolimits_{k=1}^{K} v\bigl(\mathbf{x}^{(k)}\bigr)
-B\,v\Bigl(\sum\nolimits_{\tilde k=1}^{K}\mathbf{x}^{(\tilde k)}\Bigr),\\
\Phi^{\mathrm{SW}}(\mathbf{x})&=-A\sum\nolimits_{k=1}^{K} v\bigl(\mathbf{x}^{(k)}\bigr)
-KB\,v\Bigl(\sum\nolimits_{\tilde k=1}^{K}\mathbf{x}^{(\tilde k)}\Bigr).
\end{align*}
Both are strictly convex, so the GCE and the SWM can be solved via
Theorem~\ref{thm:GCE-Structural} with $\Phi$ and $\Phi^{\mathrm{SW}}$,
respectively, and each solution is unique.
As in the proof of Lemma~\ref{lem:cha-dc}-ii), both
buses clear at the common LMP $\Gamma q\mathbf{1}^\top\mathbf{x}$, so the two
problems reduce to minimizing
$\Phi(\mathbf{x})+\tfrac{1}{2}\Gamma q^{2}(\mathbf{1}^\top\mathbf{x})^{2}$ and
its $\Phi^{\mathrm{SW}}$ counterpart over $\mathbf{x}\ge\mathbf{0}$, whose
unique minimizers are symmetric across companies and datacenters, i.e.,
$x_j^{\star(k)}=\bar x$ for all $j,k$ under the GCE and
$x_j^{\star(k)}=\bar x^{\mathrm{SW}}$ under the SWM.

Stationarity in any coordinate $x_j^{(k)}$ then requires
\begin{align}
A\Theta(\delta+\bar x)^{\beta\!-\!1}
&\!+\!B\Theta(\delta+K\bar x)^{\beta\!-\!1}
\!=\!\Gamma q^2 K N_{\mathrm D}\,\bar x,\label{eq:ces-positive-gce}\\
A\Theta(\delta+\bar x^{\mathrm{SW}})^{\beta\!-\!1}
&\!+\!KB\Theta(\delta+K\bar x^{\mathrm{SW}})^{\beta\!-\!1}\! =\!\Gamma q^2 K N_{\mathrm D}\,\bar x^{\mathrm{SW}},\label{eq:ces-positive-swm}
\end{align}
for the GCE and the SWM, respectively, where
$\Theta:=\beta N_{\mathrm D}^{\beta/\chi-1}>0$. In each equation the left-hand side is
strictly decreasing in the workload and positive at zero, while the
right-hand side is strictly increasing, so
$\bar x,\bar x^{\mathrm{SW}}>0$ are unique and both minimizers are interior.
Because $K>1$ and $B>0$, the left-hand side
of~\eqref{eq:ces-positive-swm} strictly exceeds that
of~\eqref{eq:ces-positive-gce} when $\bar x^{\mathrm{SW}}$ takes the same value
as $\bar x$, whence
$\bar x^{\mathrm{SW}}>\bar x$ and, since $\lambda_i=\Gamma q K N_{\mathrm D}\bar x$,
$\lambda_i^{\mathrm{SWM}}>\lambda_i^{\mathrm{GCE}}$ for $i=1,2$.
\end{proof}

\section{Proof of Theorem~\ref{thm:nonpotential_UE_sigma}}
\begin{proof}[Proof of Theorem~\ref{thm:nonpotential_UE_sigma}]

    % \begin{enumerate}[label=\roman*), leftmargin=*]
        % \item Merit function for NE. 
        Fix an LMP vector $\boldsymbol{\lambda}\in\mathbb{R}^{N}$ and define the augmented operator
\(
\mathbf{F}_{\boldsymbol{\lambda}}(\mathbf{x})
\;:=\;
-\mathbf{F}(\mathbf{x})\;+\;\bigl(\mathbf{A}^\mathrm{FL}\bigr)^{\!\top}\boldsymbol{\lambda}.
\)
The VI problem~\eqref{eq:vi-dc} can be denoted as  $\operatorname{VI}(\mathcal{X}, \mathbf{F}_{\boldsymbol{\lambda}})$.
Since $-\mathbf{F}$ is {strictly} monotone on the compact, convex set $\mathcal{X}$, $\mathbf{F}_{\boldsymbol{\lambda}}$ is strictly monotone as well.  
Therefore, $\operatorname{VI}(\mathcal{X}, \mathbf{F}_{\boldsymbol{\lambda}})$ has a {unique} solution according to Theorem~2.3.3 in \cite{faechinei2003finite}.  
The merit function for $\operatorname{VI}(\mathcal{X}, \mathbf{F}_{\boldsymbol{\lambda}})$ is as follows:
\[
\Psi_{\boldsymbol{\lambda}}(\mathbf{x})
\;:=\;
\sup_{\tilde{\mathbf{x}}\in\mathcal{X}}
-\mathbf{F}(\mathbf{x})^{\!\top}(\mathbf{x}-\tilde{\mathbf{x}})
\!+\!
\boldsymbol{\lambda}^{\!\top}\!\bigl(\mathbf{A}^\mathrm{FL}\mathbf{x}-\mathbf{A}^\mathrm{FL}\tilde{\mathbf{x}}\bigr).
% \max_{\tilde{\mathbf{x}}\in\mathcal{X}}
% \bigl\langle \mathbf{F}_{\boldsymbol{\lambda}}(\mathbf{x}),\,\mathbf{x}-\tilde{\mathbf{x}}\bigr\rangle .
\] $\Psi_{\boldsymbol{\lambda}}$ is non-negative and $\Psi_{\boldsymbol{\lambda}}(\mathbf{x})=0$ if and only if $\mathbf{x}$ is the solution to $\operatorname{VI}(\mathcal{X}, \mathbf{F}_{\boldsymbol{\lambda}})$; furthermore, strict monotonicity of $\mathbf{F}_{\boldsymbol{\lambda}}$ implies $\Psi_{\boldsymbol{\lambda}}$ is {strictly} convex.  
Consequently, its unique minimizer is \(\mathbf{x}^\mathrm{NE}(\boldsymbol{\lambda})\), resulting in~\eqref{eq:nonpotential_UE}.
% \item Price-load mapping. 
The optimal FL load profile $\mathbf{s}^\star$ is then the minimizer of the following problem:
\begin{subequations}
\begin{align}
\min_{\mathbf{s}\in\mathcal{S}}\min_{\mathbf{x} \in \mathcal{X}} \sup_{\tilde{\mathbf{s}}\in\mathcal{S}}\sup_{\tilde{\mathbf{x}} \in \mathcal{X}}\quad &-{\mathbf{F}(\mathbf{x})}^{\top} (\mathbf{x} -\tilde{\mathbf{x}} ) + \boldsymbol{\lambda}^\top\left(\mathbf{s} - \tilde{\mathbf{s}} \right) \label{obj-ss}
\\
\mathrm{s.t.} \quad  & \mathbf{A}^\mathrm{FL}\mathbf{x} = \mathbf{s}, \quad \mathbf{A}^\mathrm{FL}\tilde{\mathbf{x}} = \tilde{\mathbf{s}}. 
\end{align}%
\end{subequations}
The objective in \eqref{obj-ss} is convex in $\mathbf{x}$ by the standing convexity assumption and, being linear in $\tilde{\mathbf{x}}$ and maximized over the convex fibre $\{\tilde{\mathbf{x}}\in\mathcal{X}:\mathbf{A}^\mathrm{FL}\tilde{\mathbf{x}}=\tilde{\mathbf{s}}\}$, concave in $\tilde{\mathbf{s}}$; since $\{\mathbf{x}\in\mathcal{X}:\mathbf{A}^\mathrm{FL}\mathbf{x}=\mathbf{s}\}$ and $\mathcal{S}$ are compact and convex, Sion's minimax theorem allows us to interchange the operators $\min_{\mathbf{x} \in \mathcal{X}}$ and $ \sup_{\tilde{\mathbf{s}}}$.  Carrying out this swap and invoking the definition of $J(\mathbf{s},\tilde{\mathbf{s}})$ yields exactly~\eqref{non-p:price-load}.
% \item GCE characterization.

Denote $\mathbf{z} = \big(\mathbf{x}^\top,\, \mathbf{g}^\top\big)^\top$, and define the block operator
\[
\mathbf{F}_{\mathbf{C}}(\mathbf{z})
:=
\begin{bmatrix}
    -\mathbf{F}(\mathbf{x}) \\
    \nabla C(\mathbf{g})
    \end{bmatrix},
    \]
    and the convex feasible set \[
\Omega \;:=\;
\bigl\{\mathbf{z}\mid
\mathbf{z} \in \mathcal{Z} = \mathcal{X} \times \mathcal{G},\
\mathbf{p}(\mathbf{z}) = \mathbf{g}-\boldsymbol\ell-\mathbf{A}^\mathrm{FL}\mathbf{x}\!\in\!\mathcal P\bigr\}.\]
Next, we illustrate the relationship between the characterization of GCE and the solution of $\operatorname{VI}(\Omega,\mathbf{F}_{\mathbf{C}})$.
Since $C$ is convex and $-\mathbf{F}$ is monotone, $\mathbf{F}_{\mathbf{C}}$ is a monotone operator on the convex set~$\Omega$.  
By Proposition 1.3.4 of \cite{faechinei2003finite}, a pair $(\mathbf{x}^{\star},\mathbf{g}^{\star})$ solves $\operatorname{VI}(\Omega,\mathbf{F}_{\mathbf{C}})$  \emph{if and only if} there exists a multiplier $\boldsymbol{\lambda}^{\star}$ associated with the coupling constraints \(\mathbf{g}-\boldsymbol\ell-\mathbf{A}^\mathrm{FL}\mathbf{x}\!\in\!\mathcal P\) such that the triplet $(\mathbf{x}^{\star},\mathbf{g}^{\star}, \boldsymbol{\lambda}^{\star})$ satisfies the KKT conditions for the VI, which consist of three parts:
\begin{enumerate}[label=\roman*), leftmargin= *]
\item Stationarity: $\mathbf{0} \in \mathbf{F}_{\mathbf{C}}(\mathbf{z}^\star) - \nabla \mathbf{p}(\mathbf{z}^\star)^\top \boldsymbol{\lambda}^\star + N_{\mathcal{Z}}(\mathbf{z}^\star).$
\item Primal Feasibility: $\mathbf{p}(\mathbf{z}^\star) \in \mathcal{P}.$
\item Complementarity: $-\boldsymbol{\lambda}^\star \in N_{\mathcal{P}}(\mathbf{p}(\mathbf{z}^\star)).$
\end{enumerate}
These conditions map directly to the GCE definition:
\begin{enumerate}[label=\roman*), leftmargin= *]
\item FL system's best response: The pair $(\mathbf{x}^{\star},\boldsymbol{\lambda}^{\star})$ satisfies the KKT conditions for the FL system's NE, expressed as $\operatorname{VI}(\mathcal X,\mathbf{F}_{\boldsymbol{\lambda}^{\star}})$. This implies $\mathbf{x}^{\star}$ is the optimal response to the price $\boldsymbol{\lambda}^{\star}$, and the resulting load profile $\mathbf{s}^{\star}:=\mathbf{A}^\mathrm{FL}\mathbf{x}^{\star}$ satisfies the price-load mapping $\boldsymbol{\sigma}(\boldsymbol{\lambda}^{\star})=\mathbf{s}^{\star}$.
\item Classical CE conditions:
\begin{itemize}[leftmargin=*]
    \item  The pair  $(\mathbf{g}^{\star},\boldsymbol{\lambda}^{\star})$ satisfies the KKT conditions for generators' individual rationality in~\eqref{uni-generator}.
    \item Feasibility holds as $\mathbf{p}^\star = \mathbf{g}^{\star}-\boldsymbol\ell-\mathbf{A}^\mathrm{FL}\mathbf{x}^{\star} \in \mathcal{P}$.
    \item Price optimality holds as $ - \boldsymbol{\lambda}^\star \in  N_{\mathcal{P}}(\mathbf{p}^\star)$.
\end{itemize}
\end{enumerate}
The above conditions, which are the necessary and sufficient KKT conditions for a solution to $\operatorname{VI}(\Omega,\mathbf{F}_{\mathbf{C}})$, are precisely the defining criteria for a GCE.
The tuple $(\mathbf{g}^\star,\mathbf{s}^\star,\boldsymbol{\lambda}^\star)$ derived from the VI solution $(\mathbf{x}^\star, \mathbf{g}^\star, \boldsymbol{\lambda}^\star)$ thus constitutes a GCE.
Conversely, any GCE $(\mathbf{g}^{\star},\mathbf{s}^\star,\boldsymbol{\lambda}^{\star})$, along with the corresponding $\mathbf{x}^\star$ satisfying $\mathbf{s}^\star = \mathbf{A}^\mathrm{FL}\mathbf{x}^\star$, fulfills all KKT conditions above for VI. Therefore, the correspondence between the solutions of $\operatorname{VI}(\Omega,\mathbf{F}_{\mathbf{C}})$ and the set of GCEs is one-to-one.

Finally, we express $\operatorname{VI}(\Omega,\mathbf{F}_{\mathbf{C}})$ as an optimization problem:
\begin{subequations}
\begin{align}
    \min_{\mathbf{x}\in \mathcal{X},\mathbf{g}\in \mathcal{G} }\sup_{\tilde{\mathbf{x}}\in \mathcal{X}, \tilde{\mathbf{g}}\in \mathcal{G} } & \quad -{\mathbf{F}(\mathbf{x})}^\top (\mathbf{x}- \tilde{\mathbf{x}}) + {\nabla C(\mathbf{g})}^\top(\mathbf{g}-\tilde{\mathbf{g}})\\
\mathrm{s.t.} \quad &  \mathbf{p} = \mathbf{g}- \boldsymbol{\ell} - \mathbf{A}^\mathrm{FL} \mathbf{x} \in \mathcal{P}, \\ & \tilde{\mathbf{p}} = \tilde{\mathbf{g}}- \boldsymbol{\ell} - \mathbf{A}^\mathrm{FL} \tilde{\mathbf{x}} \in \mathcal{P},
\end{align}%
\end{subequations}%
which is equivalent to the following form including the variables $(\mathbf{s},\tilde{\mathbf{s}})$:
\begin{subequations}
\begin{align}
    \min_{\mathbf{g}\in \mathcal{G},\mathbf{s}\in\mathcal{S} }\min_{\mathbf{x}\in \mathcal{X}}\sup_{\tilde{\mathbf{g}}\in \mathcal{G},\tilde{\mathbf{s}}\in\mathcal{S} }\sup_{\tilde{\mathbf{x}}\in \mathcal{X}} & \quad -{\mathbf{F}(\mathbf{x})}^\top (\mathbf{x}- \tilde{\mathbf{x}}) + {\nabla C(\mathbf{g})}^\top(\mathbf{g}-\tilde{\mathbf{g}})
    \label{obj-gce-np}\\
\mathrm{s.t.} \quad &  \mathbf{s} = \mathbf{A}^{\mathrm{FL}}\mathbf{x},\\
&\tilde{\mathbf{s}} = \mathbf{A}^{\mathrm{FL}}\tilde{\mathbf{x}},\\
& \mathbf{p} = \mathbf{g}- \boldsymbol{\ell} - \mathbf{s} \in \mathcal{P}, \\ & \tilde{\mathbf{p}} = \tilde{\mathbf{g}}- \boldsymbol{\ell} -  \tilde{\mathbf{s}} \in \mathcal{P}.
\end{align}%
\end{subequations}%
By the standing convexity assumption, the objective in~\eqref{obj-gce-np} is convex in $(\mathbf{x},\mathbf{g})$ and concave in $(\tilde{\mathbf{s}},\tilde{\mathbf{g}})$ over compact convex sets, so by Sion's minimax theorem the operators $\min_{\mathbf{x}\in \mathcal{X}}$ and $\sup_{\tilde{\mathbf{g}}\in \mathcal{G},\tilde{\mathbf{s}}\in\mathcal{S} }$ can be swapped, 
which yields exactly~\eqref{eq:GCE-minmax} with the definition of  $J(\mathbf{s},\tilde{\mathbf{s}})$.
\end{proof}

\end{document}